\documentclass[twoside,11pt]{article}
\usepackage{jmlr2e}
\usepackage{bm}

\usepackage[english]{babel}

\usepackage{url}

\usepackage{imakeidx}
\makeindex[columns=2, title=Alphabetical Index]

\usepackage{stackengine}

\usepackage{enumitem}
\setenumerate[1]{itemsep=0pt,partopsep=0pt,parsep=\parskip,topsep=5pt}
\setitemize[1]{itemsep=0pt,partopsep=0pt,parsep=\parskip,topsep=5pt}
\setdescription{itemsep=0pt,partopsep=0pt,parsep=\parskip,topsep=5pt}

\usepackage{dsfont}

\newcommand{\Var}{\mathrm{Var}}

\usepackage[framemethod=tikz]{mdframed}

\usepackage{amsmath}
\usepackage{arydshln}

\numberwithin{equation}{section}

\usepackage{amsmath}
\usepackage{blkarray}

\usepackage[final]{pdfpages}

\usepackage[noframe]{showframe}
\usepackage{framed}
\usepackage{lipsum}
\definecolor{color0}  {RGB}{174,225,254} 
\definecolor{color1}  {RGB}{220,227,248} 
\definecolor{color2}  {RGB}{28,130,185} 
\definecolor{color3}  {RGB}{255,253,250} 

\definecolor{colormiddleright}  {RGB}{245,253,250} 
\definecolor{colorbottomleft}  {RGB}{255,243,250} 
\definecolor{coloruppermiddle}  {RGB}{255,253,230} 
\definecolor{colormiddleleft}  {RGB}{255,253,250}

\definecolor{colorcr}  {RGB}{249,253,232} 
\definecolor{colorreduction}  {RGB}{255,235,254} 
\definecolor{colorqr}  {RGB}{254,221,199} 
\definecolor{colorbiconjugate}  {RGB}{251,149,161} 

\definecolor{colorsvd}  {RGB}{215,247,235} 

\definecolor{colorupperright}  {RGB}{239,246,251} 
\definecolor{colorspectral}  {RGB}{206,226,243} 

\definecolor{colorbottomright}  {RGB}{220,224,236} 
\definecolor{coloreigenvalue}  {RGB}{197,203,224} 

\definecolor{colorupperleft}  {RGB}{235,243,240} 
\definecolor{colorsemidefinite}  {RGB}{217,232,226} 

\definecolor{colormiddle} {RGB}{235, 240,255}
\definecolor{colorlu}  {RGB}{220,227,255} 

\definecolor{colorals}  {RGB}{240,230,255} 
\definecolor{coloralsbkg}  {RGB}{248,243,255} 

\definecolor{canaryyellow}{rgb}{1.0, 0.75, 0.0}
\definecolor{bluepigment}{rgb}{0.0, 0.0, 1.0}

\definecolor{canarypurple}{RGB}{208, 13, 241}

{\endMakeFramed}
\definecolor{shadecolor}{gray}{0.75}

\usepackage{xcolor}

\usepackage{tcolorbox}

\usepackage{graphicx}  
\usepackage[hang]{subfigure}

\usepackage{algorithm}
\usepackage{algpseudocode}
\usepackage{amsmath}
\usepackage{graphics}
\usepackage{epsfig}

\usepackage{blkarray}

\usepackage{listings}

\usepackage[colorlinks]{hyperref}

\usepackage{color}   
\definecolor{winestain}{rgb}{0.5,0,0}
\usepackage{hyperref}
\usepackage[]{hyperref}
\hypersetup{
    colorlinks=true, 
    linktoc=all,     
    linkcolor=red,  
    anchorcolor=blue,
    citecolor=green,
}
\usepackage[hyperpageref]{backref} 

\usepackage{booktabs}  

\usepackage{tikz}
\usetikzlibrary{mindmap,trees,backgrounds}
\usetikzlibrary{arrows.meta}

\usetikzlibrary{decorations.text}

\usetikzlibrary{calc,backgrounds}

\usepackage{changepage}                 

\newlength{\offsetpage}
	{\end{adjustwidth}}

\usepackage{hhline}

\usetikzlibrary{positioning,shapes,shadows,arrows}
\tikzstyle{condition}=[rectangle, draw=black, rounded corners, fill=colorqr, drop shadow,
text centered, anchor=north, text=black, text width=3cm]
\tikzstyle{abstract}=[rectangle, draw=black, rounded corners, fill=blue!30, drop shadow,
text centered, anchor=north, text=black, text width=3cm]
\tikzstyle{comment}=[rectangle, draw=black, rounded corners, fill=color1, drop shadow,
text centered, anchor=north, text=black, text width=3cm]
\tikzstyle{myarrow}=[->, >=open triangle 90, thick]
\tikzstyle{line}=[-, thick]

\usepackage{sidecap}
\sidecaptionvpos{figure}{t}
\usepackage{verbatimbox}

\usepackage[labelfont=bf]{caption}
\newcommand\myhrulefill[1]{\leavevmode\leaders\hrule height#1\hfill\kern0pt}
\DeclareCaptionFormat{myformat}{{\color[RGB]{0,0,0}\myhrulefill{0.08em}}\\#1#2#3}
\usepackage{graphicx}
\usepackage{pythonhighlight}

\newcommand{\leadtosmall}{\,\,\underrightarrow{ \text{leads to} }\,\,}

\newcommand{\Exp}{\mathrm{E}}
\newcommand{\normal}{\mathcal{N}}
\newcommand{\gap}{\,\,\,\,\,\,\,\,}

\jmlrheading{1}{2022}{1-48}{4/00}{10/00}{Jun Lu}

\ShortHeadings{Exploring Classic Quantitative Strategies}{Jun Lu}
\firstpageno{1}

\begin{document}

\title{Exploring Classic Quantitative Strategies}

\author{
\begin{center}
\name Jun Lu \\ 
\email jun.lu.locky@gmail.com
\end{center}
}


\maketitle

\begin{abstract}
The goal of this paper is to debunk and dispel the magic behind the black-box quantitative strategies.
It aims to build a solid foundation on how and why the techniques work.
This manuscript crystallizes this knowledge by deriving from simple intuitions, the mathematics behind the strategies. 
This tutorial doesn't shy away from addressing both the formal and informal aspects of quantitative strategies.
By doing so, it hopes to provide readers with a deeper understanding of these techniques as well as the when, the how and the why of applying these techniques. 
The strategies are presented in terms of both S\&P500 and SH510300 data sets. However, the results from the tests are just examples of how the methods work; no claim is made on the suggestion of real market positions.

\end{abstract}


\begingroup
\hypersetup{linkcolor=winestain}
\tableofcontents
\endgroup

\section{Performance Measures}
In this section, we shortly review measures for the performance that will be useful for evaluating the strategies in the sequel. For more measurements, one can also refer to \citep{investopedia2022}.
\subsection{Rate of Return (RR)}
In finance, \textit{return} is a profit on an investment, and a \textit{loss} instead of a profit is described as a \textit{negative return}, assuming the amount invested is greater than zero. Then the \textit{total profit (TP)} is defined to represent the profitability of all the transactions; see Equation~\eqref{equation:total-profit}. We note that when the loss is greater than the profit, TP can be negative. 
We use RR to express the return or loss of investment in a given time period as a percentage of the investment amount; see 
Equation~\eqref{equation:rr-rate-raturn} where ``INVEST" is the initial amount of investment.
\begin{align}
\text{TP} &= \text{all returns} - \text{all losses};\label{equation:total-profit}\\
\text{RR} &=\text{TP/INVEST} \times 100. \label{equation:rr-rate-raturn}
\end{align}
We realize that the definition of the RR depends on a time period. Typically, the period of time is a year by default, in which case the RR is also called the \textit{annualized return}; and the conversion process is called \textit{annualization}. The conversion procedure can be implied from the context and we shall not give the details.

\subsection{Volatility of Returns (VOL)} In statistics, the dispersion of a sequence can be defined by its standard deviation. While, in finance, the dispersion of returns for a given security or market index is known as the \textit{volatility of returns} and is defined as the standard deviations of \textbf{returns in percentage} \footnote{In the following sections of this article, returns represent profits or losses (positive or negative); returns in percentage are the profits or losses divided by the amount of investment.} (i.e., returns or losses divided by the amount of investment). In most cases, the higher the volatility, the riskier the security. The volatility of returns measures the fluctuation of returns day over day, and it is taken as a risk measure generally. Volatility is sometimes measured by the variance among the returns in percentage. However, we will only consider the standard deviation version of it in the sequel.

\subsection{Maximum Drawdown}
A  maximum drawdown (MDD) is the maximum observed loss from a peak to a trough of a portfolio, before a new peak is attained. The MDD is an indicator of downside risk over a specified time period and is usually expressed in percentage terms. The smaller the MDD, the smaller the risk.


\subsection{Sharpe Ratio (SR)} The Sharpe ratio, originally called the \textit{reward-to-variability ratio}, measures the performance of an investment such as a security or a portfolio compared to a \textit{risk-free asset}, say, annualized return with 5\%\footnote{For simplicity, we will set the risk-free asset to be 0 in our strategy measurements.}. The Sharpe ratio discounts the expected excess returns of a portfolio by the volatility of the returns, i.e., measures the \textbf{excess return per unit of deviation} in an investment asset or a trading strategy. The \textit{ex-ante Sharpe ratio} is defined as follows
\begin{equation}\label{equation:sharpe-ratio}
	\text{SR} = \frac{\Exp[R_a- R_f]}{\sigma} =\frac{\Exp[R_a- R_f]}{\sqrt{\Var[R_a - R_f]}} ,
\end{equation}
where $\sigma = \sqrt{\Var[R_a - R_f]}$,
$R_a$ is the sequence of the asset returns, and $R_f$ is the sequence of the risk-free asset returns \footnote{Such as the U.S. Treasury security, or simply $5\%$ per year.} or the return of the benchmark asset in general as long as the index of $R_a$ and $R_f$ are consistent (e.g., in days if the return is measured daily). 
Note, if $R_f$ is a constant risk-free return through the period, $\sqrt{\Var[R_a - R_f]} = \sqrt{\Var[R_a]}$.
To see what's in the measure, we realize $\Exp[R_a- R_f]$ is the expected value of the excess of the asset return over the benchmark return, and $\sigma$ is the standard deviation of the excess return which is regarded as a proxy for risk. 
It characterizes \textbf{how well the return of an asset compensates the investor for the risk taken} and provides a \textbf{combined performance of risk and return}. When comparing two assets to a same benchmark, the one with a higher Sharpe ratio is considered to provide a better return under the same risk conditions. 
If we graph the return-risk figure with the measure of return in the vertical axis and the measure of risk in the horizontal axis, then the Sharpe ratio simply measures the gradient of the line from the risk-free rate to the combined
return and risk of each asset (or portfolio). Thus, the steeper the gradient, the higher the Sharpe ratio, and the better
the combined performance of risk and return. See also discussion in \citep{bloch2014practical}.

The \textit{ex-post Sharpe ratio} uses the same equation as the one above but with realized returns of the asset and benchmark rather than the expected returns.
\begin{remark}[Ex-Post Sharpe Ratio]
The Sharpe ratio can be recalculated at the end of the year to examine the actual return rather than the expected return:
\begin{equation}\label{equation:sharpe-ratio2}
	\text{SR} = \frac{r_a- r_f}{\sigma_a},
\end{equation}
where $r_a$ is the annualized asset return, $\sigma_a$ is the standard deviation of return, and $r_f$ is the annualized risk-free return. 

Suppose further we have the daily asset return $\widehat{r}_a$ whose standard deviation of return is $\widehat{\sigma}_a$, and daily risk-free return $\widehat{r}_f$. Let the trading days per year be $N$, then the \textit{annualization} process is to transfer the return and standard deviation to be the annual ones:
\begin{equation}\label{equation:sharpe-ratio-annualization-expost}
	\text{SR} = \frac{\widehat{r}_a- \widehat{r}_f}{\widehat{\sigma}_a} \times \sqrt{N}.
\end{equation}
The above equation results from the annualized asset return $r_a =\widehat{r}_a \cdot N$, $r_f=\widehat{r}_f \cdot N$, and annualized standard deviation $\sigma_a = \widehat{\sigma}_a \cdot  \sqrt{N}$.

Specifically, the average number of trading days for the U.S. markets is about 252 days per year, but not every year has 252 trading days. For example, the 2020 trading year consists of 253 trading days. For simplicity, we will use 252 days per year in our examples. And the Python code for calculating the annualized Sharpe ratio is shown as follows (terse comments begin with a \# or are inside ``````$\cdot$""").
\end{remark}

\begin{python}
def sharpe_ratio_annual(returns, rf=0.05/252):
	"""
	:param returns: DAILY returns in percentage
	:param rf:      DAILY risk-free return
	:return:        annualized Sharpe ratio
	"""
	er = np.mean(returns) # Calculate the expected return
	return (er - rf) / np.std(returns) * np.sqrt(252) # trading days is 252
\end{python}

\noindent In our test, the risk-free return is set to be 0 for simplicity. 
\begin{itemize}
\item Usually, any Sharpe ratio greater than 1.0 is considered good at an industrial level;
\item A ratio higher than 2.0 is rated as very good;
\item A ratio of 3.0 or higher is considered excellent;
\item A ratio under 1.0 is considered sub-optimal.
\end{itemize}
However, in our test, a Sharpe ratio greater than 0.8 (or even 0.7) is also acceptable.
 
\paragraph{History} The Sharpe ratio was introduced in 1966 by William Sharpe as an extension of the \textit{Treynor ratio} \citep{sharpe1966mutual, jobson1981performance, cadsby1986performance}. Sharpe originally called it the ``reward-to-variability" ratio and later known as the Sharpe ratio by later financial operators.
The original definition was 
\begin{equation}\label{equation:sharpe-ratio-ori}
	\text{SR} =\frac{\Exp[R_a- R_f]}{\sqrt{\Var[R_a ]}} ,
\end{equation}
where $R_f$ is a constant risk-free return throughout the time period.
And the definition in Equation~\eqref{equation:sharpe-ratio} is due to Sharpe's revision in \citep{sharpe1994sharpe} where Sharpe acknowledged that the basis of the comparison should be an applicable benchmark that changes with time, i.e., $R_f$ is not a constant.

%
%
%

\subsection{Information Ratio (IR)}

The information ratio (IR) is a measurement of portfolio returns above the returns of a benchmark, usually an index such as the S\&P500, to the volatility of those returns. In this sense, the information ratio is just the same as the revised version of the Sharpe ratio (Equation~\eqref{equation:sharpe-ratio}).

The information ratio is used to evaluate the skill of a portfolio manager at generating returns in excess of a given benchmark.
A higher IR result implies a better portfolio manager who's achieving a higher return in excess of the benchmark, given the risk taken.

As mentioned above, the information ratio is similar to the Sharpe ratio, the main difference being that the Sharpe ratio uses a risk-free return as the benchmark (such as a U.S. Treasury security) whereas the information ratio uses a risky index as the benchmark (such as the S\&P500). The Sharpe ratio is useful for an attribution of the absolute returns of a portfolio, and the information ratio is useful for an attribution of the relative returns of a portfolio.

The information ratio is a benchmark-relative statistic.
It is entirely possible for a manager to have a high
information ratio, but still exhibit significant losses if the
benchmark is down. The Python code for calculating the annualized Information ratio is shown as follows:
\begin{python}
def information_ratio_annual(returns, benchmark):
	"""
	:param returns: DAILY returns in percentage
	:param benchmark: DAILY benchmark return, e.g., S&P 500
	:return: annualized Information ratio
	"""
	diff = returns - benchmark # Calculate the difference
	return np.mean(diff) / np.std(diff) * np.sqrt(252)
\end{python}
As one can tell, the higher the information ratio, the better. If the information ratio is less than zero, it means the active
manager failed on the first objective of outperforming the benchmark. Of all the performance statistics, the information ratio is one of the most difficult hurdles to clear. Generally speaking,
\begin{itemize}
\item An information ratio between 0.40 and 0.60 is considered quite good;
\item An information ratio between 0.61 and 1 is considered a great investment;
\item An information ratio of 1.00 for a long period of time is rare.
\end{itemize}
Typical
values for information ratios vary by asset class.

\subsection{Fundamental Analysis vs Technical Analysis}
There are many different ways to assess the value of a company/security, and the methods used to analyze securities and make investment decisions fall into two very broad categories: fundamental analysis and technical analysis. 
Fundamental analysis is a method of evaluating a security that includes measuring its intrinsic value by examining related economic, financial, and other qualitative and quantitative factors. Fundamental analysts attempt to study everything that can affect the company's value, including macroeconomic factors (like the overall economy, economic period, and industry conditions) and company-specific factors (like financial condition, company size, and management). However, technical analysis takes a completely different approach. It is a method of evaluating assets by analyzing statistics generated by market signals or indicators, such as past prices, volume, and liquidity (total volumes in a specific time frame). Technical analysts do not attempt to measure a company's intrinsic value, but instead, use algorithms and other methods to identify patterns that can suggest future positions.

\begin{SCfigure}
	\centering
	\includegraphics[width=0.6\textwidth]{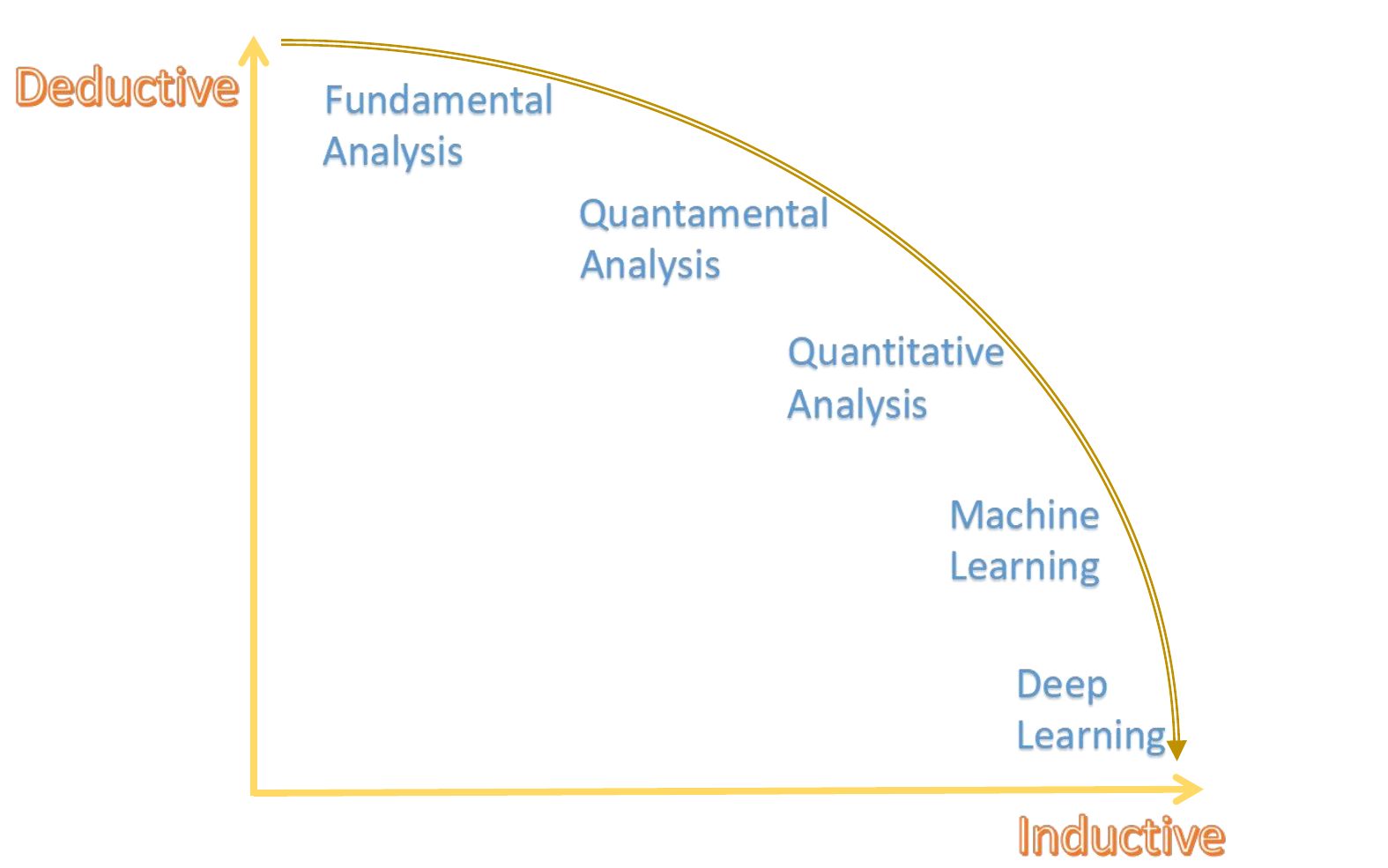}
	\caption{Demonstration of where the quantitative strategies lie in the assessment of the market.}
	\label{fig:quantanaly}
\end{SCfigure}

To be more specific, most of the technical strategies in the sequel lie in between the deductive and inductive analysis as shown in Figure~\ref{fig:quantanaly}. While we shall shed light on how to apply machine learning or deep learning techniques to find new strategies.

%

\subsection{The Kelly Criterion and Optimal Betting}
Given a gambling game, for 1 unit of investment, there is a probability of $p$ to obtain additional positive return $L$, and there is a probability of $q = 1-p$ to obtain additional negative return $M$. Suppose further the proportion of each investment is $x$ ($0\leq x\leq 1$), we want to maximize the expected logarithmic rate of return as shown in the following problem:
\begin{equation}
	\max \log f(x) = p \log(1+Lx) + q\log (1-Mx) \gap \text{s.t. }0 \leq x \leq 1. 
\end{equation}
The problem can be equivalently categorized as follows, since the log function is monotone:
\begin{equation}
	\max f(x) = (1+Lx)^p (1-Mx)^q \gap \text{s.t. }0 \leq x \leq 1. 
\end{equation}
Taking the gradient and setting to 0, it follows that 
\begin{equation}
\begin{aligned}
	\max f(x)^\prime &= Lp(1+Lx)^{p-1} (1-Mx)^q - Mq  (1+Lx)^p (1-Mx)^{q-1} = 0;\\
\leadtosmall \gap	0 &=f(x)  \left\{ \frac{Lp}{1+Lx} - \frac{M(1-p)}{1-Mx}\right\}.
\end{aligned}
\end{equation}
Since $f(x)$ is assumed to be positive, we have $\left\{ \frac{Lp}{1+Lx} - \frac{M(1-p)}{1-Mx}\right\} = 0$, and $x=\frac{Lp-Mq}{LM}\in [0,1]$. Let $b = \frac{L}{M}$, we have 
$$
x = \frac{bp-q}{bM},
$$
where $b$ can be understood as the proportion of the bet gained with a win. E.g., if betting \$10 on a 2-to-1 odds bet, (upon win you are returned \$30, winning you \$20), then $b=\$20/\$10=2.0$.
Notice that this expression reduces to the \textit{simple gambling formula} (or known as the black jack example) when $M=1=100\%$, when a loss results in full loss of the wager.

\begin{SCfigure}
	\centering
	\includegraphics[width=0.6\textwidth]{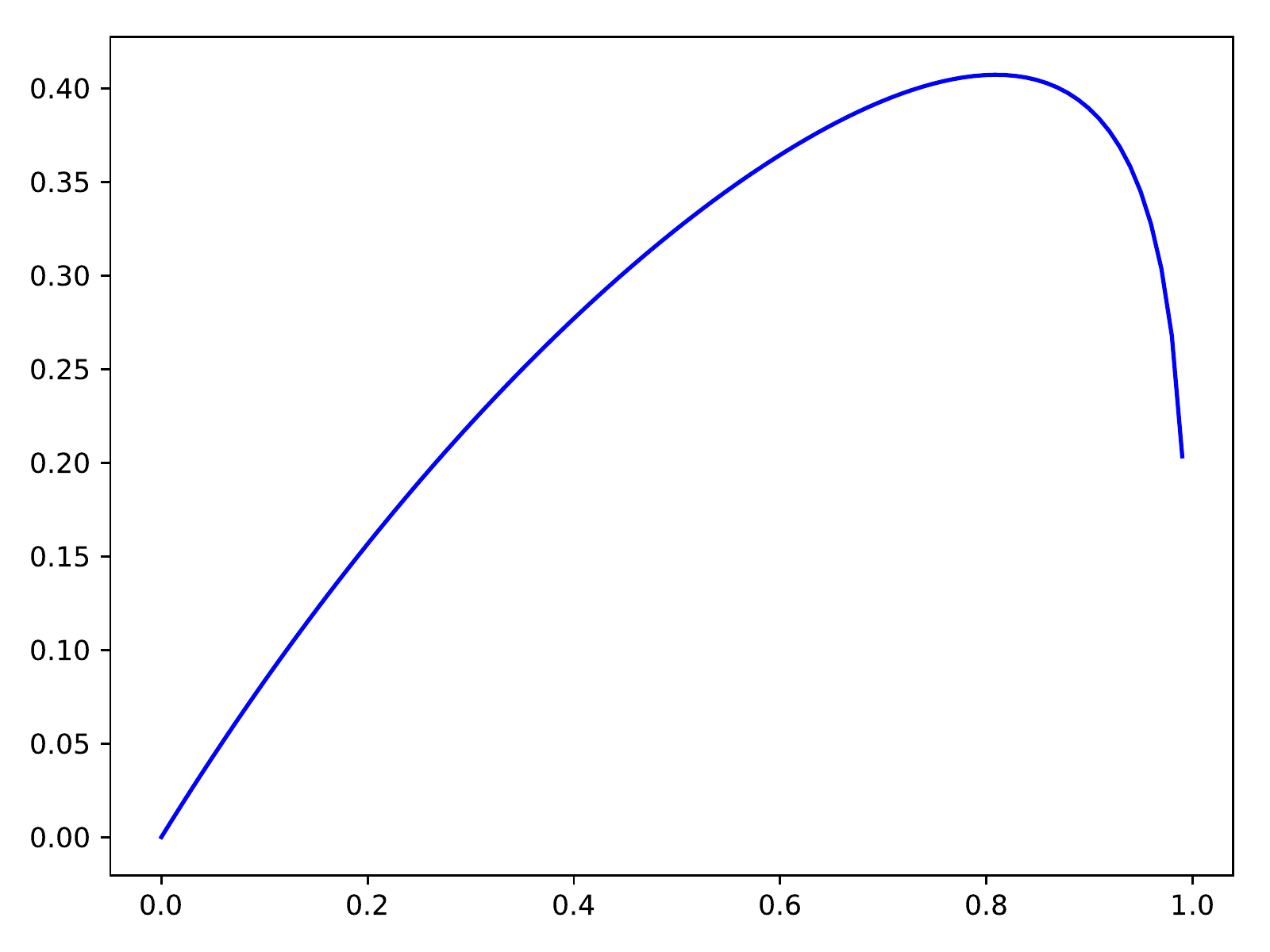}
	\caption{Logarithmic returns of different $x$'s when $p=0.9, q=0.1$, $L=1.1, M=1$.}
	\label{fig:kelly12}
\end{SCfigure}

Come to the black jack example, where $p=0.9, q=0.1$, $L=1.1, M=1$, and the optimal proportion of investment is $x=0.81$. We compare the different expected logarithmic return of different $x$'s in Figure~\ref{fig:kelly12}. The figure shows Kelly Criterion, $x=0.81$, obtains best expected return.

The Kelly Criterion is original published in \citep{kelly2011new}. As Edward O. Thorp, an American mathematician, recently suggested, if one bets half of Kelly's formula every time, it will result in about three-quarters of the return, while the volatility is only half. The could be a guidance to reduce the risk of a strategic system.


\section{Two-Average Strategy}
In the following sections, we first introduce some moving average methods that will be often useful. For a more detailed discussion on the indicators, e.g., least squares MA and DEMA, one can also refer to \citep{fmlab2020}.
\subsection{Simple Moving Average (SMA)}
Moving average (MA), an indicator that cannot be bypassed in morphological analysis. The moving averages are used to smooth the data in an array (e.g., the daily closing prices of a stock) to help eliminate noise and identify trends. If the MA line is angled up, an upward trend is underway; and if the MA line is angled down, a downward trend is ongoing. However, moving averages don't make predictions about the future value of a stock; they simply reveal what the price is doing, on average, over a period of time.

The simple moving average (SMA) is literally the simplest form of a moving average. Each output value is the average of the previous $N$ values where $N$ is known as the time period to smooth the array. 
Given the time period $N$, the $i$-th element of the SMA is defined as follows:
\begin{equation}\label{equation:SMA}
\left\{
\begin{aligned}
	\text{SMA}[i] &= \frac{\text{input}[i-N+1:i]}{N}, \gap &\text{if $i>N$};\\
	\text{SMA}[i] &= \text{not a number, or input}[i], \gap &\text{if $i\leq N$},\\
\end{aligned}
\right.
\end{equation}
where $\text{SMA}[i]$ indicates the $i$-th entry of the SMA array, input$[i-N+1:i]$ is the input array sliced  by the index ``$i-N+1:i$", i.e., an array with a length of $N$ (to avoid confusion, the $(i-N+1)$-th and the $i$-th elements are included). 
When $N$ is small, the SMA uses a small time period to indicate the trading trends that will potentially contain ``noise"; however, this signal can reflect the upward and downward trend of the market quickly. While if $N$ is large, the signal contains less noise; on the other hand, the trend is smoothed, unfortunately.

In the SMA, each value in the time period carries equal weight, and values outside of the time period are not included in the average. This makes it less responsive to recent changes in the data, which can be useful for filtering out those changes.

\paragraph{History} The moving average was first proposed by Joseph E. Granville, an American financial writer and investment seminar speaker, in the mid-20th century. It is still widely used by people and has become an important indicator to judge the trading signal. 
It is now known as the SMA line because it is usually presented in linear form, and the SMA with a time period of 5 days is usually shown as ``MA5" in various stockbroker softwares. Similar for the ``MA10, MA20".
From a statistical point of view, the moving average is the average value of historical (closing) prices, which can represent the average trend of stock prices in the past $N$ days.
In 1963, Joseph E. Granville put forward the famous eight trading rules of Granville in his book \citep{granville2018granville} whose core discussion is the well-known \textit{on-balance volume (OBV)}. The method is simple and effective. Once proposed, it is quickly sought after by the market. In particular, the \textit{golden cross} and \textit{dead cross} signals are still used today, which we will shortly see.

\subsection{Exponential Moving Average (EMA)}
Going further,
the exponential moving average (EMA) is a staple of financial analysis and is used in countless technical indicators. In an SMA, each value in the time period carries equal weight, and values outside of the time period are not included in the average. However, the EMA is a cumulative indicator, including all data. Past values have a diminishing contribution to the average, while more recent values have a greater contribution. This method allows the moving average to be more responsive to changes in the data.
Given a time period $N$, the EMA is defined as follows:
\begin{equation}\label{equation:EMA}
\left\{\begin{aligned}
	K  &=\frac{s}{N+1};\\
	\text{EMA}[i] &= K \times \text{input}[i] + (1-K)\times \text{EMA}[i-1], 
\end{aligned}
\right.
\end{equation}
where $\text{EMA}[i-1]$ is the EMA of the previous period (e.g., EMA of last day is the data in a daily manner). From the definition above, the $i$-th element of the EMA array is also regarded as the weighted average of the $\text{input}[i]$ and the previous EMA element. And the value $s$ in the above equation is referred to as the ``smoothing factor" (or ``smoothing constant") that is usually taken as 2 so that the calculation for EMA puts more emphasis on the recent data points. In rare cases, the input[$i$] in Equation~\eqref{equation:EMA} can also be obtained by the SMA of the input sequence:
$$
\text{EMA}[i] = K \times \text{\textcolor{blue}{SMA(input)}}[i] + (1-K)\times \text{EMA}[i-1].
$$
However, we will only use the definition in the original Equation~\eqref{equation:EMA}.

To conclude, the EMA assigns a relatively higher weighting to the recent data point, and as a result, it stays closer to the price action than an SMA and reduces the lag.
\subsection{Adaptive Moving Average (AMA)}
When carefully notice that, for both the SMA and the EMA, when the $N$ is small, the SMA and EMA use a small time period to indicate the trading trends that will potentially contain ``noise"; however, this signal can reflect the upward and downward trend of the market effectively. 
While if $N$ is large, the signal contains less noise; on the other hand, the trend is smoothed and the upward/downward trends may be delayed, unfortunately.
One of the disadvantages of different smoothing MA algorithms for price series is that accidental price leaps can result in the appearance of false trend signals. On the other hand, smoothing leads to the unavoidable lag of a signal about trend stop or change. 
Although EMA reduces the lag potentially, the EMA still fails to address the problem that trading signals will lead to a large number of losing trades. 

One method of addressing the disadvantages of MAs is to multiply the weighting factor by a \textit{volatility ratio}. 
Doing this would mean that the MA would be closer to the current price in an angled market (i.e., favor a short time period). This would allow the traders to keep track of the trend for either selling or buying. 
As a trend comes to an end and prices start to be smoothed or volatile, the MA would move further from the current market action and, in theory, allow the winners (either going up or going down) to run (i.e., favor a long time period).

%
%

To rephrase, it is reasonable to favor a MA with a smaller time period when the price/array is moving in a certain direction;  and a larger time period when the price/array becomes smooth or volatile. That is, when there are no certain upward/downward trends, the MA line should be moving smoothly, like what the long time period MA does; while there is a certain trend, the MA line must reflect the change quickly, like the short time period MA does.

An adaptive moving average (AMA) is one more moving average overlay, just like the EMA. It changes its sensitivity to price fluctuations. The adaptive moving average becomes more sensitive during periods when the price is moving in a certain direction and becomes less sensitive to price movement when the price is smooth or volatile.

Perry Kaufman suggested replacing the ``weight" variable in the EMA formula with a constant based on the \textit{efficiency ratio (ER)} \citep{kaufman2013trading, kaufman1995smarter}. This indicator is designed to measure the \textbf{strength of a trend}, defined within a range from -1.0 to +1.0. It is calculated with a simple formula:
\begin{equation}
\text{ER}[i]  = \frac{\text{Signal}[i]}{\text{Noise}[i]}= \frac{\text{Total price change for a period}}{\text{Sum of absolute price change for each bar}}.
\end{equation}
To avoid confusion, we call the period in the above equation an ``adaptive window length" (AdaWin) to differentiate from the time period for the MAs. Given the AdaWin=$M$, we have
\begin{itemize}
\item ER[$i$] is the current value of the efficiency ratio; 
\item Signal[$i$] = input[$i$] $-$ input[$i - M$] is the current signal value, i.e., difference between the current input price and price $M$ periods ago;
\item Noise[$i$] = Sum(ABS(input[$i$] $-$ input[$i-1$]),$M$) is current noise value, i.e., sum of absolute values of the difference between the price of the current period and price of the previous period for $M$ periods.
\end{itemize}
That is,
\begin{equation}
	\text{ER}[i]  = \frac{\text{Signal}[i]}{\text{Noise}[i]}= 
	\frac{\text{ABS}(\text{input}[i] - \text{input}[i-M]) }{\sum_{k=i-M}^{i} \text{ABS}(\text{input}[k] - \text{input}[k-1])}
\end{equation}
At a strong trend (i.e., the input price is moving in a certain direction, up or down) the ER will tend to 1; if there is no directed movement, it will be a little more than 0.


\subsection*{AMA with EMA}
 The obtained value of ER is used in the exponential smoothing formula. Given the original EMA calculation,
\begin{equation}\label{equation:EMA-fro-AMA}
\left\{\begin{aligned}
	\text{SC}  &=\frac{2}{N+1};\\
	\text{EMA}[i] &= \text{SC} \times \text{input}[i] + (1-\text{SC} )\times \text{EMA}[i-1], 
\end{aligned}
\right.
\end{equation}
where SC again is the EMA smoothing constant, $N$ is the period of the exponential moving, $\text{EMA}[i-1]$ is the previous value of EMA. \textbf{Now, what we want to go further is to set the time period $N$ to be a smaller value \footnote{Called the TimeperiodShort in the following Python code.} when the ER tends to 1 in absolute value; or a larger value \footnote{Called the TimeperiodLong in the following Python code.} when the ER moves towards 0.} When $N$ is small, $\text{SC} $ is known as a ``fast $\text{SC} $"; otherwise, $\text{SC} $ is known as a ``slow $\text{SC} $". 

For example, let the large time period be 30, and the small time period be 2.
The smoothing ratio for the fast market must be as for EMA with period 2 (``fast SC" = 2/(2+1) = 0.6667), and for the period of no trend EMA period must be equal to 30 (``slow SC" = 2/(30+1) = 0.06452). 
Thus the new changing smoothing constant is introduced, called the ``scaled smoothing constant" (SSC):

\begin{equation}
	\text{SSC}[i] = \text{ER}[i] \times  ( \text{fast SC} - \text{slow SC}) + \text{slow SC}
\end{equation}
For a more efficient influence of the obtained smoothing constant on the averaging period Kaufman recommended squaring it.
The final calculation formula then follows:
\begin{equation}
\text{AMA}[i] =  (\textcolor{blue}{\text{SSC}[i]^2})\times \text{input}[i]  +  \left(\textcolor{blue}{1-\text{SSC}[i]^2}\right)\times \text{AMA}[i-1],
\end{equation}
or after rearrangement:
$$
\text{AMA}[i] = \text{AMA}[i-1] + (\text{SSC}[i]^2) \times (\text{input}[i] -  \text{AMA}[i-1]),
$$
where AMA$[i]$ is the current value of AMA, AMA$[i-1]$ is the previous value of AMA, and SSC$[i]$ is the current value of the scaled smoothing constant.

\subsection*{AMA with SMA}
Although the AMA is extensively applied over the EMA, the SMA version of AMA also gives promising results in our tests compared to the EMA version of AMA. This is partly because the AMA from EMA has been extensively used in the industry, and the signal might disappear in the recent markets; while, AMA from SMA is not that famous at the moment. To our best knowledge, there is no article about the SMA version of AMA. We shall give the formula as follows; and readers are highly recommended to apply the AMA idea for other moving averages, such as the DEMA, TEMA, WMA, T3, and so on.

\begin{figure}[h]
	\centering  
	\vspace{-0.35cm} 
	\subfigtopskip=2pt 
	\subfigbottomskip=2pt 
	\subfigcapskip=-5pt 
	\subfigure[MAs for S\&P500.]{\label{fig:sp500-sma-ema}
		\includegraphics[width=0.47\linewidth]{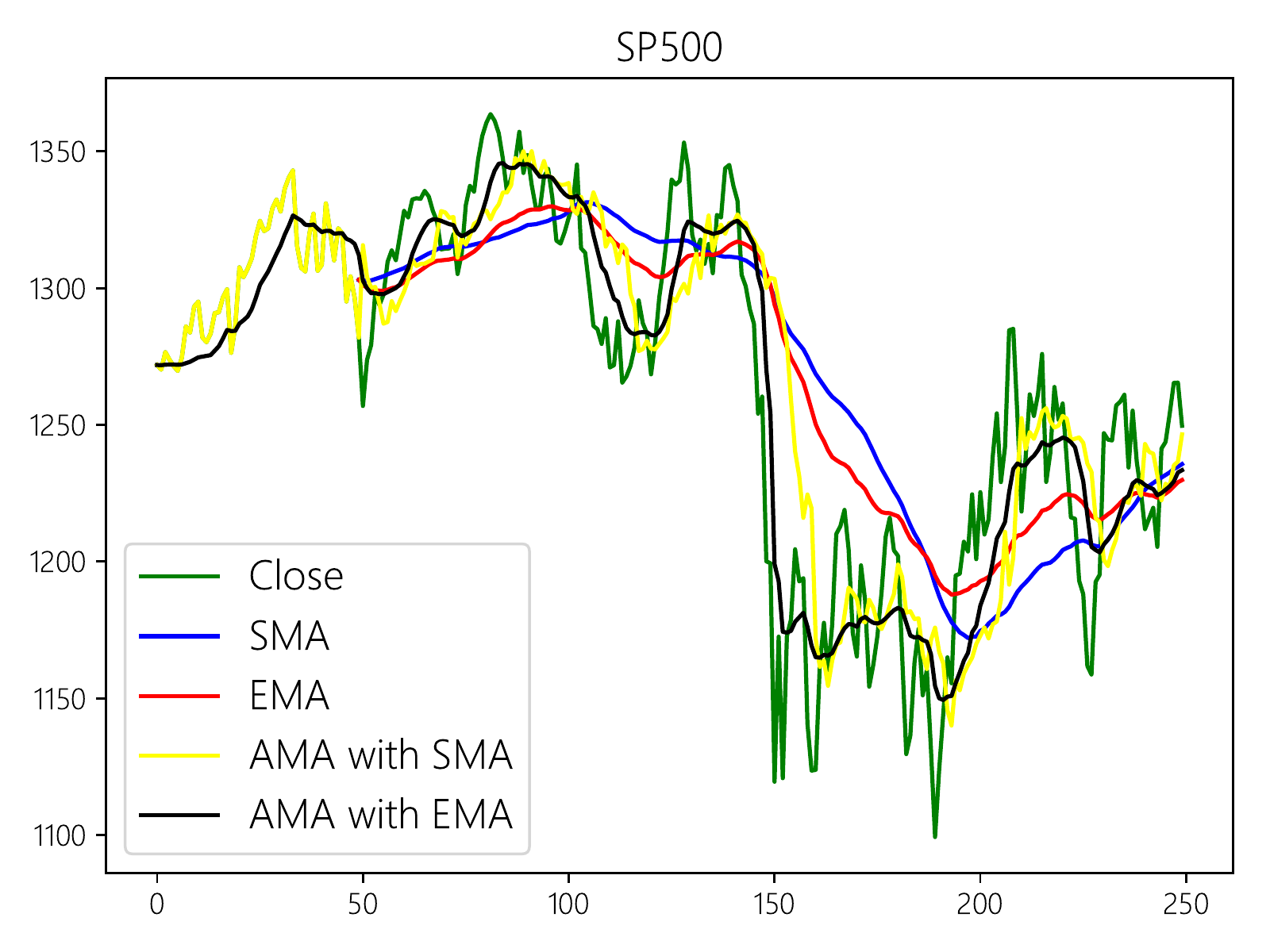}}
	\quad 
	\subfigure[MAs for SH510300.]{\label{fig:sh510300-sma-ema}
		\includegraphics[width=0.47\linewidth]{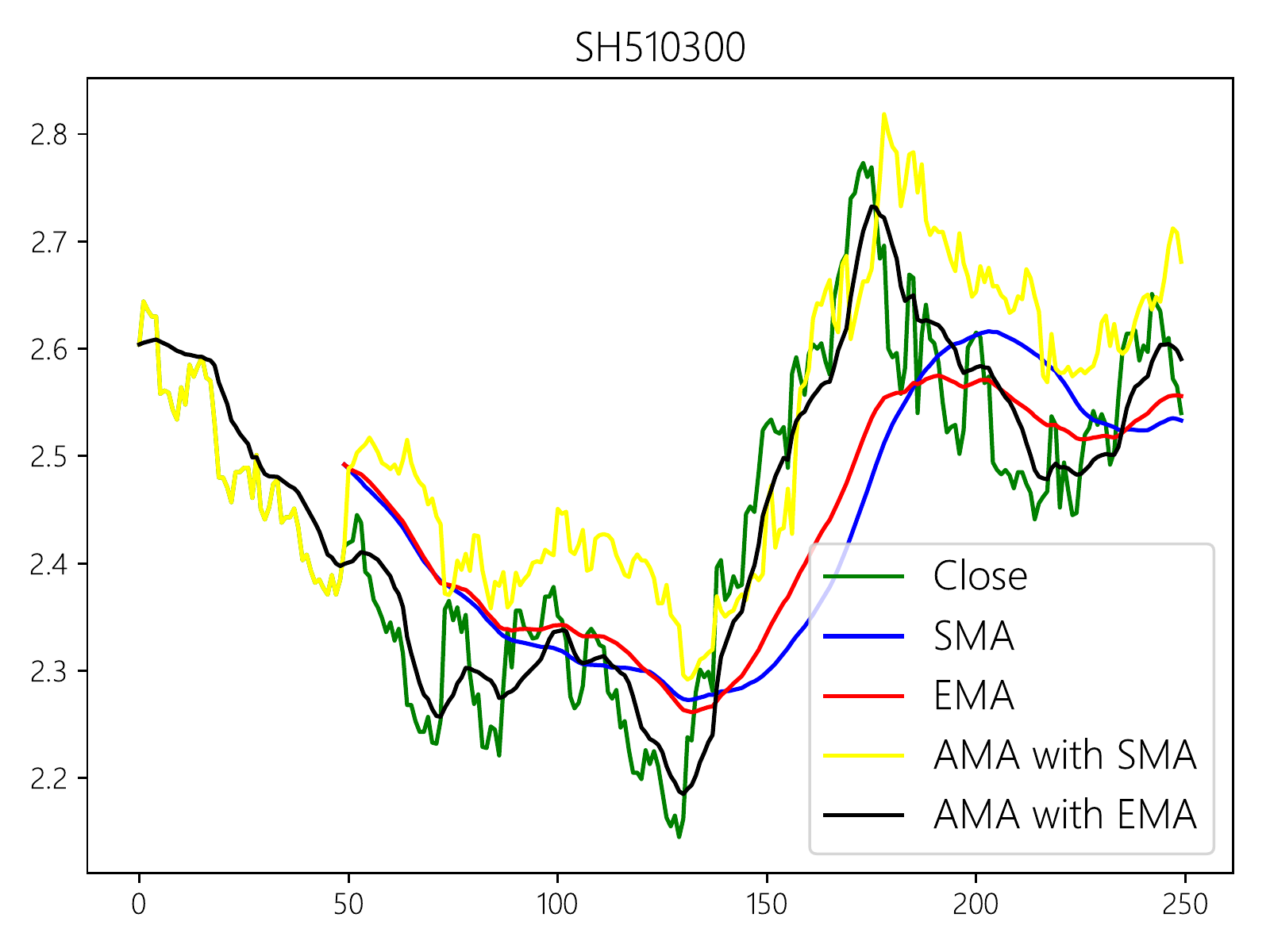}}
	\caption{MAs for S\&P500 and SH510300 where the time periods for SMA and EMA are both 50. TimeperiodLong, TimeperiodShort, AdaWin for AMA are 50, 5, 12 respectively.}
	\label{fig:sh510300-sp500-sma-ema}
\end{figure}

Given the long time period $N_1$ and short time period $N_2$, the final time period can be obtained by 
\begin{equation}\label{equation:adap-time-period}
N = N_2 + \text{ABS}(\text{ER}[i]) \times  (N_1-N_2),
\end{equation}
where $\text{ABS}(\text{ER}[i]) \in [0,1]$ such that $N_2 \leq N \leq N_1$.
And then, the AMA with SMA is obtained by (same as Equation~\eqref{equation:SMA})
\begin{equation}\label{equation:SMA-for-AMA}
	\left\{
	\begin{aligned}
		\text{SMA}[i] &= \frac{\text{input}[i-N+1:i]}{N}, \gap &\text{if $i>N$};\\
		\text{SMA}[i] &= \text{not a number, or input}[i], \gap &\text{if $i\leq N$},\\
	\end{aligned}
	\right.
\end{equation}
A comparison of different MAs on the S\&P500 and SH510300 data sets within 250 trading days (we shall shortly introduce later) is shown in Figure~\ref{fig:sh510300-sp500-sma-ema}. We find that when the price is moving towards a certain direction, both the ``AMA with SMA" and ``AMA with EMA" move similarly to the price value (e.g., around 145-th day for the S\&P500 data); while on the other hand, when the price value is volatile, the AMAs are moving smoothly (e.g., between 150-th day and 200-th day for the S\&P500 data).

The following ``adaptiveMovAvg" code is written for Python 3.7. The code is not computationally efficient but explanatory (terse comments begin with a $\#$ in green color texts).
\begin{python}
def adaptiveMovAvg(array, TimeperiodLong=30, TimeperiodShort=2, AdaWin=10,matype=1):
	"""
	:param TimeperiodLong: long time period
	:param TimeperiodShort: short time period
	:param AdaWin: adaptive window length
	:param matype == 1: AMA with EMA,
	:param matype == 2: AMA with SMA
	:return res: AMA sequence of an array
	"""
	res = np.zeros(array.shape)
	# absolute moving momentum - noise
	absshift = [0]+[abs(array[i] - array[i-1])\
									for i, closeData in enumerate(array[1:])]
	noise = [0.0001]*AdaWin + [sum(absshift[i-AdaWin:i+1])\
									for i in range(len(absshift)) if i>=AdaWin]
	# relative moving momentum - signal
	signal = [0]*AdaWin + [(array[i] - array[i-AdaWin]) \
									for i in range(len(array)) if i>=AdaWin]
	# ERs \in [-1,1]
	ER = [signal[i]/noise[i] for i in range(len(signal))]
	if matype == 1: # AMA by EMA
		slowSC = 2*1./(TimeperiodLong+1)    # e.g., 2/31
		fastSC = 2*1./(TimeperiodShort+1)   # e.g., 2/3
		diffSC = fastSC - slowSC
		for i, closeData in enumerate(array):
			if i==0:
				res[i] = array[i]
				continue
			er_this = abs(ER[i])
			# mimicking EMA
			scaledSC = pow(slowSC + er_this*diffSC, 2)
			res[i] = res[i-1] + scaledSC * (array[i] - res[i-1])
	elif matype == 2: # AMA by SMA
		for i, closeData in enumerate(array):
			if i<TimeperiodLong:
				res[i] = array[i]
				continue
			finalperiod = TimeperiodShort + \
			              		abs(ER[i]) * (TimeperiodLong-TimeperiodShort)
			finalperiod = int(finalperiod)
			res[i] = np.mean(array[i-finalperiod:i+1])
	else:
		pass
	return res
\end{python}

\subsection{The Strategy}
The Two-Average strategy, also known as the \textit{Crossover strategy}, is one of the main moving average strategies.
The first type is a price crossover, which is when the (closing) price crosses above or below a MA to signal a potential change in trend (buy and sell respectively).

The further idea on this strategy is to have two sets of MAs: one longer and one shorter. When the shorter-term MA crosses above the longer-term MA, it's a buy signal, as it indicates that the trend is shifting up. This is known as a ``\textit{golden cross}".

Meanwhile, when the shorter-term MA crosses below the longer-term MA, it's a sell signal, as it indicates that the trend is shifting down. This is known as a ``\textit{dead/death cross}".


\paragraph{Problem} One major problem is that, if the price action becomes choppy, the price may swing back and forth, generating multiple trend reversals or trade signals. When this occurs, it's best to step aside or utilize another indicator to help clarify the trend. The same thing can occur with MA crossovers when the MAs get ``tangled up" for a period of time, triggering multiple losing trades. 

Moving averages work quite well in strong trending conditions but poorly in choppy or ranging conditions. Adjusting the time period can remedy this problem temporarily, although at some point, these issues are likely to occur regardless of the time period chosen for the MAs.

\subsection{Data: S\&P500 and SH510300}
\begin{figure}[h]
	\centering  
	\vspace{-0.35cm} 
	\subfigtopskip=2pt 
	\subfigbottomskip=2pt 
	\subfigcapskip=-5pt 
	\subfigure[Closing price of S\&P500 in the time period of 11 years where the horizontal axis is the order of the dates and vertical axis is the stock price.]{\label{fig:sp500-bin-1}
		\includegraphics[width=0.47\linewidth]{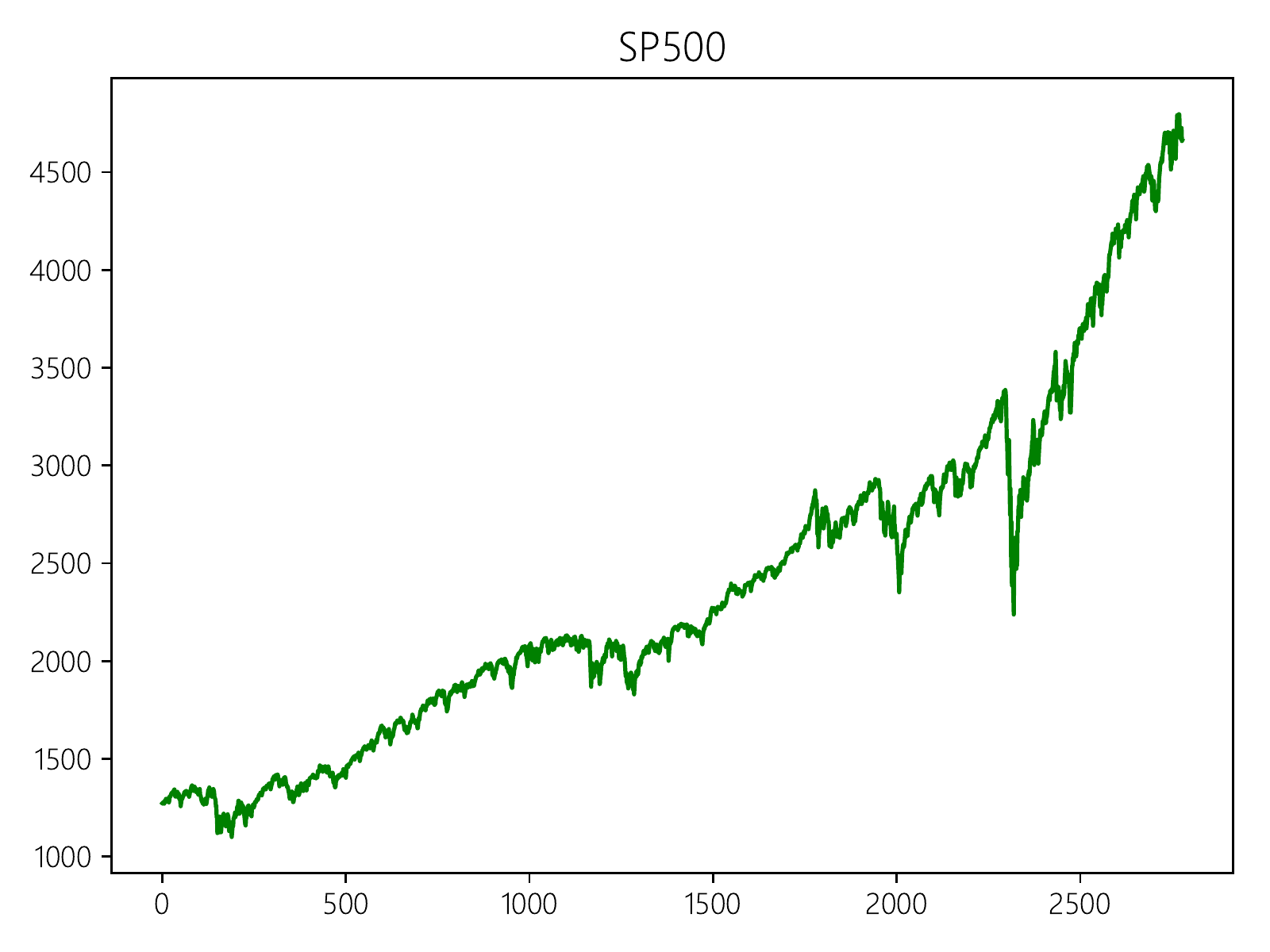}}
	\quad 
	\subfigure[Bin plot of the returns in percentage (range from -1 to 1) where the dotted line is the fitted Gaussian distribution: $\sim \normal(5.3e-4, (1.08e-2)^2)$.]{\label{fig:sp500-bin-2}
		\includegraphics[width=0.47\linewidth]{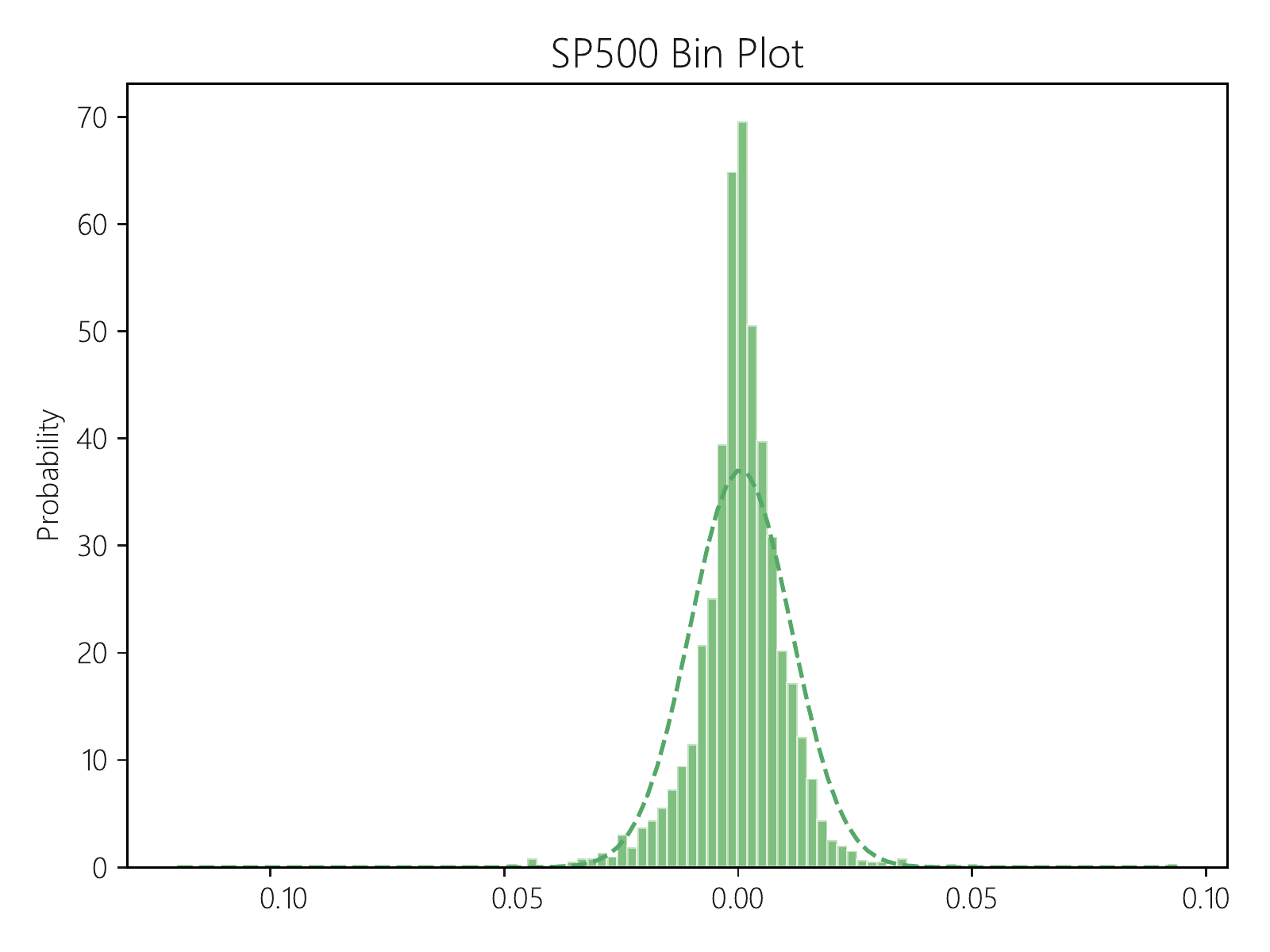}}
	\caption{S\&P500 over 11 years period.}
	\label{fig:sp500-bin-12}
\end{figure}

A typical method for obtaining measurements about quantitative/trading strategies is to run a simulation (that is, backtest) and measure characteristics of the result, such as the Sharpe ratio.
To evaluate the strategy, we then obtain the S\&P500 index data, which is a market-capitalization-weighted index of 500 leading publicly traded companies in the U.S. from Yahoo Finance\footnote{The data is from Yahoo Finance: \url{https://finance.yahoo.com/quote/\%5EGSPC/history?}.} with a time period of 11 years (between Jan. 14, 2011
and Jan. 14, 2022). The S\&P500 index uses a market-cap weighting method, giving a higher percentage allocation to companies with the larger market capitalizations. The weighting of each company in the index is calculated by taking the company's market cap and dividing it by the total market cap of the index:
$$
\text{Company weighting in S\&P500}= \frac{\text{Company market cap}}{\text{Total market caps}},
$$
where the market cap of a company is calculated by taking the current stock price and multiplying it by the company's outstanding shares.\footnote{The constituents list of S\&P500 can be obtained at \url{https://en.wikipedia.org/wiki/List_of_S\%26P_500_companies}.}
The daily closing price of the data and its daily rate of returns are shown in Figure~\ref{fig:sp500-bin-12} where we observe that the distribution of return in percentage is close to a Gaussian distribution, $\normal(5.3e-4, (1.08e-2)^2)$, with almost a zero-mean. We also observe a big drawdown in Figure~\ref{fig:sp500-bin-1} (from Feb. 21, 2020 to Mar. 23, 2020) which is largely due to the outbreak of the COVID-19 and we shall further discuss this case in the results.

\begin{figure}[h]
	\centering  
	\vspace{-0.35cm} 
	\subfigtopskip=2pt 
	\subfigbottomskip=2pt 
	\subfigcapskip=-5pt 
	\subfigure[Closing price of SH510300 in the time period of 9 years where the horizontal axis is the order of the dates and vertical axis is the stock price.]{\label{fig:sh510300-bin-1}
		\includegraphics[width=0.47\linewidth]{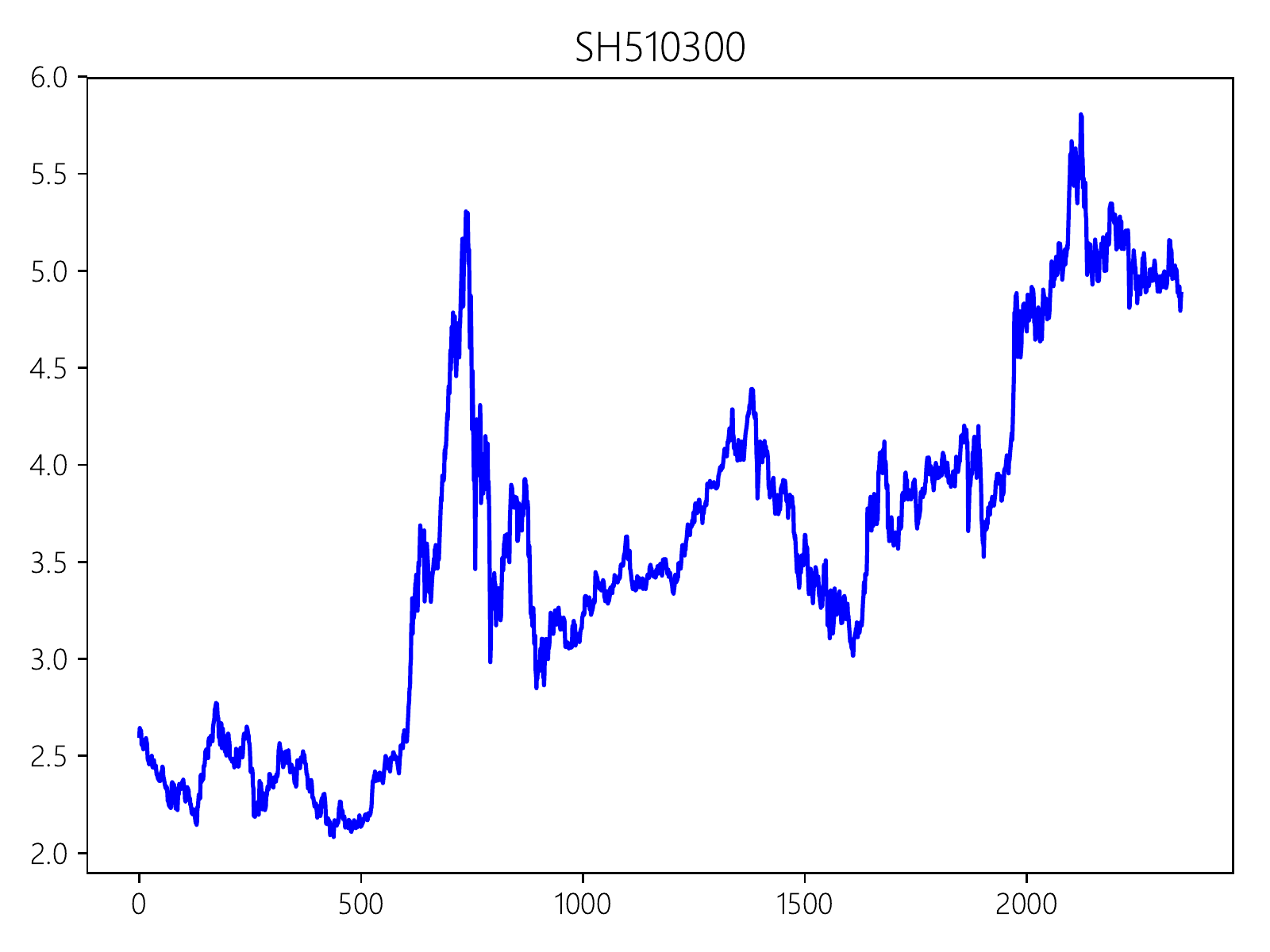}}
	\quad 
	\subfigure[Bin plot of the returns in percentage (range from -1 to 1) where the dotted line is the fitted Gaussian distribution: $\sim \normal(3.8e-4, (1.5e-2)^2)$.]{\label{fig:sh510300-bin-2}
		\includegraphics[width=0.47\linewidth]{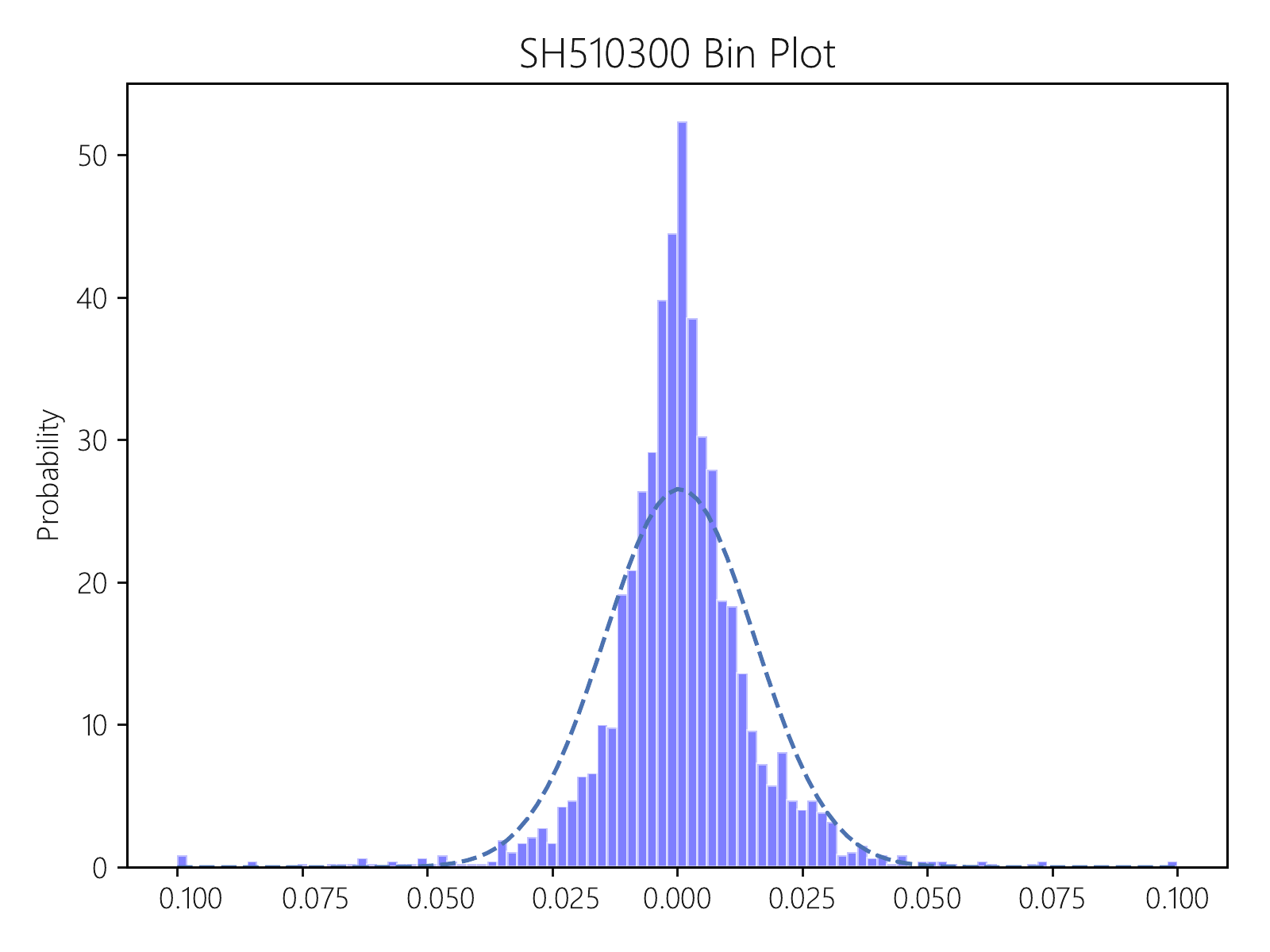}}
	\caption{SH510300 over 9 years period.}
	\label{fig:sh510300-bin-12}
\end{figure}

Further, we obtain the SH510300 index data, a Chinese alternative for the S\&P500 (``similar" to S\&P500, but still has a large difference, which is a market-capitalization-weighted index of 300 leading publicly traded companies in China), from Sina Finance\footnote{The data is due to Sina Finance: \url{https://money.finance.sina.com.cn/quotes_service/api/json_v2.php/CN_MarketData.getKLineData?symbol=sh510300&scale=240&ma=5&datalen=3000}.} with a time period of 9 years (between Jan. 4, 2013
and Jan. 14, 2022).
Similarly, the daily closing price of the data and its daily rate of returns are shown in Figure~\ref{fig:sh510300-bin-12}. We observe that, different from the S\&P500 data, there are more downtrends in the SH510300 data so that exploration on this data can provide more information on the strategies, e.g., whether the strategy can find out the downtrend and avoid large drawdown. Similar distribution for the returns in percentage, both are close to a Gaussian distribution, whilst the mean and standard deviation are close as well: $\normal(5.3e-4, (1.08e-2)^2)$ vs $\normal(3.8e-4, (1.5e-2)^2)$.

There is also an important theorem in statistical theory, the central limit theorem (CLT). 
For an arbitrarily distributed population, $N$ samples are randomly selected from the population each time, and a total number of $M$ sampling are undertaken.
Then taking the average of these $M$ groups of samples and the distribution of these averages is close to the Gaussian distribution (also known as the normal distribution).
The CLT is important for distribution theory, e.g., the large sample property in linear models \citep{lu2021rigorous}. Rigorously, the theorem is discussed as follows:
\begin{theorem}[Central Limit Theorem (CLT)]
	Let $X_1, X_2, \cdots, X_n$ be iid random variables such that $\Exp[X_i]=\mu < \infty$ and $\Var[X_i]=\sigma^2 < \infty$. Let $\bar{X}_n = \frac{1}{n} \sum_{i=1}^{n}X_i$. Then
	$$
	\sqrt{n}(\bar{X}_n - \mu) \stackrel{d}{\longrightarrow} \normal(0, \sigma^2).
	$$
	Similarly, this result can be extended to multi-dimensional case, see \citep{lu2021rigorous} and its weighted version. 
\end{theorem}
Come back to the stock data. Although the distribution of asset (closing) prices is unknown and is not a realization of a Gaussian distribution, according to the CLT, their rate of returns is always close to a Gaussian distribution. This feature provides a theoretical guide on financial risk control.

\paragraph{Find the index yourself } In real quantitative applications, one should find a set of stocks based on his/her own algorithms, i.e., a new asset index, just like how the 500 stocks in the S\&P500 are selected; but with different methods. Potentially, the weight of each stock is not dependent on the market cap anymore but on other signals. One more thing to notice, there should be a lower limit for the number of stocks selected since if the index contains a large number of stocks, its behaviors in the future will be more consistent with its historical behaviors. While, if the index contains only a small amount of stocks, it can be less dependable. Just one example of a criterion to select the stocks is to have a diversified list of the stocks with low correlation so that idiosyncratic risks can be avoided. On the other hand, when the number of stocks is too large, a sub-selection method can be applied so that a relatively small amount of them can be chosen to represent the whole. E.g., the CR decomposition, the interpolative decomposition, or the skeleton decomposition can be employed \citep{lu2021numerical, lu2022matrix}.

\begin{SCfigure}
	\centering
	\includegraphics[width=0.6\textwidth]{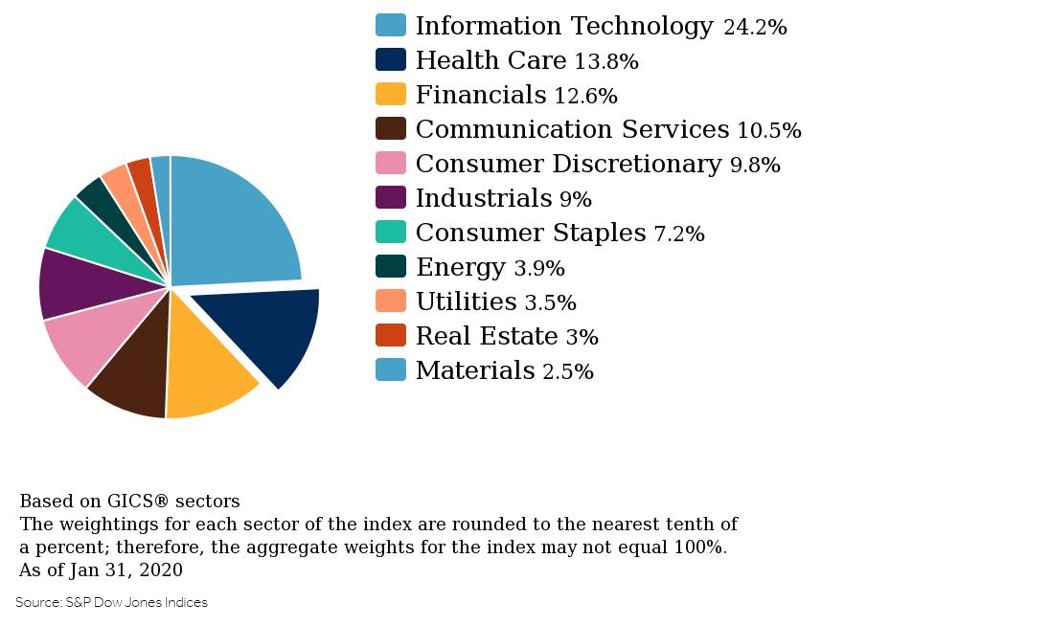}
	\caption{Demonstration on the idea of sector neutral.}
	\label{fig:sp500sector}
\end{SCfigure}

\paragraph{Sector neutral or industry neutral} Further note should be taken care of is known as the sector neutral or industry neutral.
Sector neutral means not being overweight or underweight any given sector relative to what the index weight is.  For example, according to S\&P Dow Jones Indices – a division of S\&P Global, as of Jan. 31, 2020, the ``Communication Services" was 10.5\% and the ``Health Care" was 13.8\% of the S\&P500 Index (Figure~\ref{fig:sp500sector}). A neutral index means the index should put similar weight on each sector/industry.
This is to eliminate the exposure to sectors/industries so that the quantitative strategy won't have big swings when the one of the sectors changes rapidly, say the ``oil price'' surges or drops sharply. See \citep{schumaker2009quantitative} for a discussion on this sector/industry classification.

And all in all, innovative and functional methods are discovered from data by quantitative researchers.

\subsection{Results with AMA}
\begin{figure}[H]
\centering  
\vspace{-0.35cm} 
\subfigtopskip=2pt 
\subfigbottomskip=2pt 
\subfigcapskip=-5pt 
\subfigure[TimeperiodLong=51, TimeperiodShort=5, AdaWin=12, Matype=2.]{\label{fig:sp500-twoaverage-ama1}
	\includegraphics[width=0.47\linewidth]{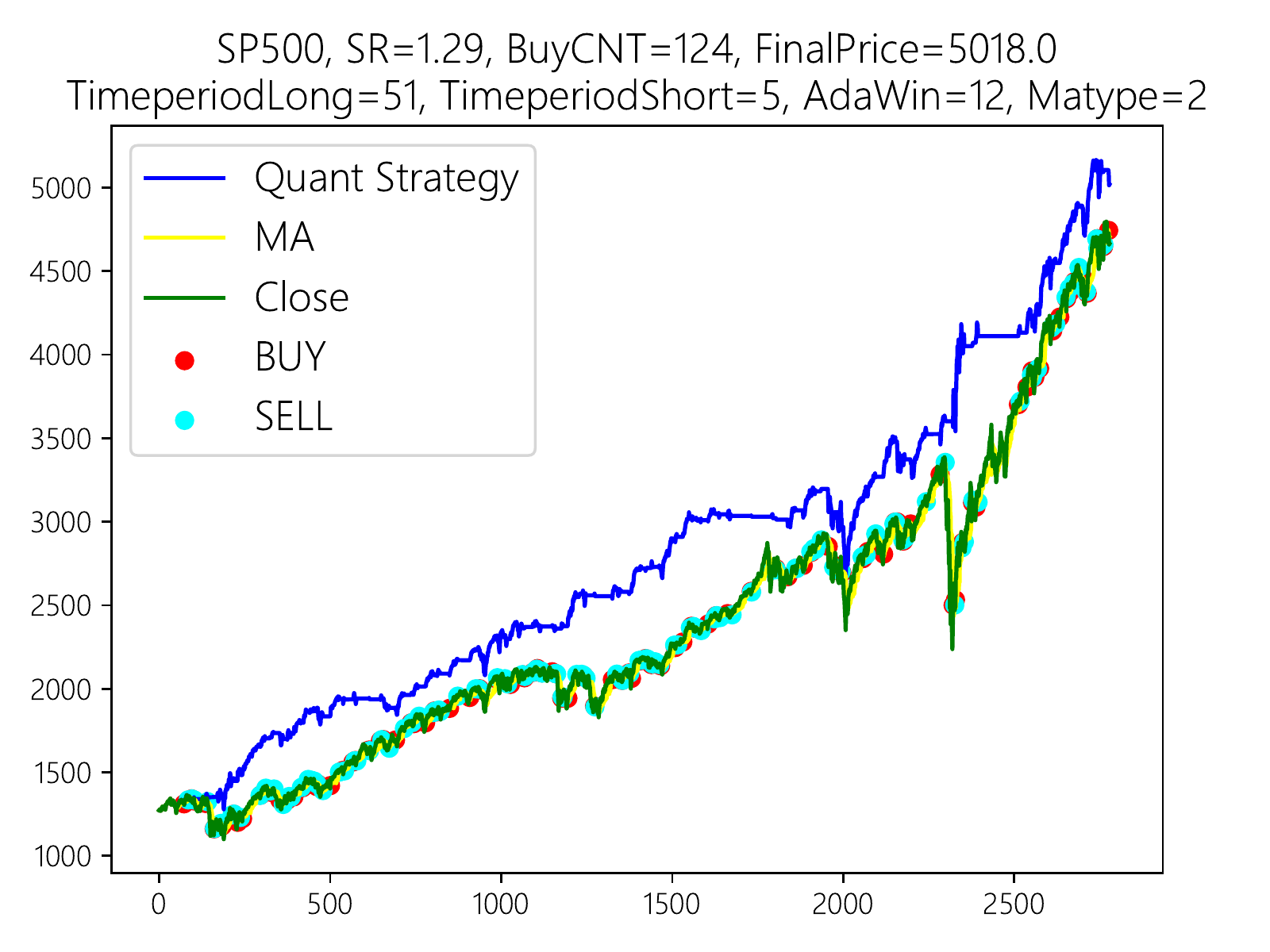}}
\quad 
\subfigure[TimeperiodLong=53, TimeperiodShort=3, AdaWin=12, Matype=2.]{\label{fig:sp500-twoaverage-ama2}
	\includegraphics[width=0.47\linewidth]{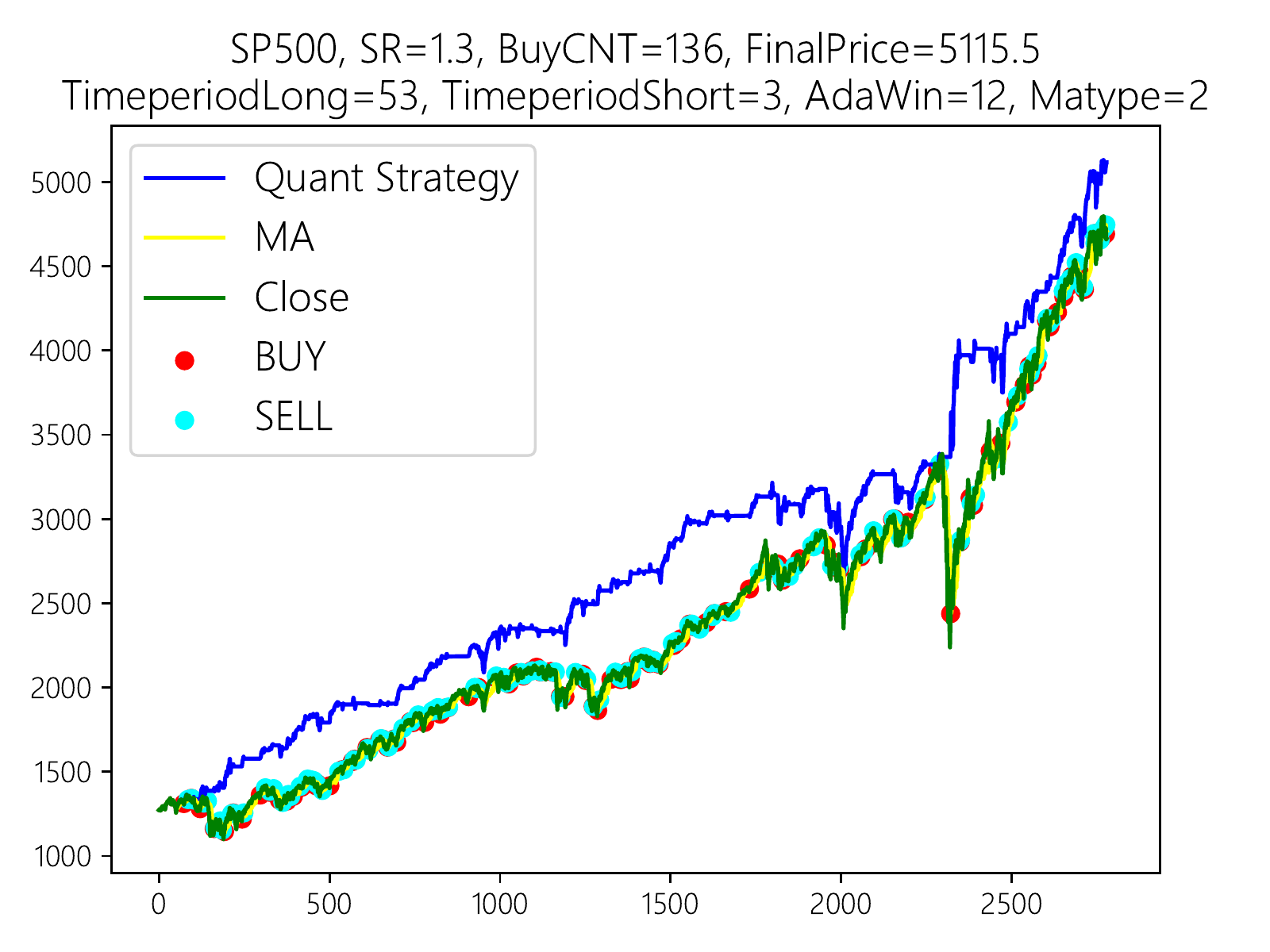}}
\caption{Two-Average strategy on S\&P500 with AMA where the blue line is the ``stock price" of the strategy, the green line is the closing price of S\&P500 on each day, and the yellow line is a specific moving average whose parameter is shown in the title of each figure. The red dots indicate when to buy and the cyan dots indicate when to sell. ``FinalPrice" in the title represents the final ``stock price" of the strategy; and ``BuyCNT" in the title counts the number of buying. ``Matype=2" means the AMA is based on SMA; while, ``Matype=1" means the AMA is based on EMA (here, we only observe good results when Matype=2).}
\label{fig:sp500-twoaverage-ama12}
\end{figure}

The result of the Two-Average strategy with AMA on the S\&P500 is shown in Figure~\ref{fig:sp500-twoaverage-ama12} where the horizontal axis is the order of the dates and the vertical axis is the stock price.

The detailed measures in Figure~\ref{fig:sp500-twoaverage-ama1} are given as follows where ``MDD" is the max drawdown, ``SR" is the Sharpe ratio, and ``IR" is the information ratio whose benchmark is set to be the S\&P500 itself. 
The ``Initial Price" is the stock price of the first buy and the ``Final Price" represents the final price of the strategy. ``RR" is short for rate of return; and minimal and maximal annualized RR are provided as a reference. 

The ``Total number of buy count" (i.e., ``BuyCNT" in the titles of Figure~\ref{fig:sp500-twoaverage-ama1} and \ref{fig:sp500-twoaverage-ama2}) can be understood as the turnover of the strategy. The higher the turnover, the higher the trading costs and the market impact. However, high turnover strategies usually have stronger trading signals that can help follow the upward trend of the markets. Therefore, turnover should not be too high or too low. And different traders may not easily agree with which turnover is the best and we shall not discuss this issue where the ``BuyCNT" is just served as a reference. 

\begin{figure}[H]
	\centering  
	\vspace{-0.35cm} 
	\subfigtopskip=2pt 
	\subfigbottomskip=2pt 
	\subfigcapskip=-5pt 
	\subfigure[TimeperiodLong=19, TimeperiodShort=3, AdaWin=14, Matype=2.]{\label{fig:sh510300-twoaverage-ama1}
		\includegraphics[width=0.47\linewidth]{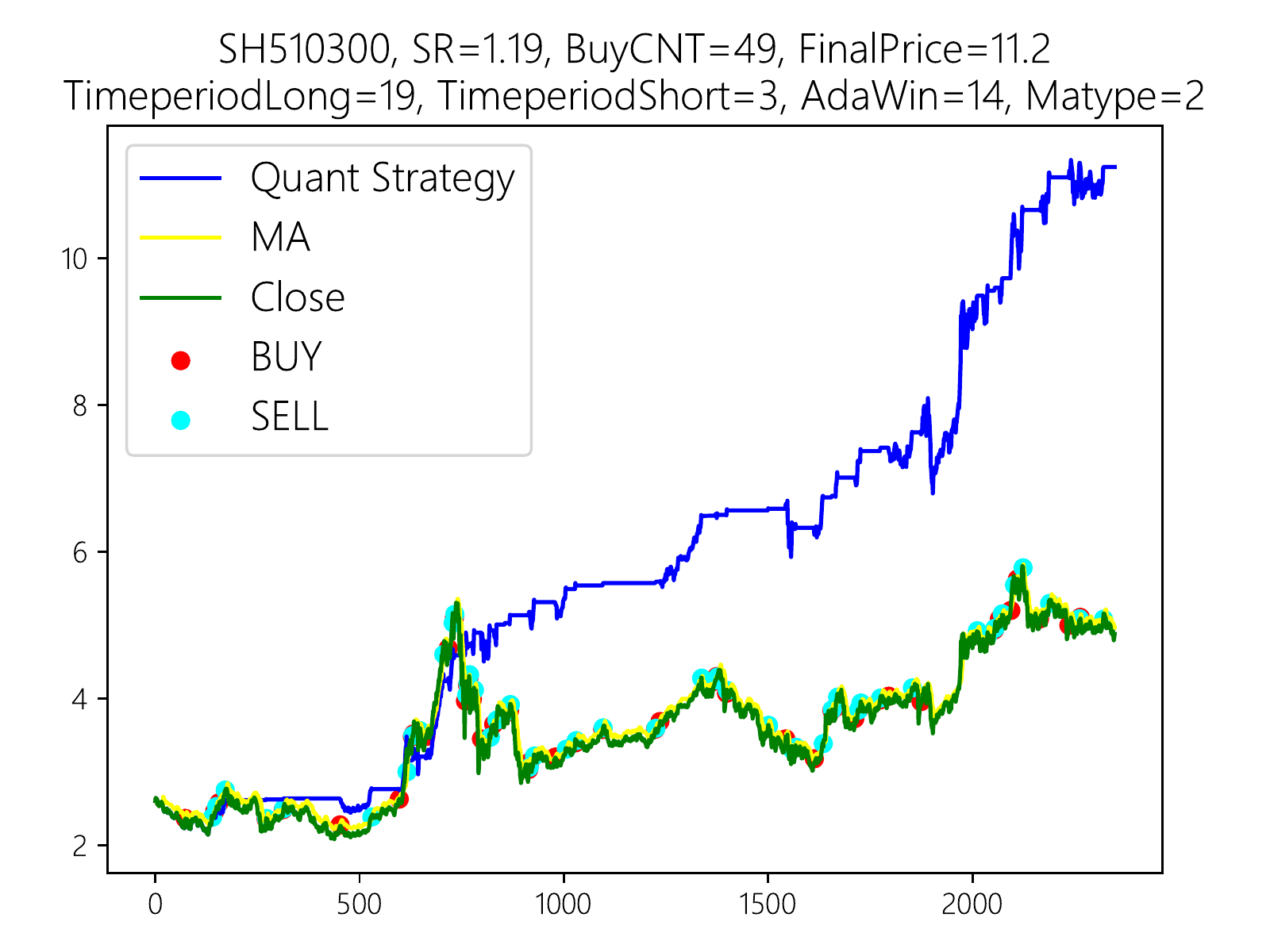}}
	\quad 
	\subfigure[TimeperiodLong=61, TimeperiodShort=5, AdaWin=46, Matype=2.]{\label{fig:sh510300-twoaverage-ama2}
		\includegraphics[width=0.47\linewidth]{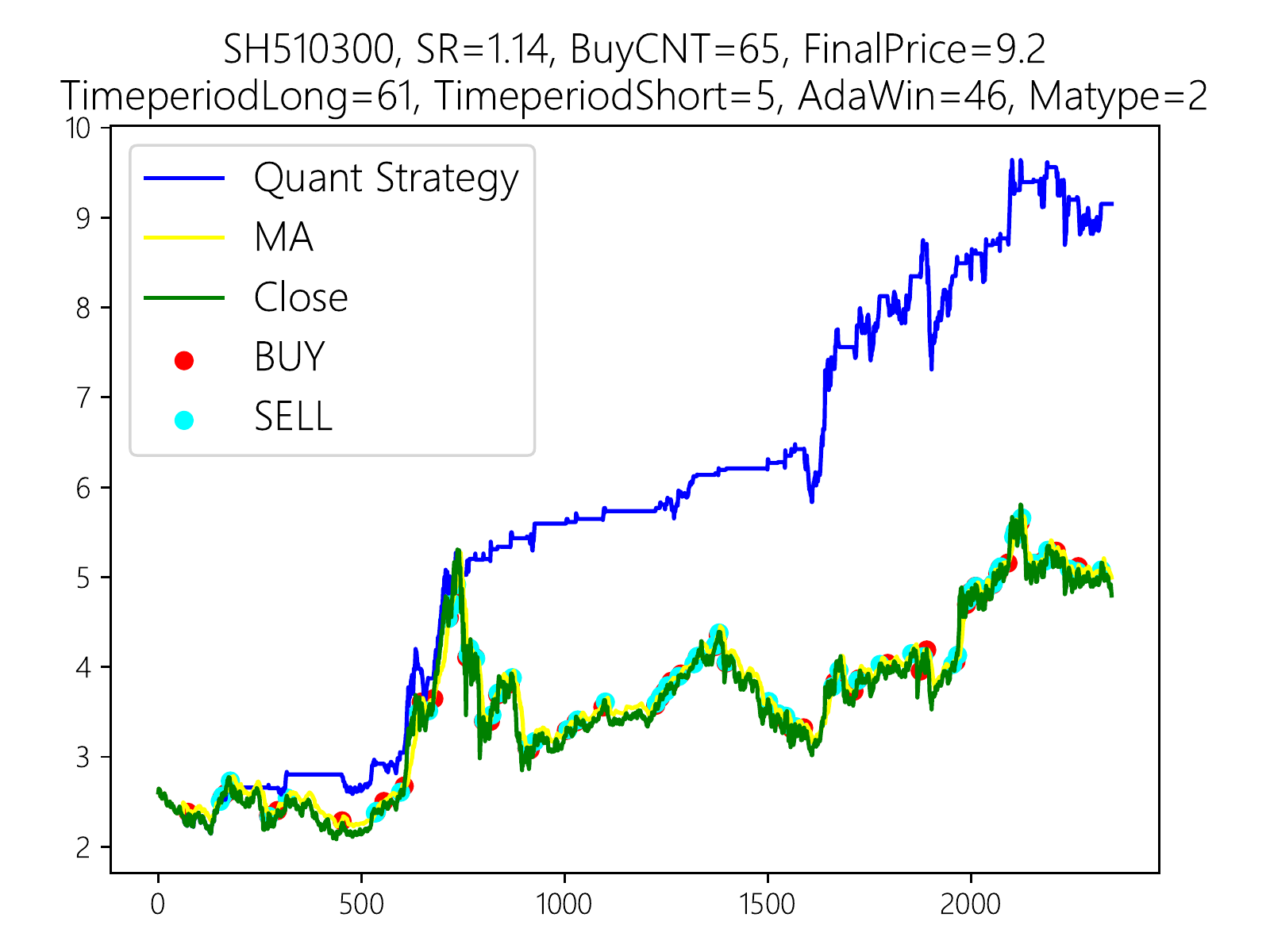}}
	\caption{Two-Average strategy with AMA on the SH510300 where the green line is the closing price of SH510300 on each day.}
	\label{fig:sh510300-twoaverage-ama12}
\end{figure}

We notice that, in this case, the IR is negative; the reason is partly from that the strategy does not work well from the 2400-th day (i.e., the outbreak of the COVID-19). The ``big drawdown" of the S\&P500 series due to the COVID-19 issue also causes interesting results as we shall see in the Keltner strategy (Section~\ref{section:keltner-strategy}, p.~\pageref{section:keltner-strategy}). Therefore, different strategies may be applied before and after the cutoff.
\begin{python}
Initial Price:        1271.87
Final Price:          5017.952801940561
RR of whole period:   3.945334666232053
RR/year:              1.132538418927603
RR of year-1: 1.2052468859894903
RR of year-2: 1.209821429404758
RR of year-3: 1.0925606788248348
RR of year-4: 1.1625258120500706
RR of year-5: 1.1196006019227314
RR of year-6: 1.150334996114205
RR of year-7: 1.0566055909800327
RR of year-8: 1.02915151309352
RR of year-9: 1.1501585844135516
RR of year-10: 1.153891406484533
RR of year-11: 1.2548521976045632
Total number of buy count:  124
MAX rate:  1.2548521976045632  MIN rate:  1.02915151309352
MDD:  0.1782640232818626
SR:  1.289569812149183
IR:  -0.021788463873900706
\end{python}
Moreover, the result of the Two-Average strategy with AMA on the SH510300 is shown in Figure~\ref{fig:sh510300-twoaverage-ama12}. As we mentioned previously, there are more drawdowns in the SH510300 data set, the results on this data set may provide more information, e.g., whether the strategy can avoid losing profit and follows the upward trend when it's coming.

The detailed measures in Figure~\ref{fig:sh510300-twoaverage-ama1} are given as follows where again the ``IR" is the information ratio whose benchmark is set to be the SH510300 itself. Figure~\ref{fig:sh510300-twoaverage-ama1} shows a promising result as the strategy can shy away from the drawdown periods, e.g., the period around 1500-th day; and more importantly, it chases the upward trends. 
\begin{python}
Initial Price:        2.604
Final Price:          11.246482441341051
RR of whole period:   4.318925668717761
RR/year:              1.1699381785623744
RR of year-1: 1.0044238376549013
RR of year-2: 0.9454947007322866
RR of year-3: 1.8294921696975432
RR of year-4: 1.1414544503541364
RR of year-5: 1.074509559922823
RR of year-6: 1.1744891266255308
RR of year-7: 1.1231252139480294
RR of year-8: 1.262463201126989
RR of year-9: 1.2130668826994588
Total number of buy count:  49
MAX rate:  1.8294921696975432  MIN rate:  0.9454947007322866
MDD:  0.18872919818456887
SR:  1.1888577990184008
IR:  0.36914729565256765
\end{python}
One thing we need to keep in mind is the difference between the in-sample results and the out-of-sample results. Before applying the strategy into live trading, careful monitoring of the results should be applied for a certain period of time. And here, we use a long period of time to evaluate the performance of the strategy(11 years for the S\&P500 data and 9 years for the SH510300 data). In practice, evaluation on a 3-year period should be acceptable since the ``old data" may not reflect the markets sufficiently and induce an overfitting on the data.

\paragraph{Key takeaway}
Comparing the results on S\&P500 and SH510300 with AMA, we find the adaptive version of SMA works better than the adaptive version of EMA. This is partly because the AMA from EMA has been extensively used in the industry, and the signal might disappear in the recent markets; while, AMA from SMA is not that famous at the moment and shows new ways to do the strategy. Readers are recommended to apply the AMA idea for other moving averages, such as the DEMA, T3, and so on. To simply put, the final time period in the adaptive version of an MA is something like the Equation~\eqref{equation:adap-time-period}. And we shall not give the details for simplicity. 

\subsection{Results with non-adaptive MA}
\begin{figure}[H]
	\centering  
	\vspace{-0.35cm} 
	\subfigtopskip=2pt 
	\subfigbottomskip=2pt 
	\subfigcapskip=-5pt 
	\subfigure[Timeperiod=136.]{\label{fig:sh510300-twoaverage-mas1}
		\includegraphics[width=0.47\linewidth]{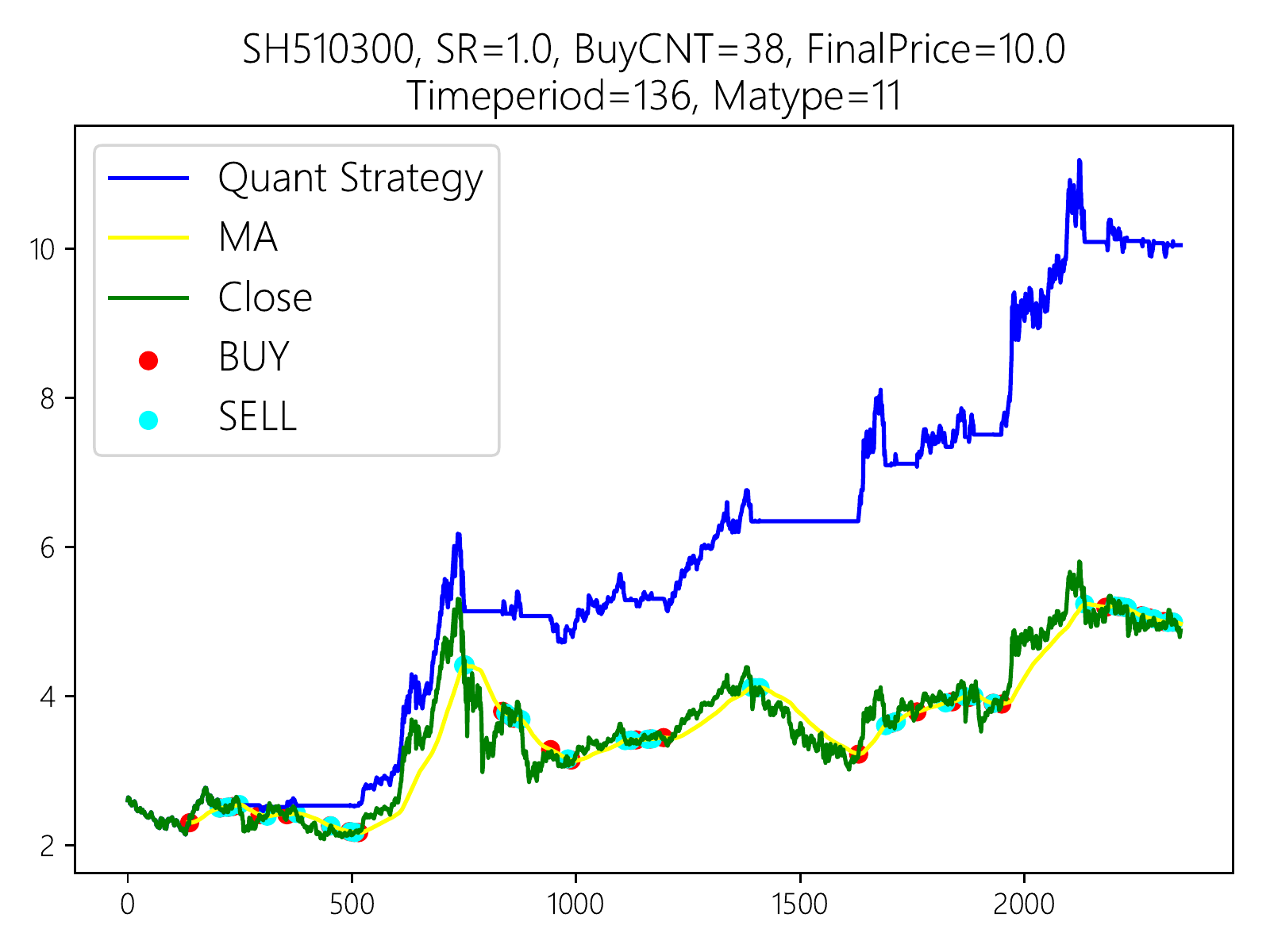}}
	\quad 
	\subfigure[Timeperiod=107.]{\label{fig:sh510300-twoaverage-mas2}
		\includegraphics[width=0.47\linewidth]{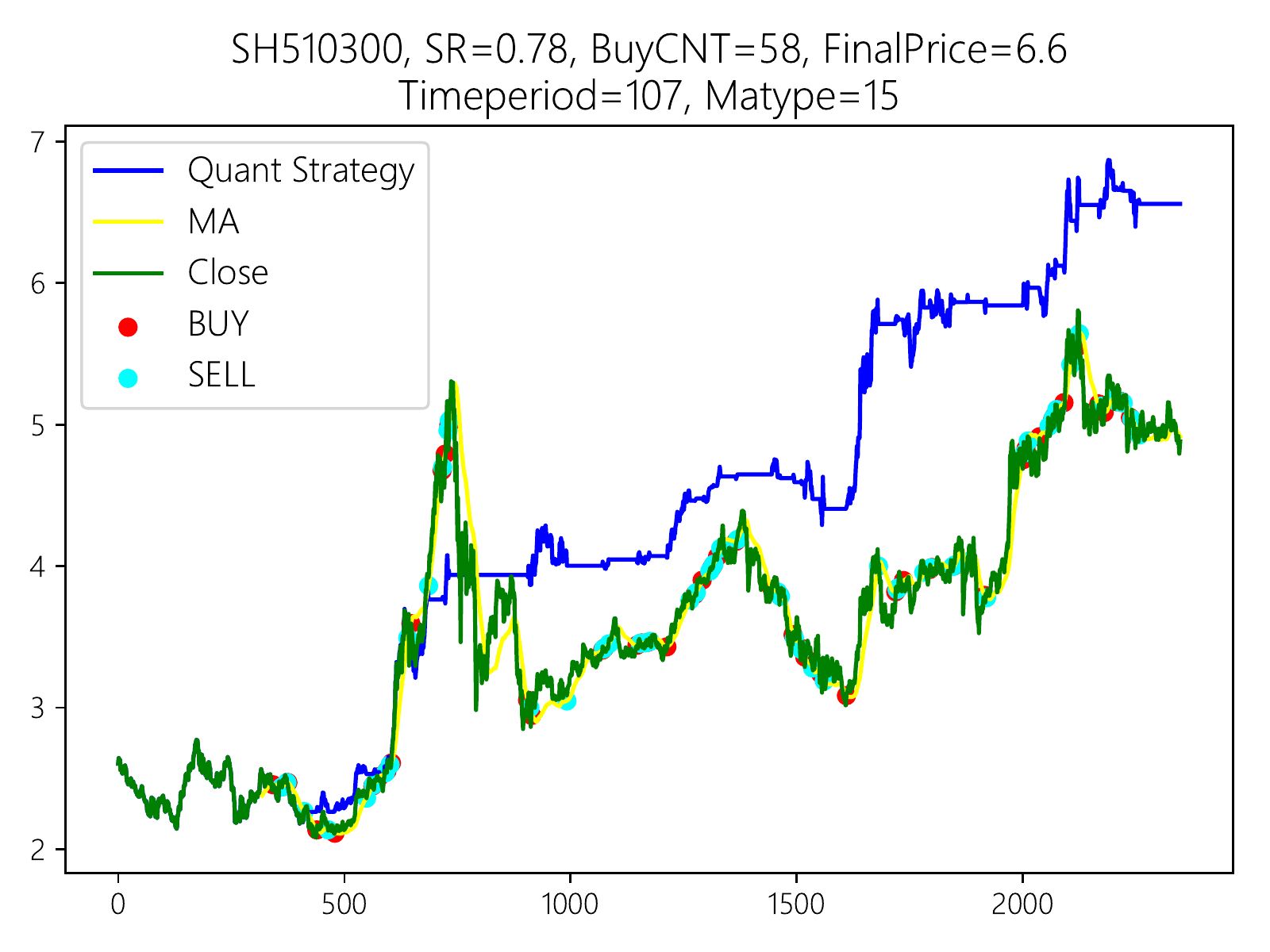}}
	\caption{Two-Average strategy with non-adaptive MA on the SH510300. Here the ``Matype=" in the title represents one of the moving average methods (e.g., SMA, EMA, T3, TEMA, and so on) and we will not give the details for simplicity.}
	\label{fig:sh510300-twoaverage-mas12}
\end{figure}

For the Two-Average strategy with non-adaptive MA on the S\&P500 data, the strategy works poor and we shall not give the details. However, it still gives promising results on the SH510300 data as shown in Figure~\ref{fig:sh510300-twoaverage-mas12}.

The detailed measures in Figure~\ref{fig:sh510300-twoaverage-mas1} are given as follows where again the ``IR" is the information ratio whose benchmark is set to be the SH510300 itself.
\begin{python}
Initial Price:        2.604
Final Price:          10.049906506216752
RR of whole period:   3.8594111006976775
RR/year:              1.1559040824113815
RR of year-1: 0.9744485970971373
RR of year-2: 0.9972268008802848
RR of year-3: 2.161406247343376
RR of year-4: 0.8644925474432263
RR of year-5: 1.165505303355888
RR of year-6: 1.1184080266316072
RR of year-7: 1.1212847717423118
RR of year-8: 1.3077094873031774
RR of year-9: 1.0867892863143571
Total number of buy count:  38
MAX rate:  2.161406247343376  MIN rate:  0.8644925474432263
MDD:  0.23630255563820962
SR:  1.0021273764071357
IR:  0.34673115120162523
\end{python}

\paragraph{Drawbacks} Moving averages are calculated based on historical data, and nothing about the calculation is predictive in nature. Therefore, results using moving averages can be random. At times, the market seems to respect MA support, and at other times, it shows these indicators no respect.

\section{Keltner Strategy}\label{section:keltner-strategy}

\subsection{Keltner Channels}
Keltner channels are volatility-based bands that are placed on either side of an asset's price and can aid in determining the direction of a trend. 
The Keltner channel was first introduced by Chester Keltner in the 1960s \citep{keltner1960make}. The original formula used SMA and the high-low price range to calculate the bands. In the 1980s, a new formula was introduced, the TR and ATR, that are commonly used today.

\paragraph{ATR} The Keltner channel uses the average true range (ATR). The ATR is a technical analysis indicator, introduced by market technician J. Welles Wilder  in his book \citep{wilder1978new}, that measures market volatility by decomposing the entire range of an asset price for that period $N$:
\begin{equation}
\left\{
\begin{aligned}
	\text{TR}[i]&= \text{Max}\bigg\{\text{high}[i]-\text{low}[i], \text{high}[i]-\text{close}[i-1], \text{close}[i-1]-\text{low}[i]\bigg\}; \\
	\text{ATR}&= \frac{\sum_{k=i-N+1}^{i} \text{TR}[i]}{N},
\end{aligned}
\right.
\end{equation}
where $\text{high}[i]$ is the highest price of the day, $\text{close}[i-1]$ is the closing price of the previous day. The above definition of volatility takes into account the jump at the opening of the day, which can more accurately reflect the volatility. While in rare cases, the TR can be also obtained without the jump:
$$
\text{TR}[i]= \text{high}[i]-\text{low}[i].
$$
\paragraph{Keltner channels}
Given the ATR, the Keltner channel is obtained as follows:
\begin{equation}
\left\{\begin{aligned}
	\text{Keltner channel middle line}&=\text{MA} \left(\frac{\text{high+close+low}}{3}\right);  \\
	\text{Keltner channel upper band}&=\text{Middle}+2\times \text{ATR};\\
	\text{Keltner channel lower band}&=\text{Middle}-2\times \text{ATR},
\end{aligned}
\right.
\end{equation}
where MA$(\cdot)$ indicates a kind of moving average on the sequence, e.g., SMA, EMA, or AMA.

\paragraph{The strategy} Wilder originally developed the ATR for commodities. However, the indicator can also be used for stocks and indices.﻿﻿ A stock experiencing a high level of volatility has a higher ATR, and a low volatility stock has a lower ATR. 
The Keltner strategy considers an upward or a downward trend happens when the price crosses over or below the band respectively. 
Therefore, if the price action breaks above the upper band, the trader should consider initiating long/buy positions while liquidating short/sell positions. If the price action breaks below the band, the trader should consider initiating short/sell positions while exiting long/buy positions.

\subsection{Results with non-adaptive MA}

\begin{figure}[H]
	\centering  
	\vspace{-0.35cm} 
	\subfigtopskip=2pt 
	\subfigbottomskip=2pt 
	\subfigcapskip=-5pt 
	\subfigure[Timeperiod=8.]{\label{fig:sp500-keltner-mas1}
		\includegraphics[width=0.47\linewidth]{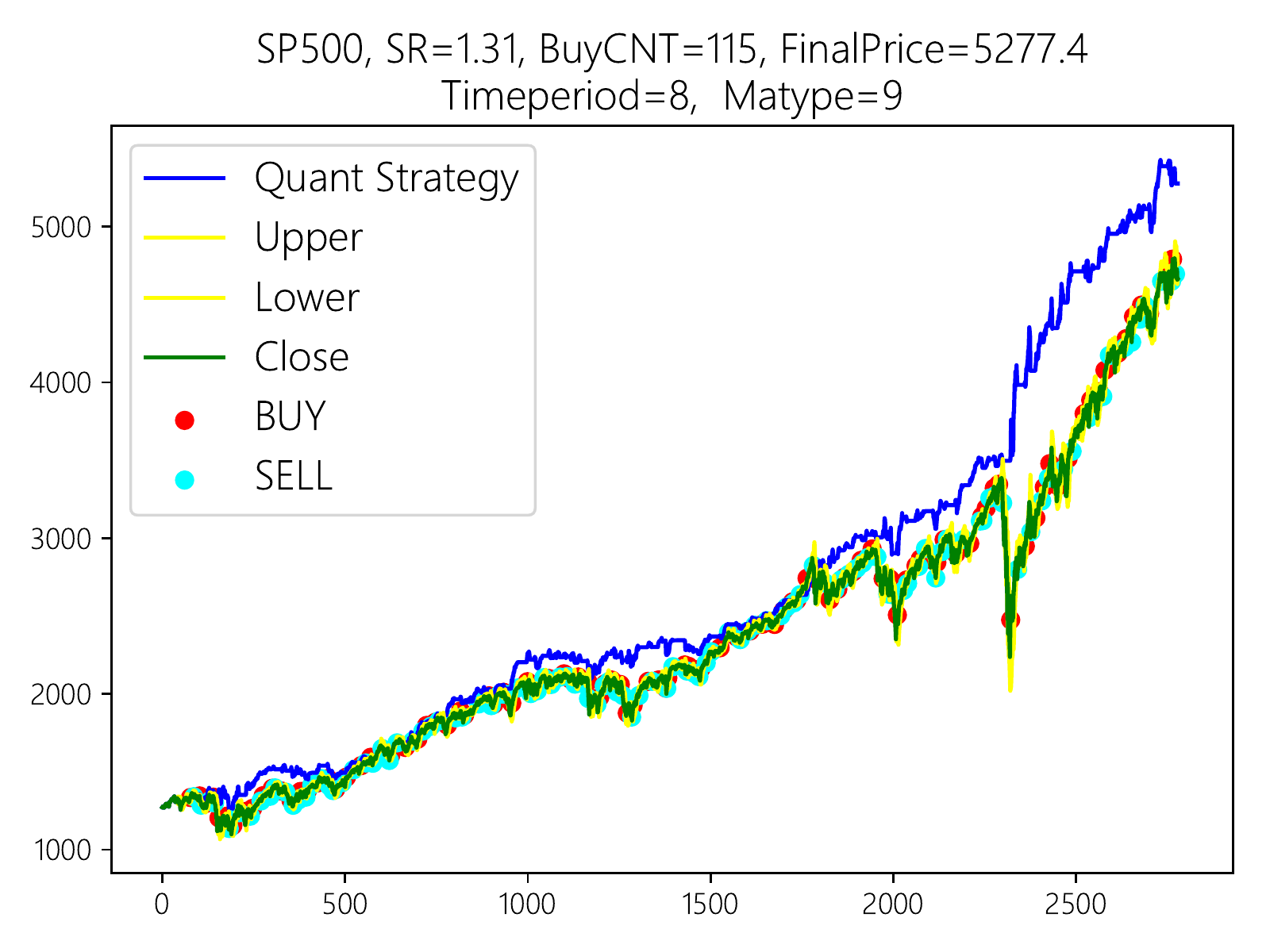}}
	\quad 
	\subfigure[Timeperiod=6.]{\label{fig:sp500-keltner-mas2}
		\includegraphics[width=0.47\linewidth]{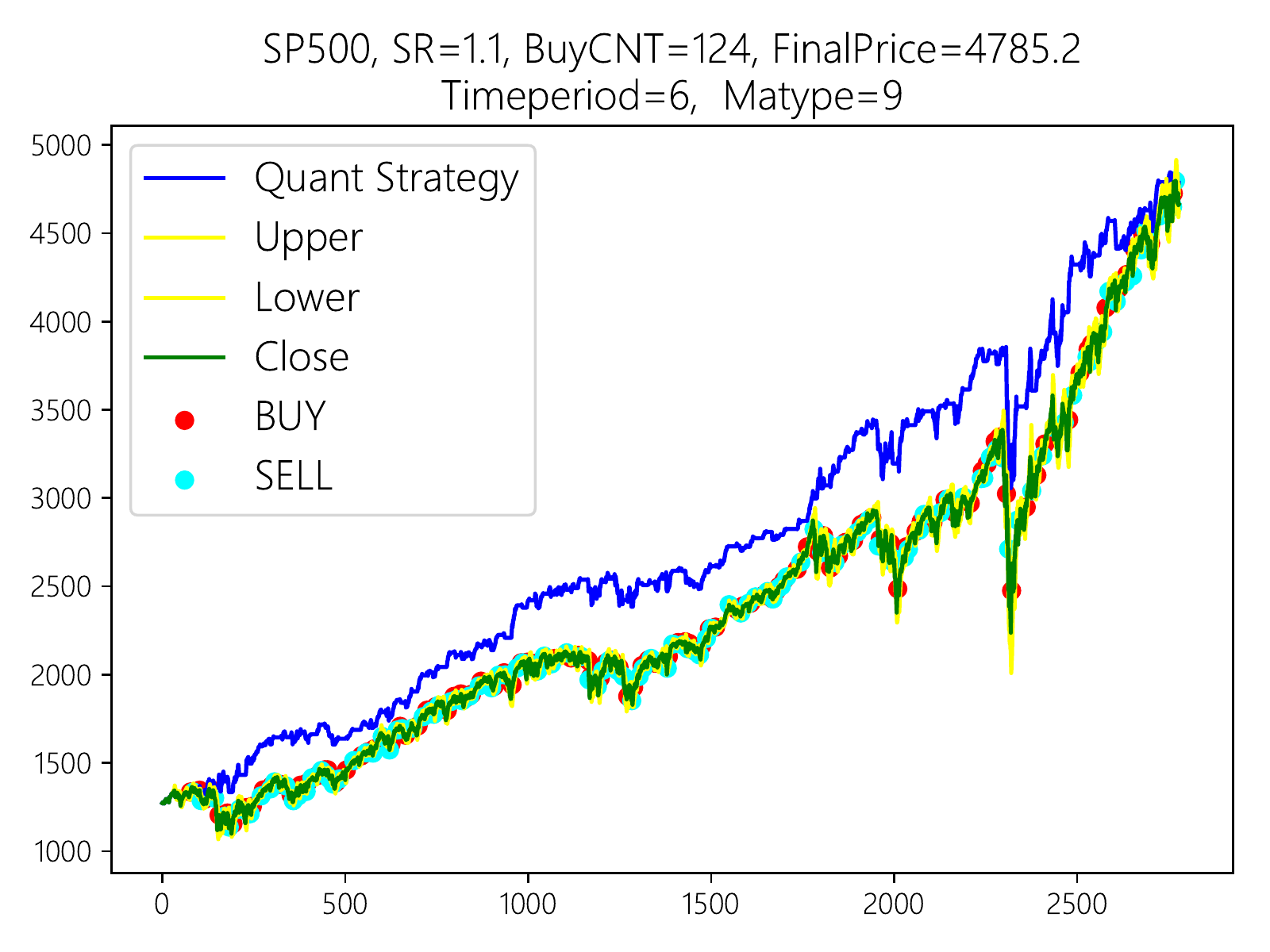}}
	\caption{Keltner strategy on S\&P500 with non-adaptive MA. Here the ``Matype=" in the title represents one of the moving average methods (e.g., SMA, EMA, T3, TEMA, and so on) and we shall not give the details for simplicity.}
	\label{fig:sp500-keltner-mas12}
\end{figure}

The result of the Keltner strategy with non-adaptive MA on the S\&P500 is shown in Figure~\ref{fig:sp500-keltner-mas12}.

The detailed measures in Figure~\ref{fig:sp500-keltner-mas1} is given as follows where again the ``IR" is the information ratio whose benchmark is set to be the S\&P500 itself. Though both the results in Figure~\ref{fig:sp500-keltner-mas1} and Figure~\ref{fig:sp500-keltner-mas2} are acceptable, we observe that the first one works better after the ``big drawdown" due to the outbreak of the COVID-19; and the second one works better before the ``big drawdown". Therefore, different strategies may be applied before and after the cutoff (in theory).
\begin{python}
Initial Price:        1271.87
Final Price:          5277.378092162411
RR of whole period:   4.149306212240568
RR/year:              1.1365625260709509
RR of year-1: 1.1209917702418386
RR of year-2: 1.0590707254487217
RR of year-3: 1.241576085383799
RR of year-4: 1.1943157812476106
RR of year-5: 1.0157534362807321
RR of year-6: 1.0560342266720353
RR of year-7: 1.122138960294929
RR of year-8: 1.1645196768443067
RR of year-9: 1.1528355526922454
RR of year-10: 1.3422459085639766
RR of year-11: 1.143119783030126
Total number of buy count:  115
MAX rate:  1.3422459085639766  MIN rate:  1.0157534362807321
MDD:  0.10047291580736482
SR:  1.312384211147428
IR:  0.013201664414390562
\end{python}

\begin{figure}[H]
	\centering  
	\vspace{-0.35cm} 
	\subfigtopskip=2pt 
	\subfigbottomskip=2pt 
	\subfigcapskip=-5pt 
	\subfigure[Timeperiod=6.]{\label{fig:sh510300-keltner-mas1}
		\includegraphics[width=0.47\linewidth]{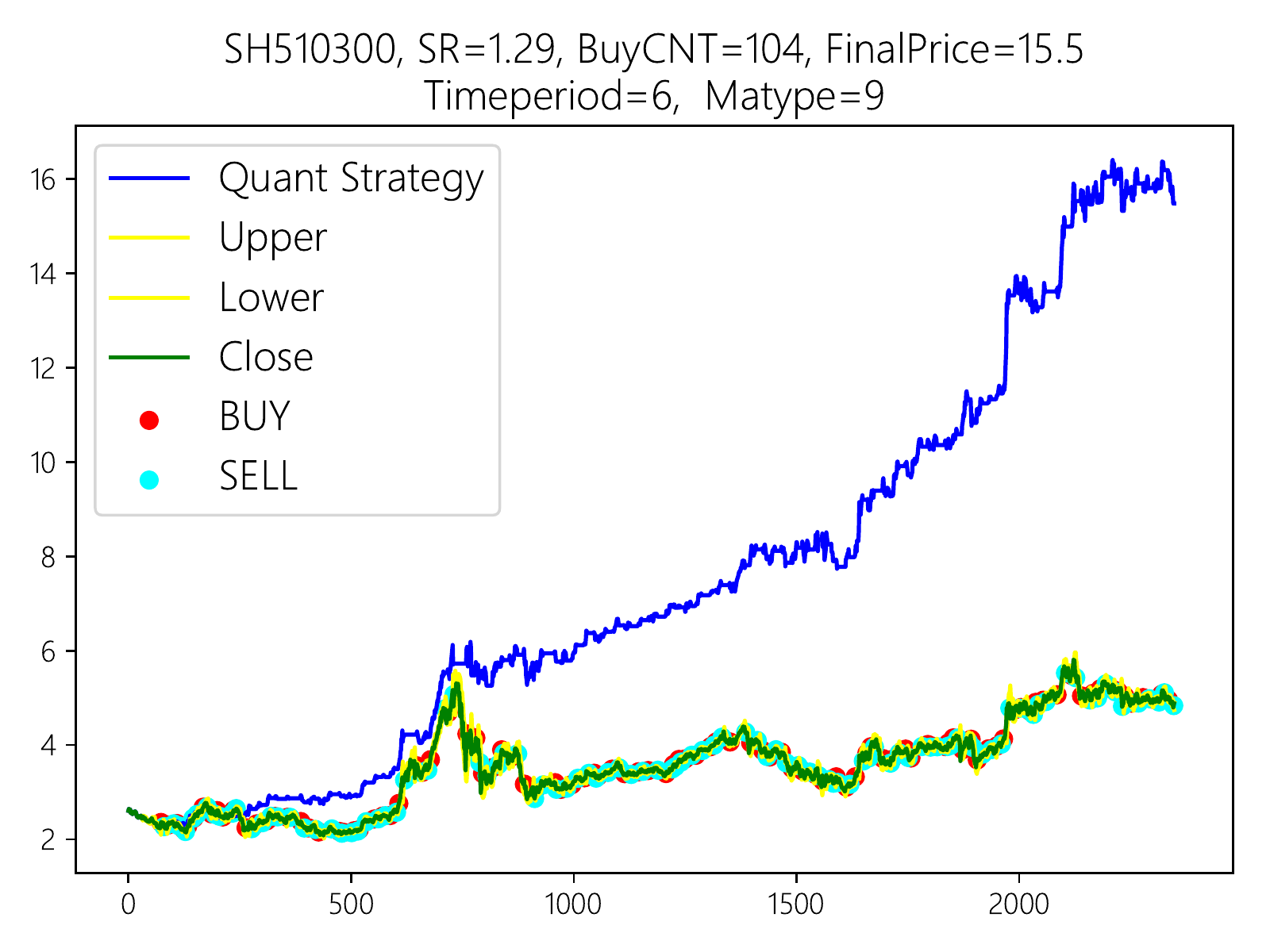}}
	\quad 
	\subfigure[Timeperiod=12.]{\label{fig:sh510300-keltner-mas2}
		\includegraphics[width=0.47\linewidth]{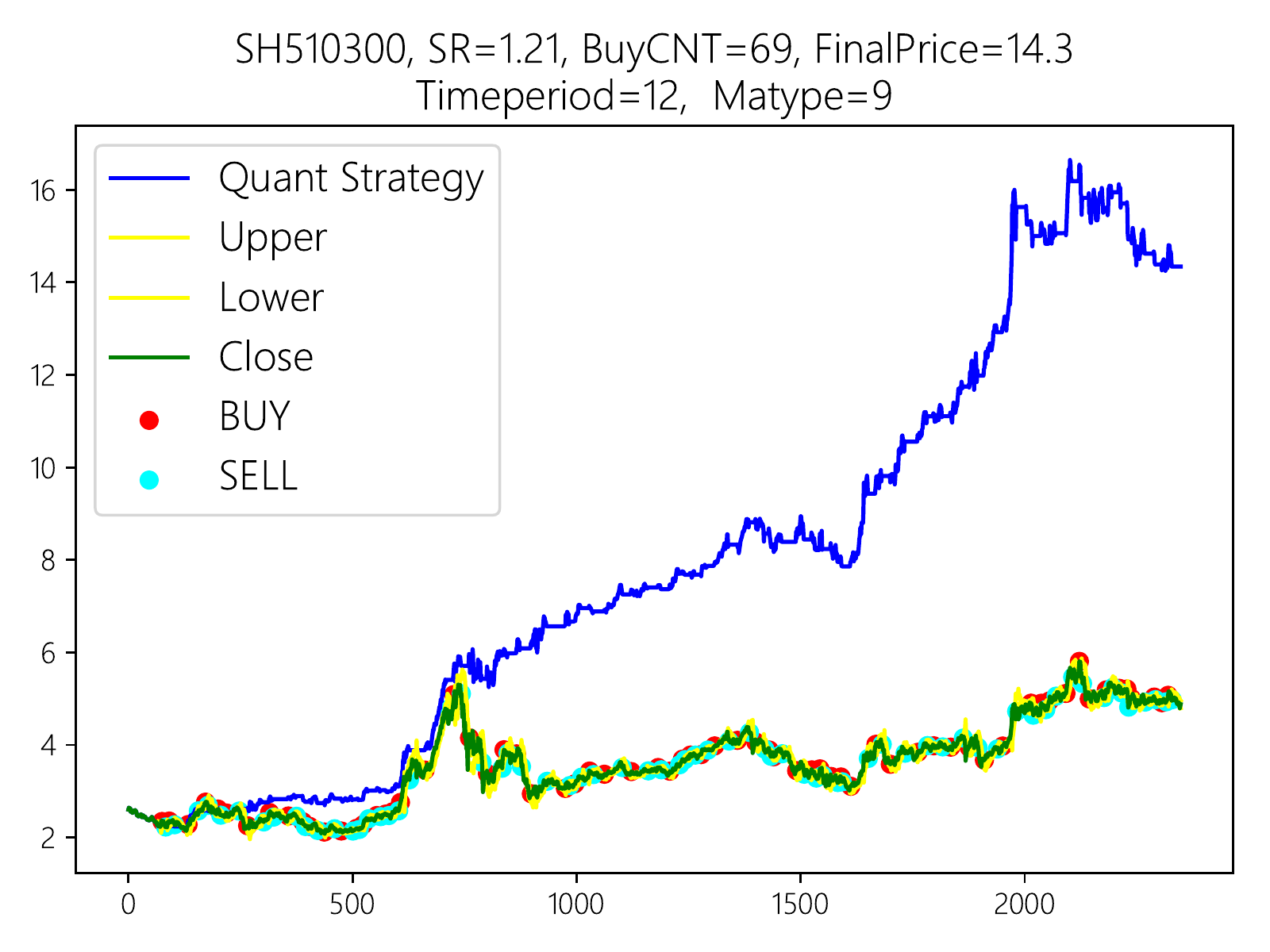}}
	\caption{Keltner strategy on SH510300 with non-adaptive MA.}
	\label{fig:sh510300-keltner-mas12}
\end{figure}

The result of the Keltner strategy with non-adaptive MA on the SH510300 is shown in Figure~\ref{fig:sh510300-keltner-mas12}.

The detailed measures in Figure~\ref{fig:sh510300-keltner-mas1} are given as follows where again the ``IR" is the information ratio whose benchmark is set to be the SH510300 itself.
\begin{python}
Initial Price:        2.604
Final Price:          15.481375580472795
RR of whole period:   5.945228717539476
RR/year:              1.2089084666981096
RR of year-1: 0.9591281739149894
RR of year-2: 1.1556931564748096
RR of year-3: 1.9551016478949061
RR of year-4: 1.0178450078822818
RR of year-5: 1.1858471067498066
RR of year-6: 1.1532814626720256
RR of year-7: 1.240845339361789
RR of year-8: 1.4062147816871258
RR of year-9: 1.1399219015987205
Total number of buy count:  104
MAX rate:  1.9551016478949061  MIN rate:  0.9591281739149894
MDD:  0.15582450832072614
SR:  1.2876574619127727
IR:  0.6059468181557052
\end{python}

\subsection{Results with AMA}

\begin{figure}[H]
	\centering  
	\vspace{-0.35cm} 
	\subfigtopskip=2pt 
	\subfigbottomskip=2pt 
	\subfigcapskip=-5pt 
	\subfigure[TimeperiodLong=54, TimeperiodShort=2, AdaWin=18, Matype=2.]{\label{fig:sh510300-keltner-ama1}
		\includegraphics[width=0.47\linewidth]{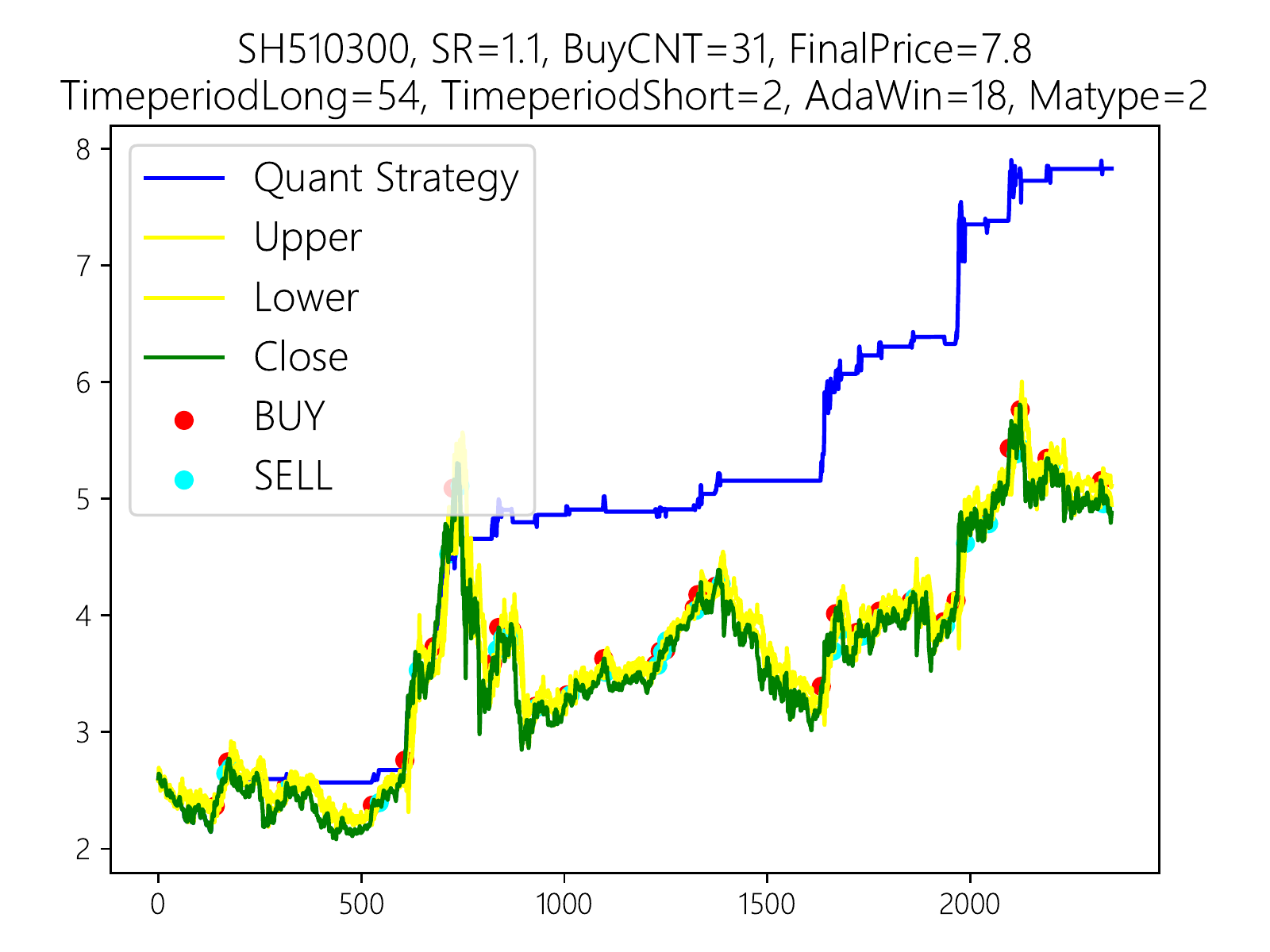}}
	\quad 
	\subfigure[TimeperiodLong=46, TimeperiodShort=2, AdaWin=16, Matype=2.]{\label{fig:sh510300-keltner-ama2}
		\includegraphics[width=0.47\linewidth]{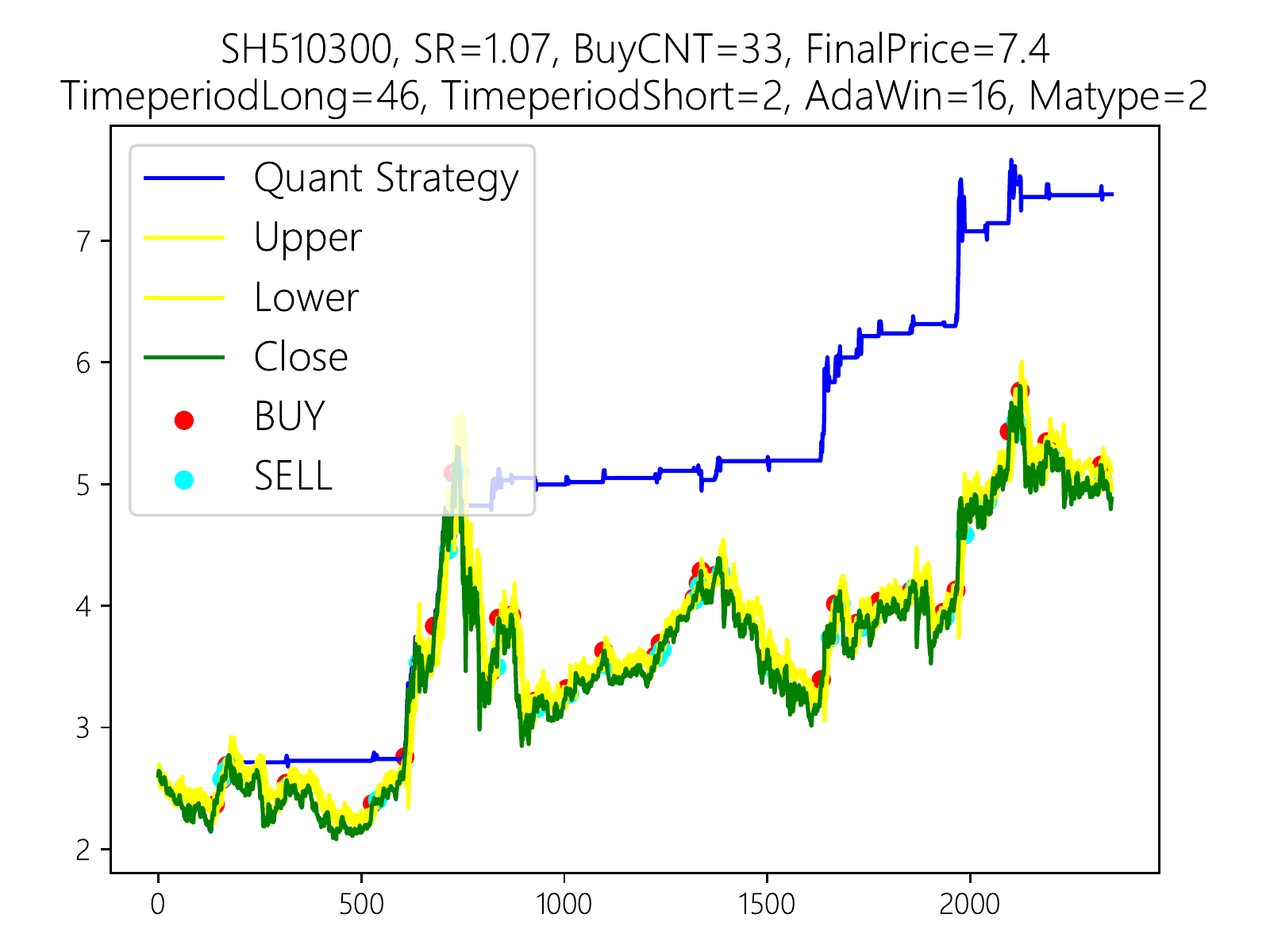}}
	\caption{Keltner strategy on SH510300 with AMA where ``Matype=2" in the title indicates the AMA is from SMA; see the ``adaptiveMovAvg" code on p.~\pageref{fig:sh510300-sp500-sma-ema}.}
	\label{fig:sh510300-keltner-ama12}
\end{figure}

The result of the Keltner strategy with AMA on the SH510300 is shown in Figure~\ref{fig:sh510300-keltner-ama12}.

The detailed measures in Figure~\ref{fig:sh510300-keltner-ama1} are given as follows where again the ``IR" is the information ratio whose benchmark is set to be the SH510300 itself.

\begin{python}
Initial Price:        2.604
Final Price:          7.83078199364342
RR of whole period:   3.007212747174892
RR/year:              1.124320637334687
RR of year-1: 0.9974343484257681
RR of year-2: 0.9892031706228506
RR of year-3: 1.812199489656705
RR of year-4: 1.04416668679696
RR of year-5: 1.0104317126809472
RR of year-6: 1.0494680691869664
RR of year-7: 1.2081837140293505
RR of year-8: 1.18035738803435
RR of year-9: 1.064563078825207
Total number of buy count:  31
MAX rate:  1.812199489656705  MIN rate:  0.9892031706228506
MDD:  0.18872919818456887
SR:  1.1016017674998995
IR:  0.1333340167631022
\end{python}

\section{RSI Overbought and Oversold Strategy}

\subsection{Relative Strength Index (RSI)}
	
The relative strength index (RSI) calculates a ratio of the recent upward price movements to the absolute price movement such that the RSI ranges from 0 to 100. The RSI is a momentum indicator used in technical analysis that measures the magnitude of recent price changes to evaluate overbought or oversold conditions in the price of a stock or other asset.
Specifically, the RSI is interpreted as an overbought/oversold indicator when the value is over 70/below 30. You can also look for divergence with price. If the price is making new highs/lows, and the RSI is not, it indicates a reversal.
The RSI was developed by J. Welles Wilder and was first introduced in his article in the June, 1978 issue of Commodities magazine, now known as the Futures magazine, and is detailed in his book \citep{wilder1978new}.  Given the time frame $N$, the definition of the RSI is shown in the following Algorithm~\ref{alg:rsi}.
\begin{algorithm}[H] 
\caption{Compute the $i$-th element of the RSI sequence} 
\label{alg:rsi} 
\begin{algorithmic}[1] 
\Require For the $i$-th element of RSI sequence, ``upavg, dnavg" are from last element.
\State Given the time period $N$;
\If{close[$i$]$>$ close[$i-1$]}
\State up = close[$i$] - close[$i-1$];
\State dn = 0;
\Else
\State up = 0;
\State dn = close[$i-1$]-close[$i$];
\EndIf
\State upavg = $\frac{\text{upavg}\times \text{($N-$1)+up}}{N}$;
\State dnavg = $\frac{\text{dnavg}\times \text{($N-$1)+dn}}{N}$;
\State RSI[i] = $100\times \frac{\text{upavg}}{\text{upavg+dnavg}}$;
\end{algorithmic} 
\end{algorithm}

\paragraph{The strategy} When the market is in overbought status, it may be primed for a trend reversal or corrective pullback in price such that the price will eventually go down. Under the overbought condition, when the rate of upward trend is smaller than a specific threshold (we call it a ``DiffRate" later), we consider the uptrend goes to an end, and initiate a short/sell position. On the contrary, under the oversold condition, when the rate of downward trend is smaller than a specific threshold, it represents the downtrend goes to an end, and we shall initiate a long/buy position. We call the strategy ``\textit{RSI Strategy without constraints}".

However, the RSI alone may not be an sufficient indicator for the overbought/oversold status. We further consider that if the closing price of the day is larger than the SMA of the closing price to some extent (captured by the ``SMArate" in the following Python code), it indicates a ``stronger" overbought signal. Similarly, if the closing price of the day is lower than the SMA of the closing price to some extent, it represents a ``stronger" oversold signal. We call this strategy ``\textit{RSI Strategy with constraints}".

The following ``algOverBoughtSoldRSI" code is written for Python 3.7. When ``rsitype=1", the algorithm computes the ``RSI Strategy without constraints"; while ``rsitype=2", the algorithm calculates the ``RSI Strategy with constraints".
\begin{python}
def algOverBoughtSoldRSI(data, position_buy, rsi, downThres=30, \
				upperThres=70, DiffRate=0.0024, rsitype=1, SMA=[], SMArate=0.001):
	"""
	:param data: data dictionary contains 'close' as data['close']
	:param position_buy: in the buy position if True, and False otherwise
	:param rsi: RSI of the asset
	:param downThres: RSI lower than downThres is considered as oversold
	:param upperThres: RSI larger than upperThres is considered as overbought
	:param DiffRate: the threshold for the end of a trend
	:param rsitype: 1 for RAW RSI; 2 for RSI+SMArate
	:param SMA: SMA of the closing price
	:param SMArate: band parameter for the second condition on overbought/oversold, in 0.001~0.02
	:return:
	"""
	for i in range(60, len(data['close'])-1):
		if rsitype == 1:  # RAW RSI
			condition1 = True
			condition2 = True
		elif rsitype == 2:  # RSI+SMArate
			condition1 = data['close'][i] < ((1-SMArate)*SMA[i])
			condition2 = data['close'][i] > ((1+SMArate)*SMA[i])
		# Oversold - BUY
		if rsi[i]<downThres and condition1:
			downrate = (data['close'][i-1] - data['close'][i])/data['close'][i-1]
			if downrate<=DiffRate and downrate>=0:
				if not position_buy:
					# TODO: BUY
					position_buy = True
					# Overbought - SELL
		elif rsi[i]>upperThres and condition2:
			uprate = (data['close'][i] - data['close'][i-1])/data['close'][i-1]
			if uprate<=DiffRate and uprate>=0:
				if position_buy:
					# TODO: SELL
					position_buy = False
\end{python}

\subsection{Result without constraints}
\begin{figure}[H]
	\centering  
	\vspace{-0.35cm} 
	\subfigtopskip=2pt 
	\subfigbottomskip=2pt 
	\subfigcapskip=-5pt 
	\subfigure[TimeperiodRSI=6, DiffRate=0.00043793.]{\label{fig:sh510300-rsioverbought-1}
		\includegraphics[width=0.47\linewidth]{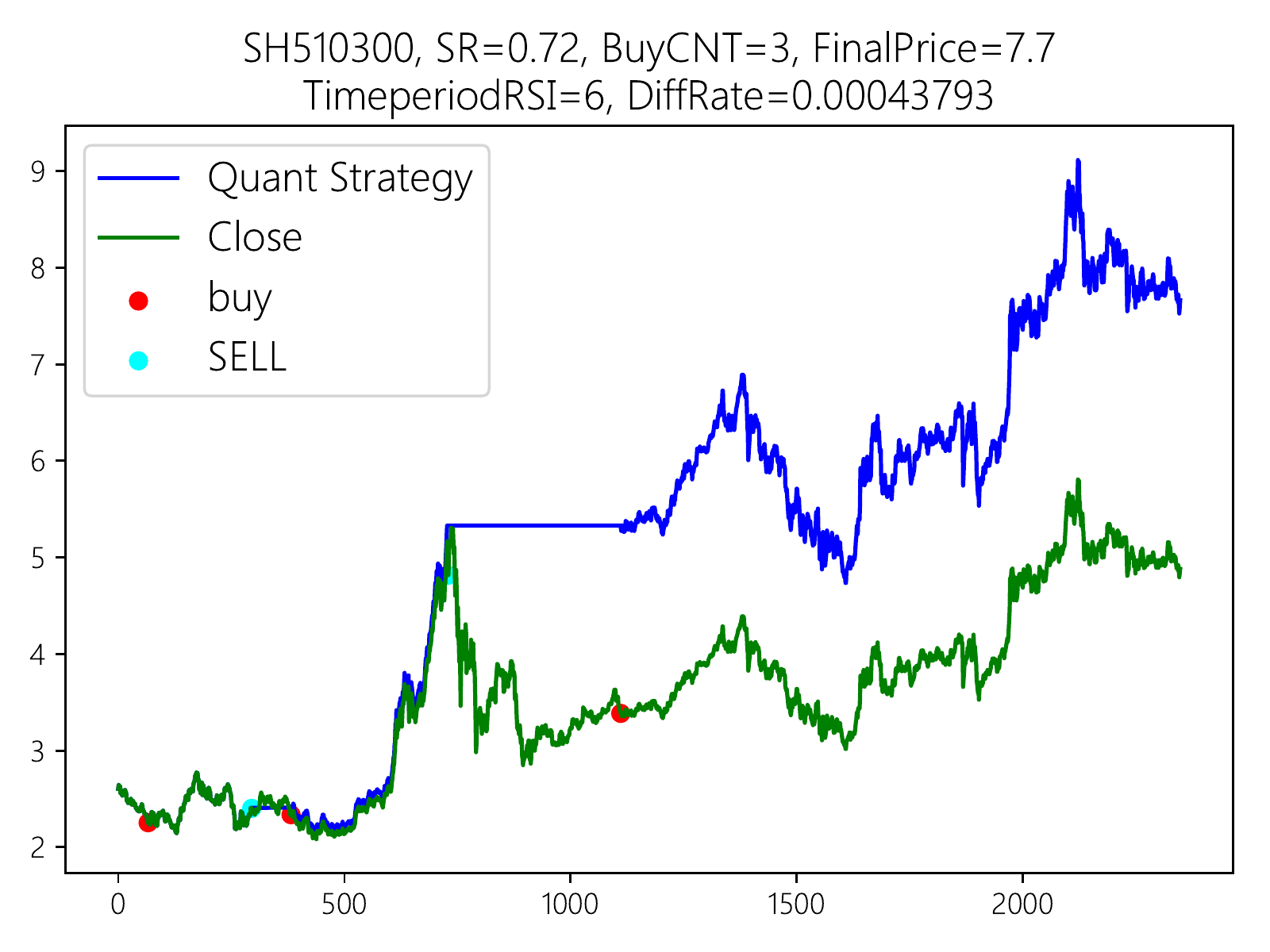}}
	\quad 
	\subfigure[TimeperiodRSI=6, DiffRate=0.0006069.]{\label{fig:sh510300-rsioverbought-2}
		\includegraphics[width=0.47\linewidth]{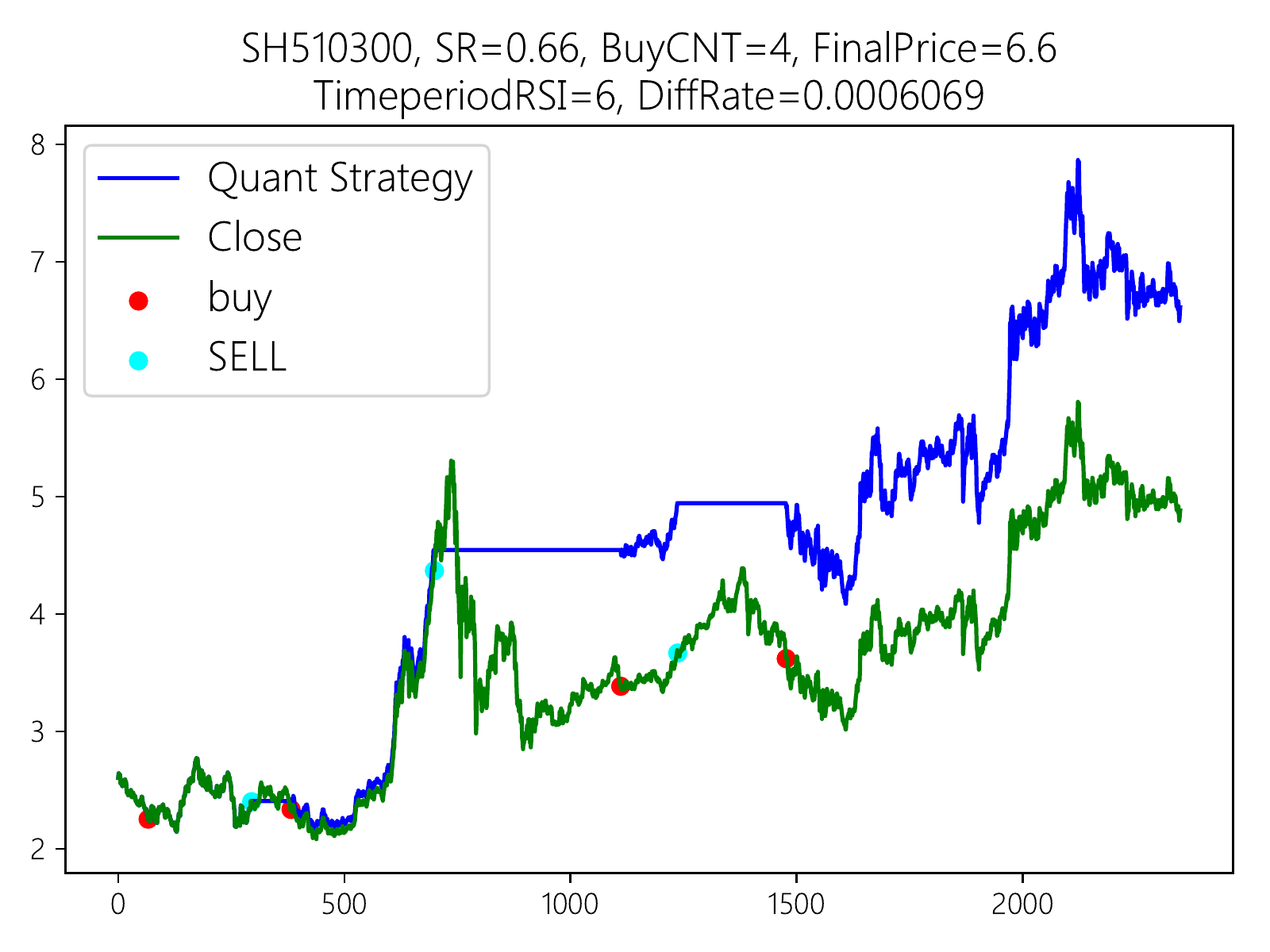}}
	\caption{RSI strategy without constraints on SH510300.}
	\label{fig:sh510300-rsioverbought12}
\end{figure}

The result of the \textit{RSI Strategy without constraints} on the SH510300 is shown in Figure~\ref{fig:sh510300-rsioverbought12}.
The detailed measures in Figure~\ref{fig:sh510300-rsioverbought-1} are given as follows where again the ``IR" is the information ratio whose benchmark is set to be the SH510300 itself. Though SR=0.72 is acceptable in some sense, the result in Figure~\ref{fig:sh510300-rsioverbought-1} is still risky as it only finds the first drawdown, and totally follows the second drawdown (around 1300-th day). While the result in Figure~\ref{fig:sh510300-rsioverbought-2} is relatively poor in SR, it actually avoids the second drawdown. On the other hand, the BuyCNT for both of the above two results are low, which may lose some trading signals.

\begin{python}
Initial Price:        2.604
Final Price:          7.6612689111278485
RR of whole period:   2.942115557268759
RR/year:              1.1217049662146583
RR of year-1: 0.9577572964669739
RR of year-2: 0.8672124701729041
RR of year-3: 2.384579870729455
RR of year-4: 1.0
RR of year-5: 1.0939617083946982
RR of year-6: 0.9492399565689466
RR of year-7: 1.1102585961920999
RR of year-8: 1.2471590909090908
RR of year-9: 1.052435233160622
Total number of buy count:  3
MAX rate:  2.384579870729455  MIN rate:  0.8672124701729041
MDD:  0.31291277613299934
SR:  0.7181964777285531
IR:  0.24693874680366934
\end{python}

\subsection{Result with constraints}
\begin{figure}[H]
	\centering  
	\vspace{-0.35cm} 
	\subfigtopskip=2pt 
	\subfigbottomskip=2pt 
	\subfigcapskip=-5pt 
	\subfigure[TimeperiodRSI=2, TimeperiodMA=38, DiffRate=0.0001, SMArate=0.02.]{\label{fig:sh510300-rsioverbought-constraint-1}
		\includegraphics[width=0.47\linewidth]{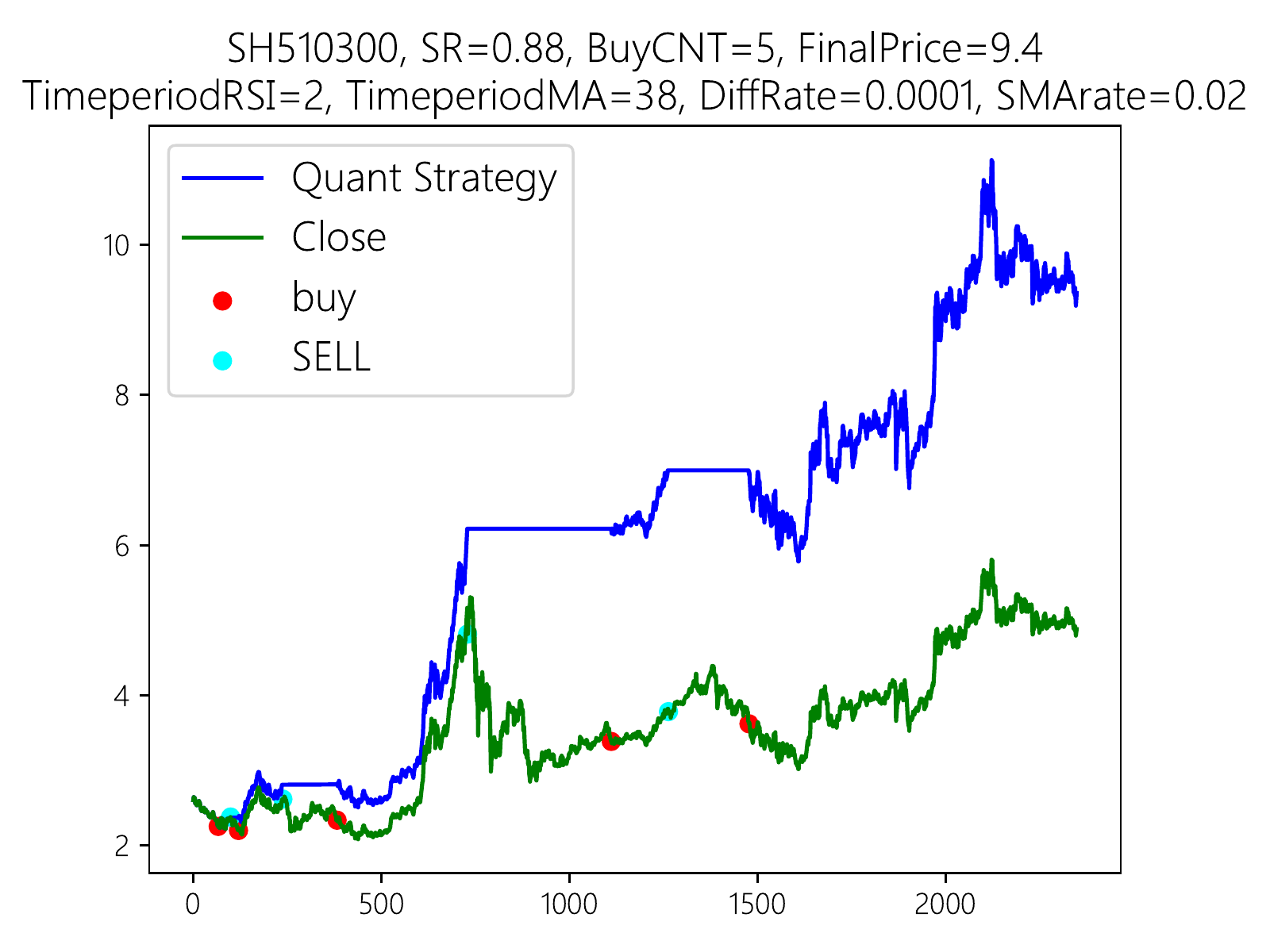}}
	\quad 
	\subfigure[TimeperiodRSI=2, TimeperiodMA=6, DiffRate=0.000644, SMArate=0.0136.]{\label{fig:sh510300-rsioverbought-constraint-2}
		\includegraphics[width=0.47\linewidth]{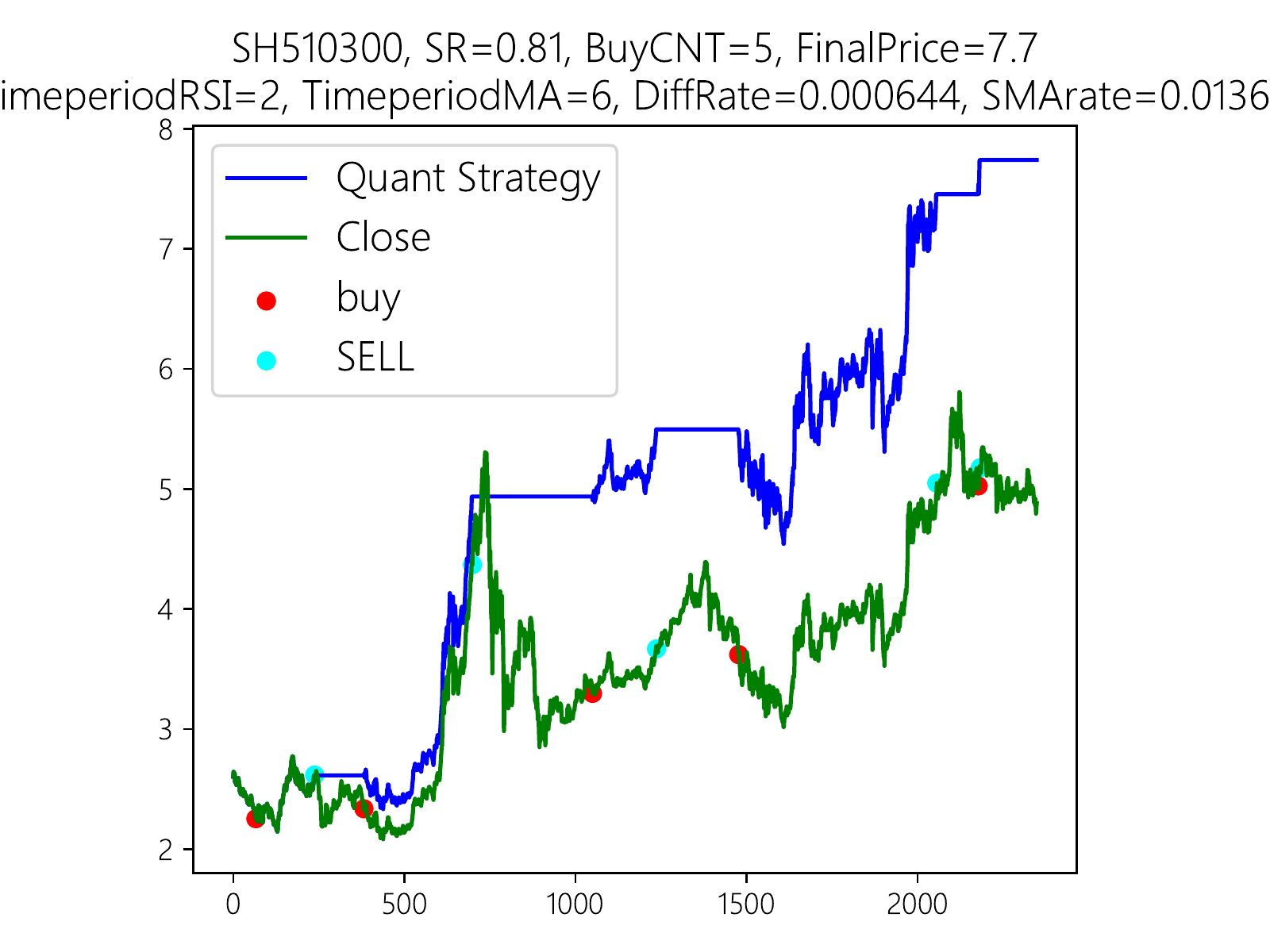}}
	\caption{RSI strategy with constraints on SH510300 where ``TimeperiodMA" in the title is the time frame for the SMA.}
	\label{fig:sh510300-rsioverbought-constraint12}
\end{figure}

The result of the \textit{RSI Strategy with constraints} on the SH510300 is shown in Figure~\ref{fig:sh510300-rsioverbought-constraint12}.
The detailed measures in Figure~\ref{fig:sh510300-rsioverbought-constraint-1} are given as follows where again the ``IR" is the information ratio whose benchmark is set to be the SH510300 itself. Interestingly, though the turnover is small (low ``BuyCNT"), every sell signal from the strategies in both Figure~\ref{fig:sh510300-rsioverbought-constraint-1} and \ref{fig:sh510300-rsioverbought-constraint-2} tells a lot about the ``big" downtrend.
\begin{python}
Initial Price:        2.604
Final Price:          9.355288065508722
RR of whole period:   3.5926605474303845
RR/year:              1.1458083483941202
RR of year-1: 1.078502286731385
RR of year-2: 0.9151307329618517
RR of year-3: 2.3845798707294548
RR of year-4: 1.0
RR of year-5: 1.0939617083946982
RR of year-6: 0.9934511326283224
RR of year-7: 1.1102585961920999
RR of year-8: 1.2471590909090908
RR of year-9: 1.052435233160622
Total number of buy count:  5
MAX rate:  2.3845798707294548  MIN rate:  0.9151307329618517
MDD:  0.1744446357843981
SR:  0.8804066887285559
IR:  0.3434084355517314
\end{python}

\subsection{More to Go}
The relative momentum index (RMI) is a variation on the RSI. To determine up and down days, the RSI uses the closing price compared to the previous closing price. While, the RMI uses the closing price compared to the closing price $M$ days ago (known as the ``look back frame"). Apparently, an RMI with a time period of 1 is equal to the RSI. Likewise, the RMI ranges from 0 to 100 and the RMI can also be interpreted as an overbought/oversold indicator when the value is over 70/below 30 respectively. We can also look for divergence with price. If the price is making new highs/lows, and the RMI is not, it indicates a reversal signal. The RMI provides a more flexible algorithm on the above strategy and we shall not provide more tests for simplicity.

Given the time frame $N$ and look back frame $M$, the definition of the RMI is shown in the following Algorithm~\ref{alg:rsi-rmi}.

\begin{algorithm}[H] 
	\caption{Compute the $i$-th element of the RMI sequence} 
	\label{alg:rsi-rmi} 
	\begin{algorithmic}[1] 
		\Require For the $i$-th element of RMI sequence, ``upavg, dnavg" are from last element.
		\State Given the time period $N$, and look back frame $M$;
		\If{close[$i$]$>$ close[$i-M$]}
		\State up = close[$i$] - close[$i-M$];
		\State dn = 0;
		\Else
		\State up = 0;
		\State dn = close[$i-M$]-close[$i$];
		\EndIf
		\State upavg = $\frac{\text{upavg}\times \text{($N-$1)+up}}{N}$;
		\State dnavg = $\frac{\text{dnavg}\times \text{($N-$1)+dn}}{N}$;
		\State RMI[i] = $100\times \frac{\text{upavg}}{\text{upavg+dnavg}}$;
	\end{algorithmic} 
\end{algorithm}

\section{Aroon Strategy}

\subsection{Aroon}
The word ``Aroon" is Sanskrit for ``dawn's early light". The Aroon indicator attempts to show when a new trend is dawning. Similar to the Keltner channel, the indicator consists of two lines (Up and Down) that measure how long it has been since the highest high/lowest low has occurred within an $N$ period frame.

The Aroon indicator was developed by Tushar S. Chande and first described in the September 1995 issue of Technical Analysis of Stocks \& Commodities magazine \citep{chande1995time}. 
Given the time period $N$, the Aroon is defined as follows:
\begin{equation}\label{equation:aroon}
\left\{
\begin{aligned}
\text{Aroon Up} &= 100\times \left(\frac{N-\text{PeriodsSinceHighestHigh}}{N}\right);\\
\text{Aroon Down} &= 100\times \left(\frac{N-\text{PeriodsSinceLowestLow}}{N}\right);\\
\text{Aroon Oscillator} &= \text{Aroon Up} - \text{Aroon Down}.\\
\end{aligned}
\right.
\end{equation}
The ``Aroon Up"/``Aroon Down" line measures the strength of the upward/downward trend: 
\begin{itemize}
\item When the ``Aroon Up" is staying between 70 and 100 then it indicates an upward trend;
\item When the ``Aroon Down" is staying between 70 and 100 then it indicates a downward trend; 
\item A strong upward trend is indicated when the ``Aroon Up" is above 70 while the ``Aroon Down" is below 30;
\item A strong downward trend is indicated when the ``Aroon Down" is above 70 while the ``Aroon Up" is below 30.
\end{itemize}

The \textit{Aroon oscillator} is calculated by subtracting the ``Aroon Down" from the ``Aroon Up". The resultant number will oscillate between 100 and -100. The ``Aroon oscillator" will be high when the ``Aroon Up" is high and the ``Aroon Down" is low, indicating a strong upward trend. The ``Aroon oscillator" will be low when the ``Aroon Down" is high and the ``Aroon Up" is low, indicating a strong downward trend. When the Up and Down are approximately equal, the ``Aroon Oscillator" will hover around zero, indicating a weak trend or consolidation. 

\paragraph{The strategy} Also look for crossovers as those in the Two-Average strategy. When the ``Aroon Down" crosses above the ``Aroon Up", it indicates a weakening of the upward trend (and vice versa). The strategy then goes to buy if the ``Aroon Up" crosses above the ``Aroon Down"; to sell if the ``Aroon Down" crosses above the ``Aroon Up". This is called the \textit{unconditioned Aroon strategy}.

The \textit{conditioned Aroon strategy} goes further that, when the ``Aroon Up" crosses above the ``Aroon Down", i.e., the upward trend might be coming; if the downward trend is in a weak signal, this should reveal a \textbf{stronger} buy signal. And the ``weak down-trend" can be identified that the ``Aroon Down" is smaller than a threshold, e.g., 45 in our test. Likewise, when the ``Aroon Down" crosses above the ``Aroon Up", the further ``weak up-trend" is identified by a small ``Aroon Up", e.g., 45 in our test.

The following ``algAroon" code is written for Python 3.7. When ``aroonType=1", the algorithm computes the ``unconditioned Aroon strategy"; while ``aroonType=2", the algorithm calculates the ``conditioned Aroon strategy".
\begin{python}
def algAroon(data, position_buy, aroon_up, aroon_down, aroonType=1):
	"""
	:param data: data dictionary contains 'close' as data['close']
	:param position_buy: in the buy position if True, and False otherwise
	:param aroon_up:   Aroon Up line
	:param aroon_down: Aroon Down line
	:param aroonType: 1 for RAW Aroon; 2 for conditioned Aroon
	:return:
	"""
	for i in range(60, len(data['close'])-1):
		if aroonType==1:
			condition1 = True
			condition2 = True
		elif aroonType==2:
			condition1 = (aroon_down[i]<45)
			condition2 = (aroon_up[i]<45)
	# aroon_up cross above down - BUY
	if (aroon_up[i-1]<aroon_down[i-1]) \
		and (aroon_up[i]>aroon_down[i]) \
		and condition1 \
		and (not position_buy):
		# TODO: BUY
		position_buy = True
	# aroon_down cross above up - SELL
	elif (aroon_up[i-1]>aroon_down[i-1]) \
		and (aroon_up[i]<aroon_down[i]) \
		and condition2
		and position_buy:
		# TODO: SELL
		position_buy = False
\end{python}

\subsection{Results without and with condition}
\begin{figure}[H]
	\centering  
\vspace{-0.35cm} 
\subfigtopskip=2pt 
\subfigbottomskip=2pt 
\subfigcapskip=-5pt 
\subfigure[Timeperiod=25, unconditioned Aroon.]{\label{fig:sh510300-aroon-1}
	\includegraphics[width=0.47\linewidth]{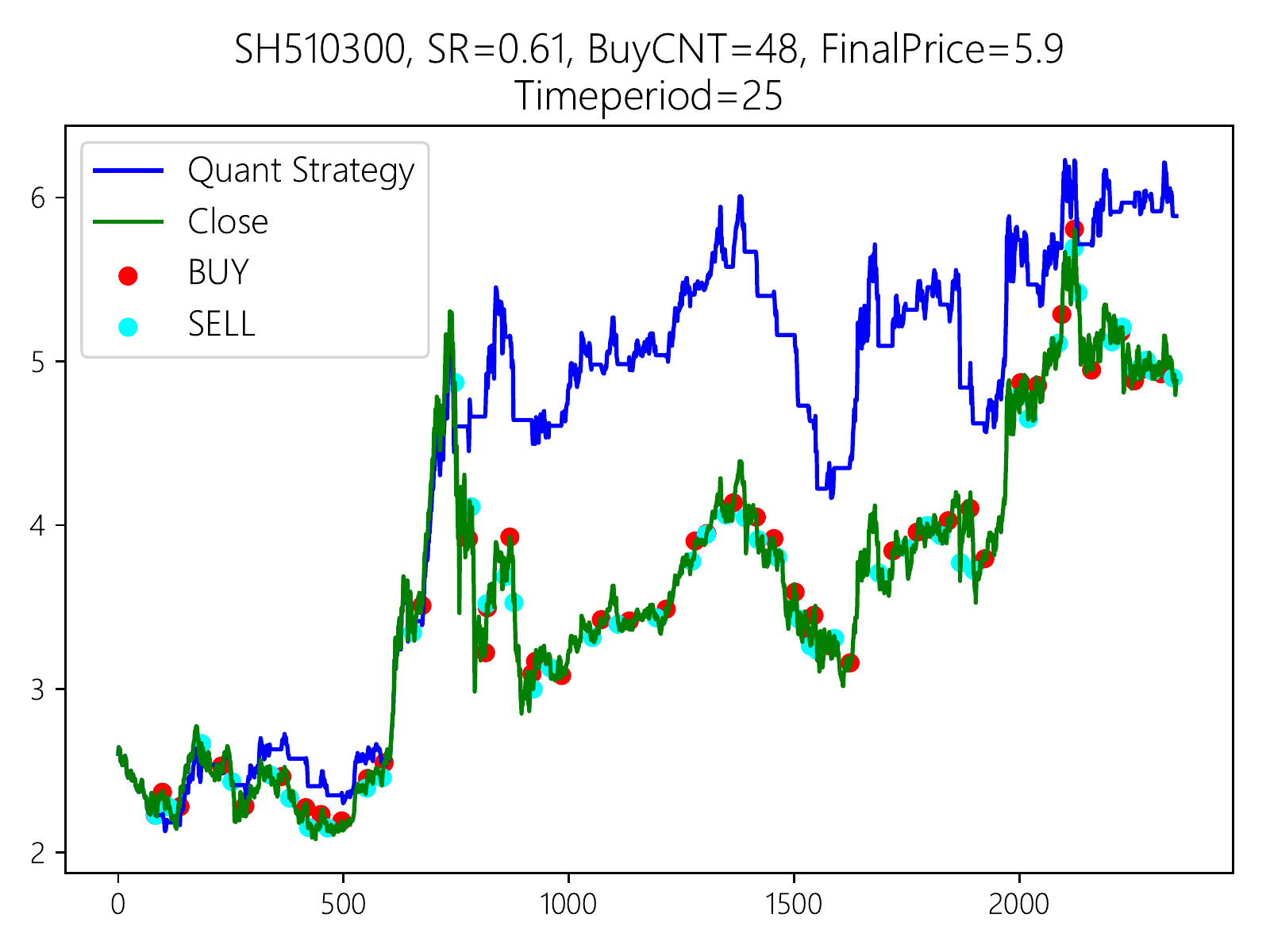}}
\quad 
\subfigure[Timeperiod=6, conditioned Aroon.]{\label{fig:sh510300-aroon-2}
	\includegraphics[width=0.47\linewidth]{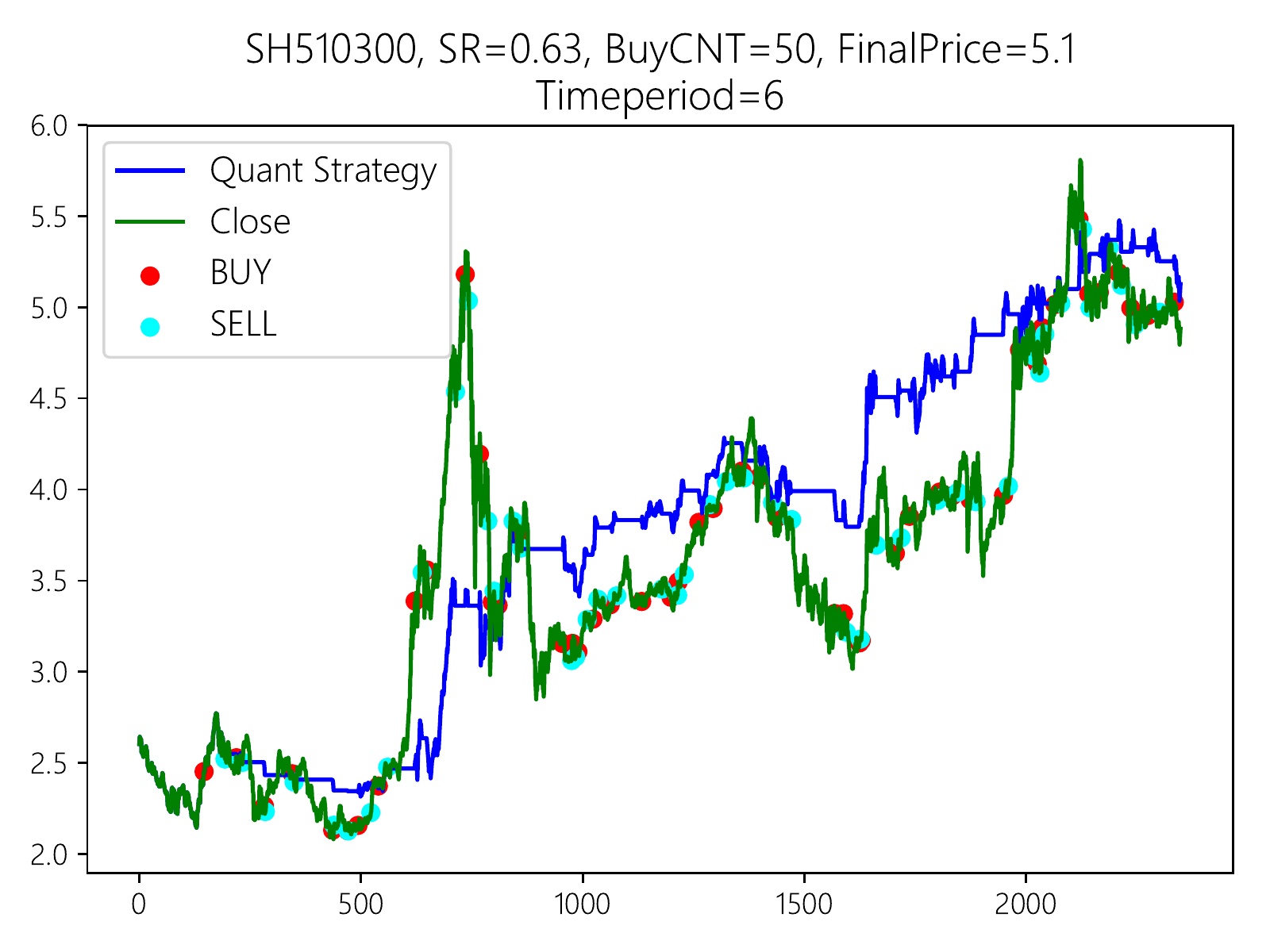}}
\caption{Aroon overbought and oversold strategy on SH510300 without and with constraints.}
\label{fig:sh510300-aroon-12}
\end{figure}

The Aroon strategy does not give many promising results and 
the result of the Aroon strategy on the SH510300 is shown in Figure~\ref{fig:sh510300-aroon-12}.
The detailed measures in Figure~\ref{fig:sh510300-aroon-1} are given as follows where again the ``IR" is the information ratio whose benchmark is set to be the SH510300 itself.

\begin{python}
Initial Price:        2.604
Final Price:          5.887990726795886
RR of whole period:   2.261133151611323
RR/year:              1.090712566635823
RR of year-1: 0.9556453450783325
RR of year-2: 0.9082324091273398
RR of year-3: 1.9710427015497842
RR of year-4: 1.0297552354919766
RR of year-5: 1.1289329700532444
RR of year-6: 0.9714393615034962
RR of year-7: 1.0325934356179702
RR of year-8: 1.101722320850014
RR of year-9: 1.0265693427413187
Total number of buy count:  48
MAX rate:  1.9710427015497842  MIN rate:  0.9082324091273398
MDD:  0.3068100965375994
SR:  0.614348285853909
IR:  0.03005677887500972
\end{python}

\section{Bollinger Bands Strategy}

\subsection{Bollinger Bands}
Bollinger bands consist of three lines. The middle band is an SMA (generally 20 periods) of the typical price (TP).\footnote{Again, one can also explore other moving average methods in this sense.} The upper and lower bands are $dev$ times standard deviations (generally $dev=$2) above and below the middle band. The bands widen and narrow when the volatility of the price becomes higher or lower, respectively.

Bollinger bands do not, in themselves, generate buy or sell signals; they are an indicator of overbought or oversold conditions. When the price is near the upper or lower band it indicates that a reversal may be imminent. The middle band becomes a support or resistance level. The upper and lower bands can also be interpreted as price targets. When the price bounces off of the lower band and crosses the middle band, then the upper band becomes the price target.

Bollinger bands were developed and copyrighted by John Bollinger, a famous technical trader \citep{bollinger1992using, bollinger2002bollinger}. See also the empirical study on the Bollinger band in various regions \citep{leung2003empirical}. The formula to compute the band is given by:
\begin{equation}\label{equation:bollinger-band}
\left\{
\begin{aligned}
	\text{TP} &=  \frac{\text{high+low+close}}{3};\\
	\text{Middle Band} &= \text{SMA(TP,  }N);\\
	\text{Upper Band} &= \text{Middle}+ dev \times \sigma(\text{TP}, N);\\
	\text{Lower Band} &= \text{Middle}- dev \times \sigma(\text{TP}, N),\\
\end{aligned}
\right.
\end{equation}
where $N$ is the given time period and $\sigma(\cdot)$ is the function to calculate the standard deviation.

\paragraph{Keltner channels vs. Bollinger bands}
These two indicators are quite similar. Keltner channels use ATR to calculate the upper and lower bands while Bollinger bands use standard deviation instead. Both of them are indicators of overbought or oversold.

\paragraph{The strategy} The indicator can also be used for stocks and indices.﻿﻿ A stock experiencing a high level of volatility has a higher standard deviation, and a low volatility stock has a lower standard deviation. As mentioned above, the band length given by the standard deviation tells us that the asset price should be probably in the bands. If the price crosses above the upper band, it will still move into the bands again; similarly when it crosses below the lower band. The strategy then acts to sell when it crosses above the upper band; and to buy when it crosses below the lower band. The method is different from the Keltner channel strategy, although they are both indicators of overbought or oversold. To be specific,  when the price cross above the upper band, we buy in the Keltner strategy; while we sell in the Bollinger strategy. The difference is in that the standard deviation tells a lot about the distribution; like it in the Gaussian distribution, 95\% percent within two standard deviations and 99.7\% within three standard deviations from the mean.

\subsection{Results with non-adaptive MA}
\begin{figure}[H]
	\centering  
	\vspace{-0.35cm} 
	\subfigtopskip=2pt 
	\subfigbottomskip=2pt 
	\subfigcapskip=-5pt 
	\subfigure[Timeperiod=5, dev=1.1.]{\label{fig:sh510300-bollinger-mas-1}
		\includegraphics[width=0.47\linewidth]{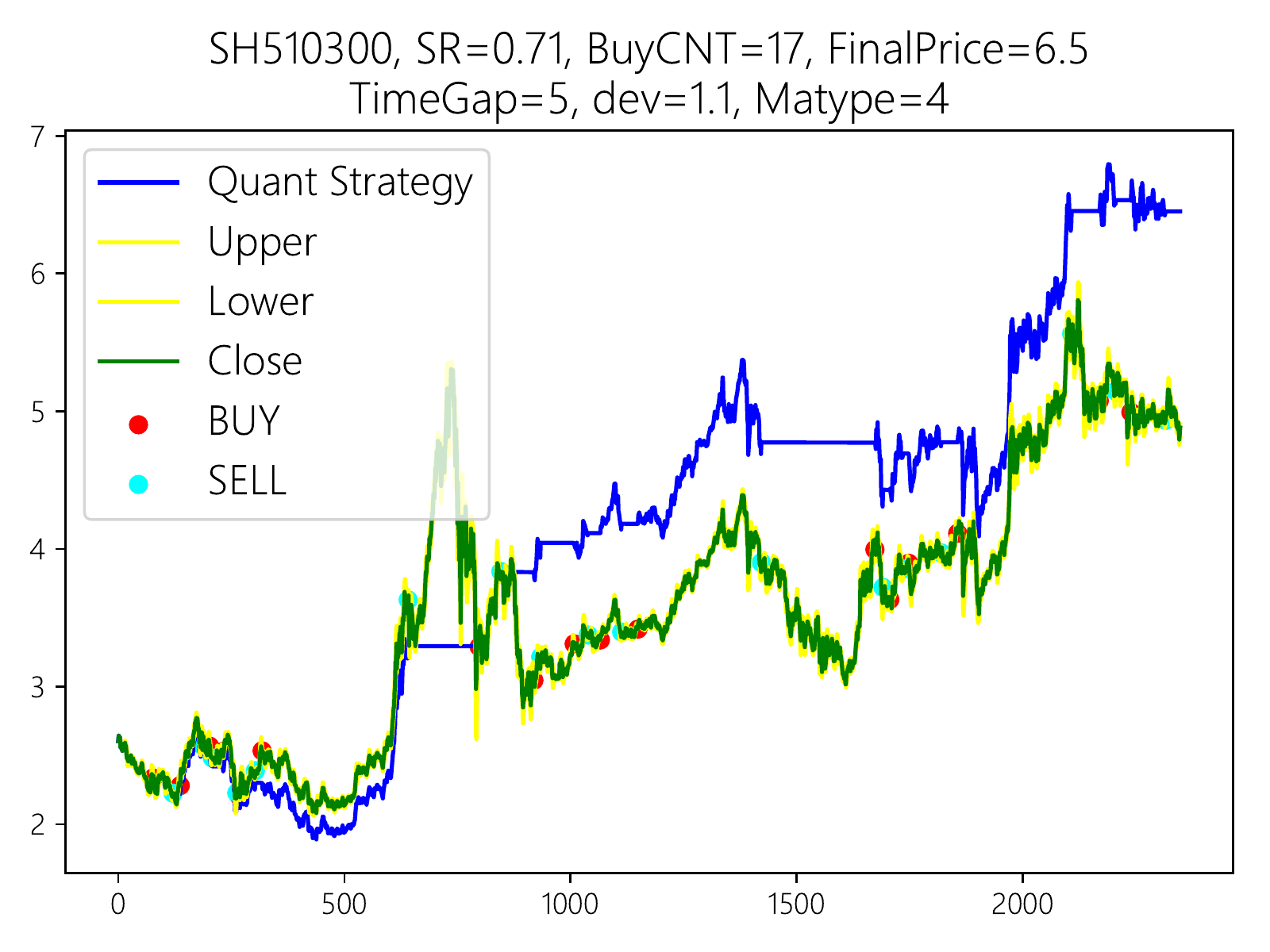}}
	\quad 
	\subfigure[Timeperiod=2, dev=3.0.]{\label{fig:sh510300-bollinger-mas-2}
		\includegraphics[width=0.47\linewidth]{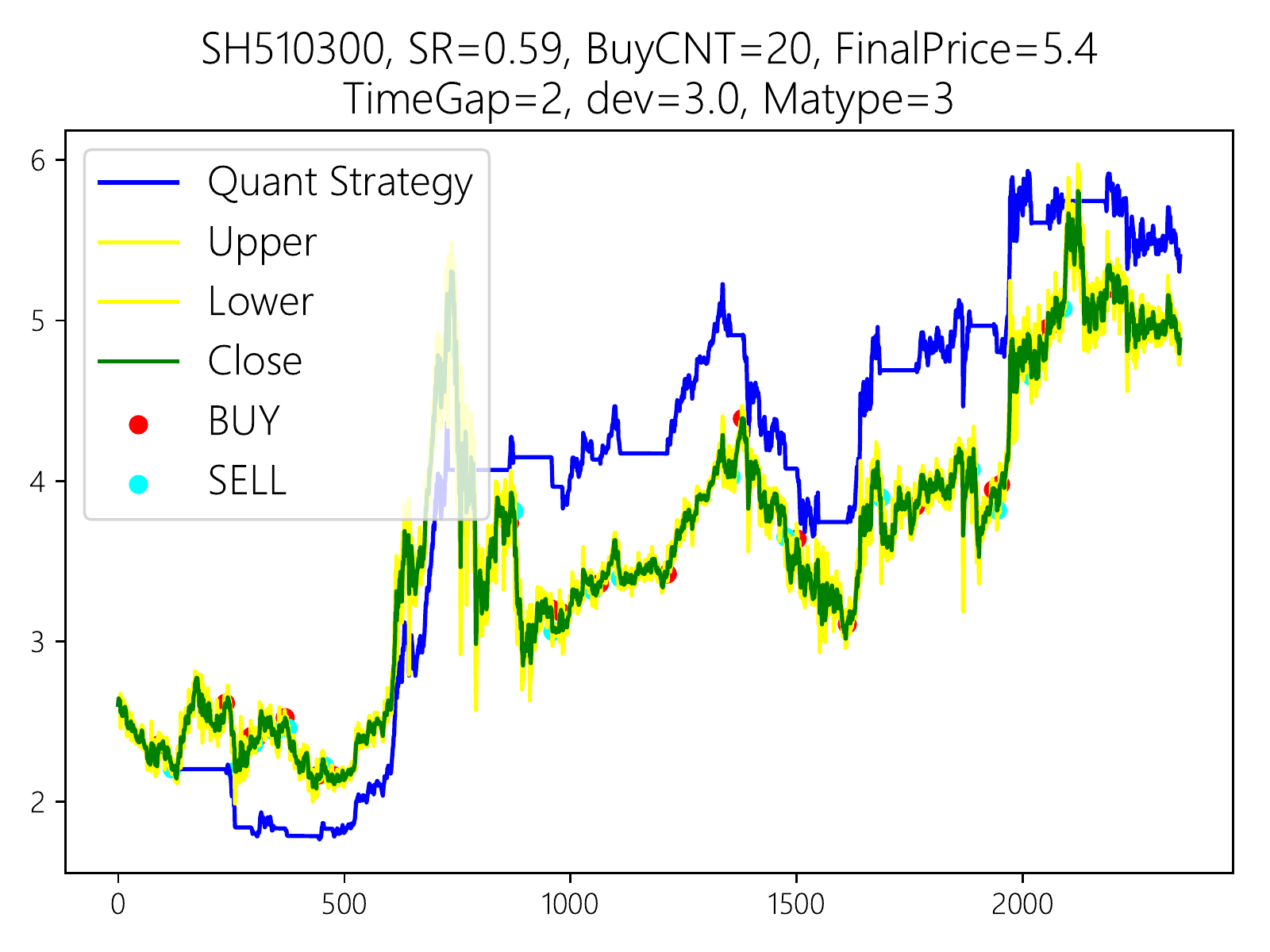}}
	\caption{Bollinger band strategy on SH510300 with non-adaptive MA.}
	\label{fig:sh510300-bollinger-mas-12}
\end{figure}

The result of the Bollinger bands strategy with non-adaptive MA on the SH510300 is shown in Figure~\ref{fig:sh510300-bollinger-mas-12}.
The detailed measures in Figure~\ref{fig:sh510300-bollinger-mas-1} are given as follows where again the ``IR" is the information ratio whose benchmark is set to be the SH510300 itself.
\begin{python}
Initial Price:        2.604
Final Price:          6.451714254225132
RR of whole period:   2.477616841100281
RR/year:              1.101377934198916
RR of year-1: 0.9167068001328132
RR of year-2: 0.7972897922117375
RR of year-3: 1.6750900346654767
RR of year-4: 1.2276852370378257
RR of year-5: 1.1238603905387452
RR of year-6: 1.0586927715683763
RR of year-7: 0.9832706903117197
RR of year-8: 1.1940637458782737
RR of year-9: 1.1859827767807352
Total number of buy count:  17
MAX rate:  1.6750900346654767  MIN rate:  0.7972897922117375
MDD:  0.2999781605155163
SR:  0.7058010527061538
IR:  0.07404386090219181
\end{python}

\subsection{Results with AMA}
\begin{figure}[H]
	\centering  
	\vspace{-0.35cm} 
	\subfigtopskip=2pt 
	\subfigbottomskip=2pt 
	\subfigcapskip=-5pt 
	\subfigure[TimeperiodLong=24, TimeperiodShort=8, AdaWin=18, dev=2.6, Matype=1.]{\label{fig:sh510300-bollinger-ama-1}
		\includegraphics[width=0.47\linewidth]{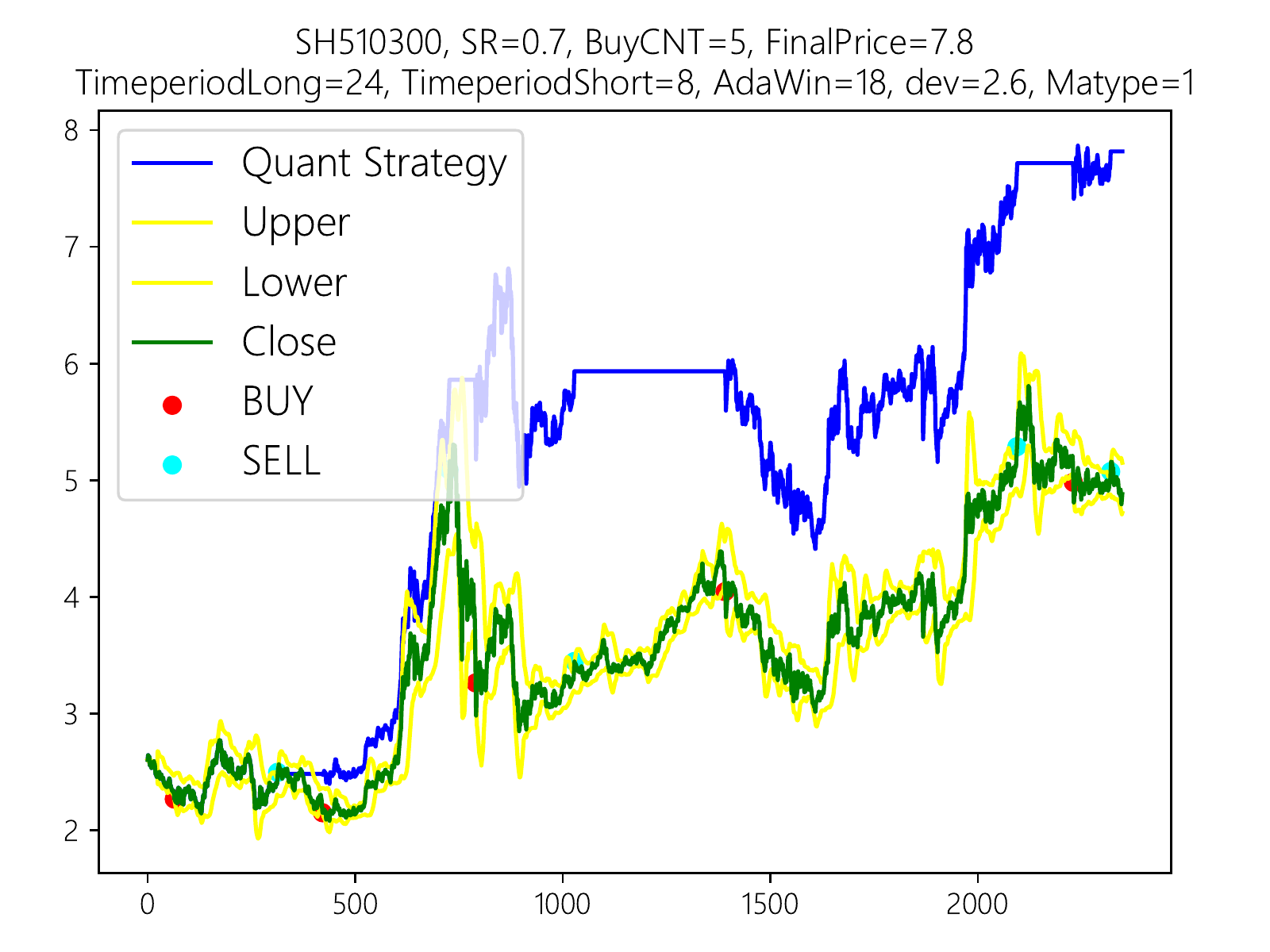}}
	\quad 
	\subfigure[TimeperiodLong=22, TimeperiodShort=8, AdaWin=16, dev=2.6, Matype=1.]{\label{fig:sh510300-bollinger-ama-2}
		\includegraphics[width=0.47\linewidth]{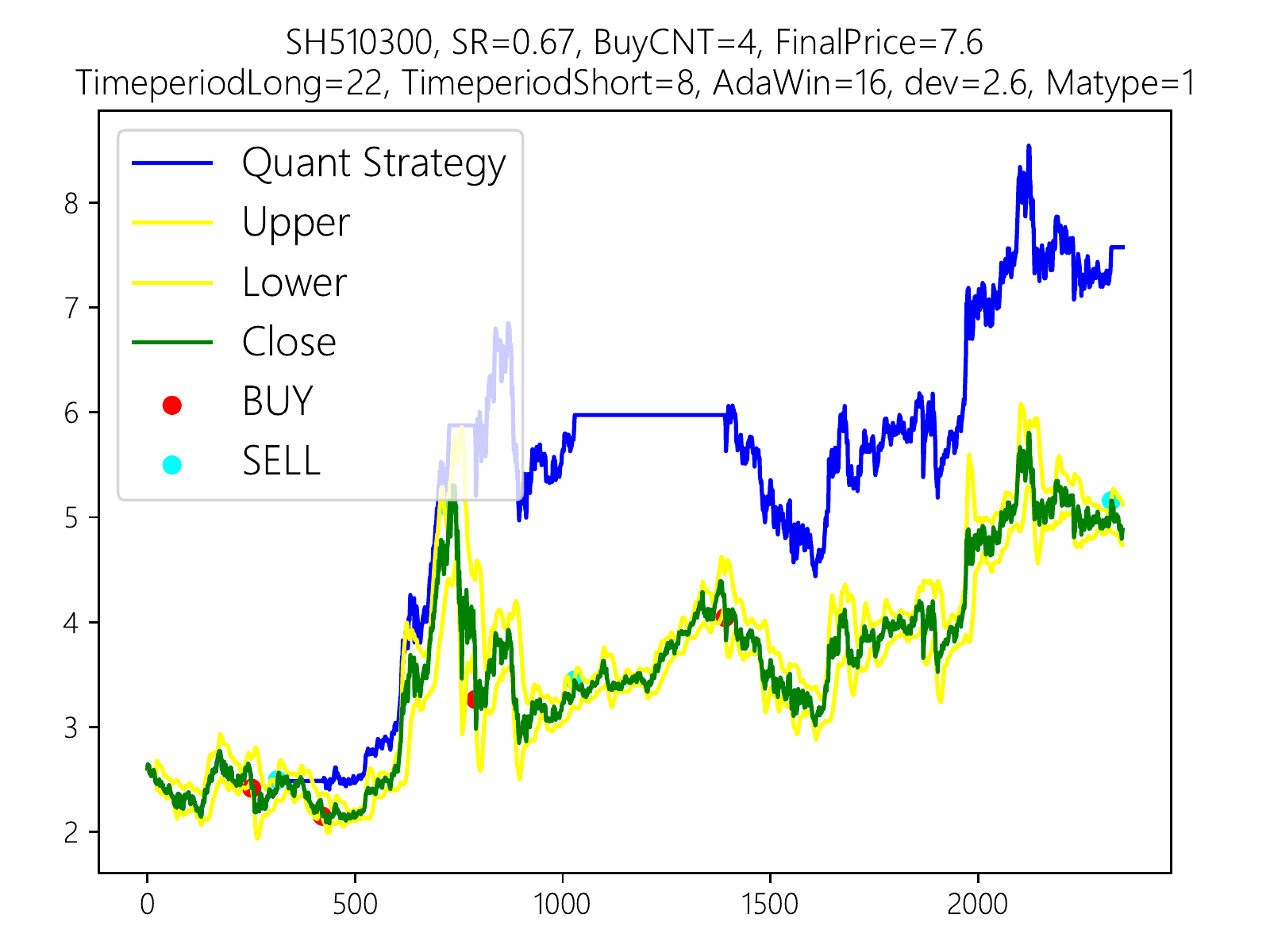}}
	\caption{Bollinger band strategy on SH510300 with AMA.  ``Matype=1" means the AMA is based on EMA; while, ``Matype=2" means the AMA is based on SMA (here, we only observe good results when Matype=1); see the ``adaptiveMovAvg" code on p.~\pageref{fig:sh510300-sp500-sma-ema}.}
	\label{fig:sh510300-bollinger-ama-12}
\end{figure}

The result of the Bollinger bands strategy with AMA on the SH510300 is shown in Figure~\ref{fig:sh510300-bollinger-ama-12}.

The detailed measures in Figure~\ref{fig:sh510300-bollinger-ama-1} are given as follows where again the ``IR" is the information ratio whose benchmark is set to be the SH510300 itself.
\begin{python}
Initial Price:        2.604
Final Price:          7.817242080105559
RR of whole period:   3.0020130875981406
RR/year:              1.1241135785454597
RR of year-1: 0.957757296466974
RR of year-2: 0.968041542757221
RR of year-3: 2.349213742419991
RR of year-4: 0.9352175867671978
RR of year-5: 1.0816596789654878
RR of year-6: 0.8618010877268545
RR of year-7: 1.1102585961920999
RR of year-8: 1.2471590909090908
RR of year-9: 1.1091469535133192
Total number of buy count:  5
MAX rate:  2.349213742419991  MIN rate:  0.8618010877268545
MDD:  0.3526473187294988
SR:  0.7003306913660954
IR:  0.3098106077858669
\end{python}
Different from the results of the Two-Average strategy or the Keltner strategy, we find the AMA's for the EMA (Matype=1 in the titles of Figure~\ref{fig:sh510300-bollinger-ama-12}) work better than that for the SMA (Matype=2). However, the AMA from SMA seems promising since it outputs more acceptable results. This is again partly because the AMA from SMA has not been extensively used in quantitative strategies till now.

\section{MACD Strategy}
\subsection{Moving Average Convergence Divergence (MACD)}
The moving average convergence divergence (MACD) is the difference between two EMAs (one with a short time period, and one with a long time period). The signal line is simply an SMA (or EMA) of the MACD.
The MACD was developed by Gerald Appel \citep{appel2007understanding, appel2008quick} and the formula is given by:
\begin{equation}
\left\{\begin{aligned}
	\text{MACD} &= \text{EMA}(\text{TimeperiodShort})- \text{EMA}(\text{TimeperiodLong}); \\
	\text{MACDsignal} &= \text{SMA}(\text{MACD, TimeperiodSignal});\\
	\text{MACDhist} &= \text{MACD} - \text{MACDsignal}.
\end{aligned}
\right.
\end{equation}
The parameters for the MACD of ``TimeperiodShort", ``TimeperiodLong", and ``TimeperiodSignal" are often given as 12, 26, and 9 respectively by default. 

The MACD signals trend changes and indicates the start of a new trend direction. High values indicate overbought conditions, low values indicate oversold conditions. Divergence with the price indicates an end to the current trend, especially if the MACD is at extreme high or low values. When the MACD line crosses above the signal line a buy signal is generated. When the MACD crosses below the signal line a sell signal is generated. To confirm the signal, the MACD should be above zero for a buy, and below zero for a sell.

\begin{figure}[H]
	\centering  
	\vspace{-0.35cm} 
	\subfigtopskip=2pt 
	\subfigbottomskip=2pt 
	\subfigcapskip=-5pt 
	\subfigure[S\&P500 with TimeperiodLong=26, TimeperiodShort=12, TimeperiodSignal=9.]{\label{fig:sh510300-macdtest-1}
		\includegraphics[width=0.47\linewidth]{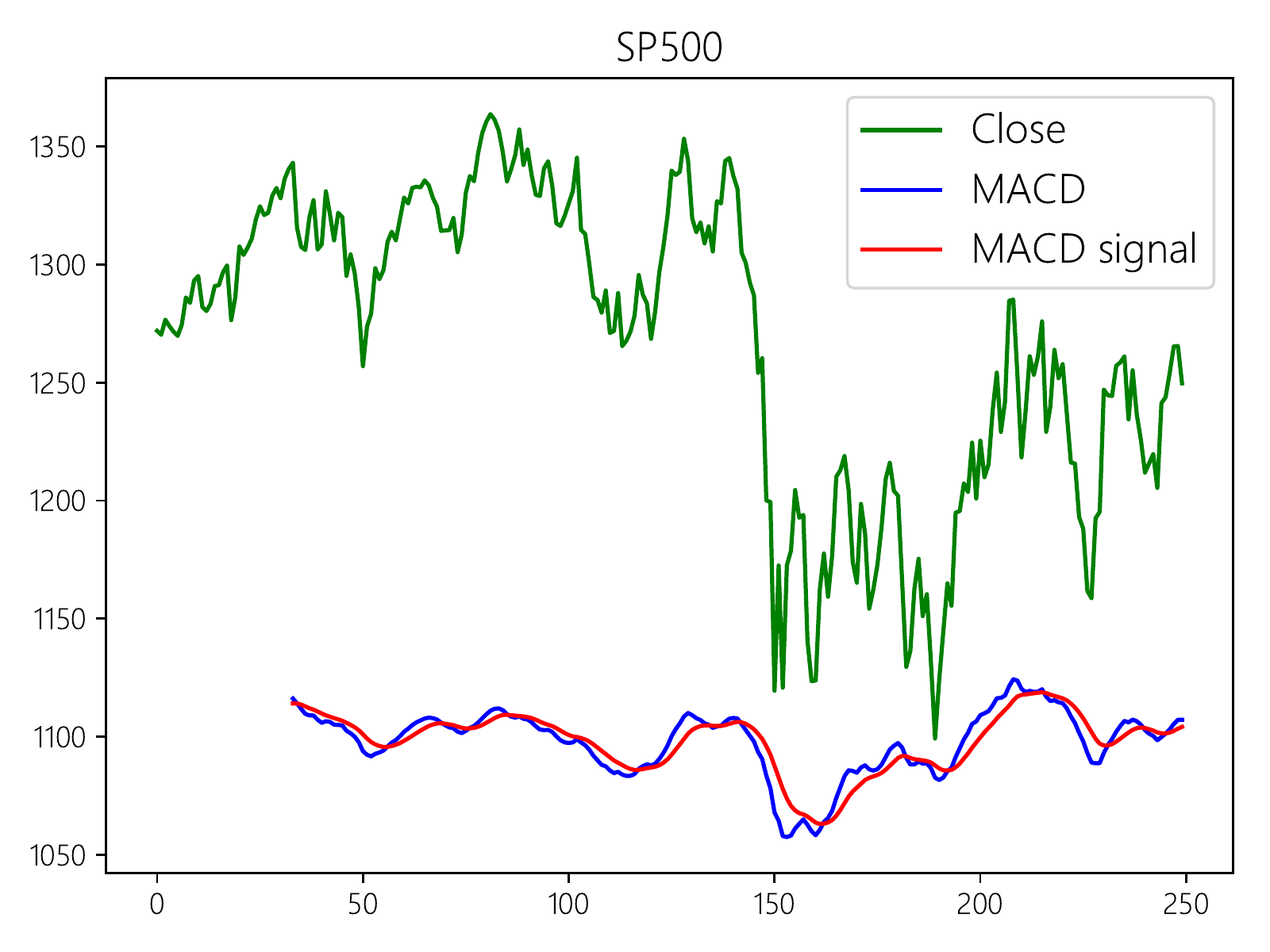}}
	\quad 
	\subfigure[SH510300 with TimeperiodLong=26, TimeperiodShort=12, TimeperiodSignal=9.]{\label{fig:sh510300-macdtest-2}
		\includegraphics[width=0.47\linewidth]{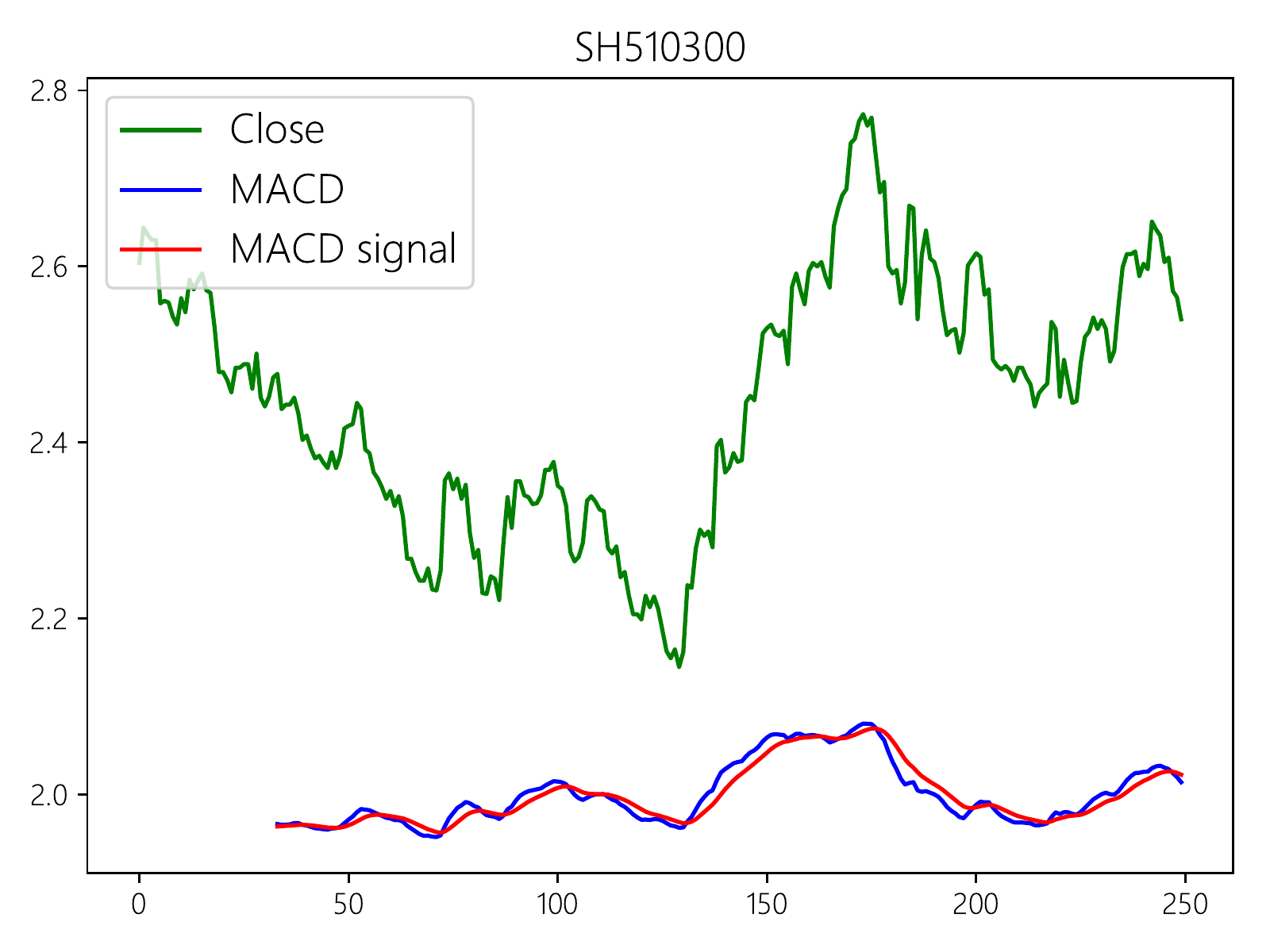}}
	\caption{MACD figures for S\&P500 and SH510300. Note the value of MACD and MACDsignal are usually around 0. The above lines are shifted for comparing.}
	\label{fig:sh510300-sp500-macd-ama-12}
\end{figure}

\paragraph{The strategy} Similar to the Two-Average strategy, when the MACD crosses above the MACDsignal, buy; when the MACD crosses below the MACDsignal, sell. Figure~\ref{fig:sh510300-sp500-macd-ama-12} shows the MACD and MACDsignal for S\&P500 and SH510300 in a 250 days frame. A clear implication of the MACD strategy can be observed (after a moment of reflexion).

\subsection{Results}

\begin{figure}[H]
	\centering  
	\vspace{-0.35cm} 
	\subfigtopskip=2pt 
	\subfigbottomskip=2pt 
	\subfigcapskip=-5pt 
	\subfigure[TimeperiodLong=4, TimeperiodShort=2, TimeperiodSignal=18.]{\label{fig:sh510300-macd-ama-1}
		\includegraphics[width=0.47\linewidth]{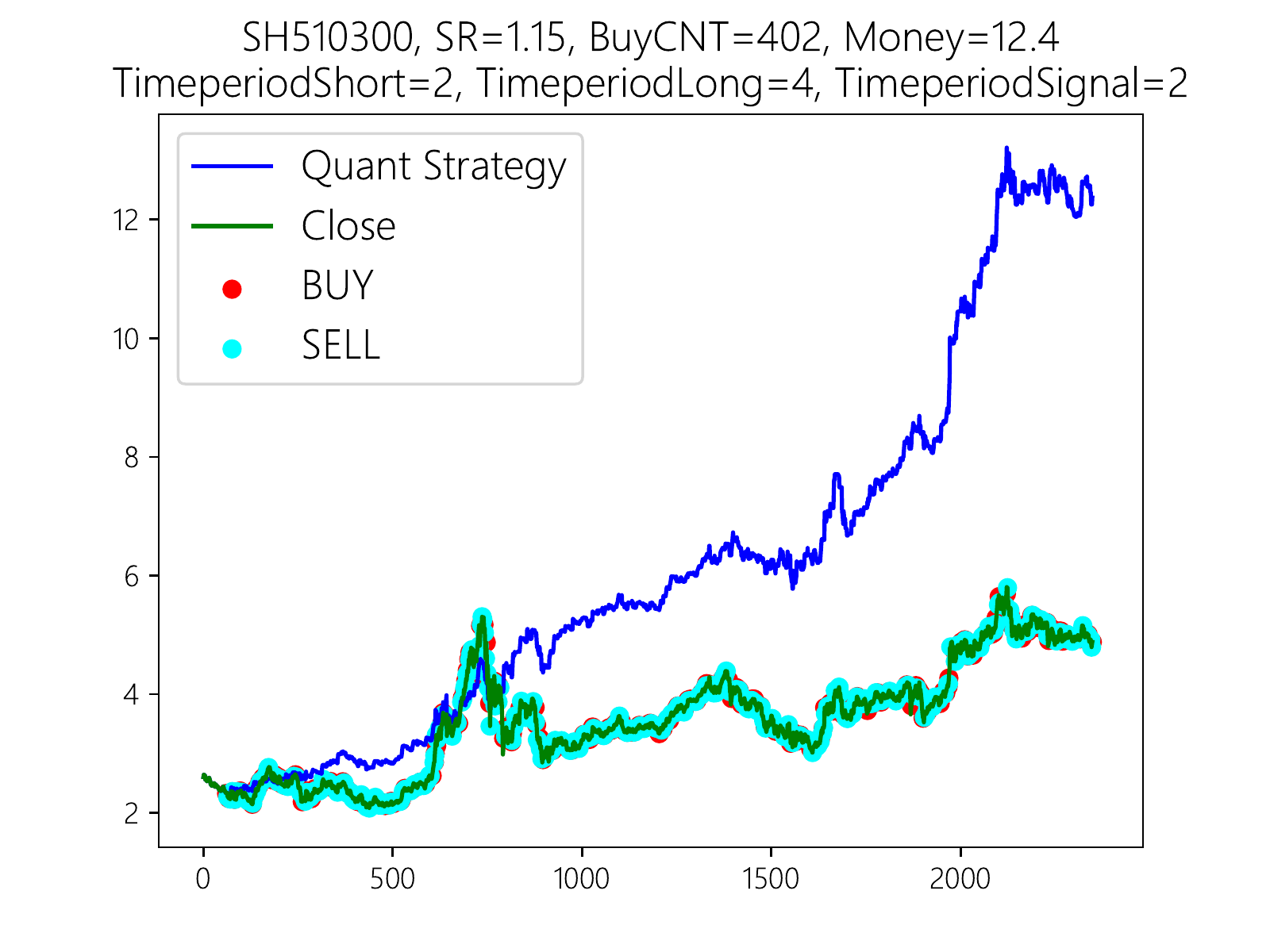}}
	\quad 
	\subfigure[TimeperiodLong=10, TimeperiodShort=4, TimeperiodSignal=18.]{\label{fig:sh510300-macd-ama-2}
		\includegraphics[width=0.47\linewidth]{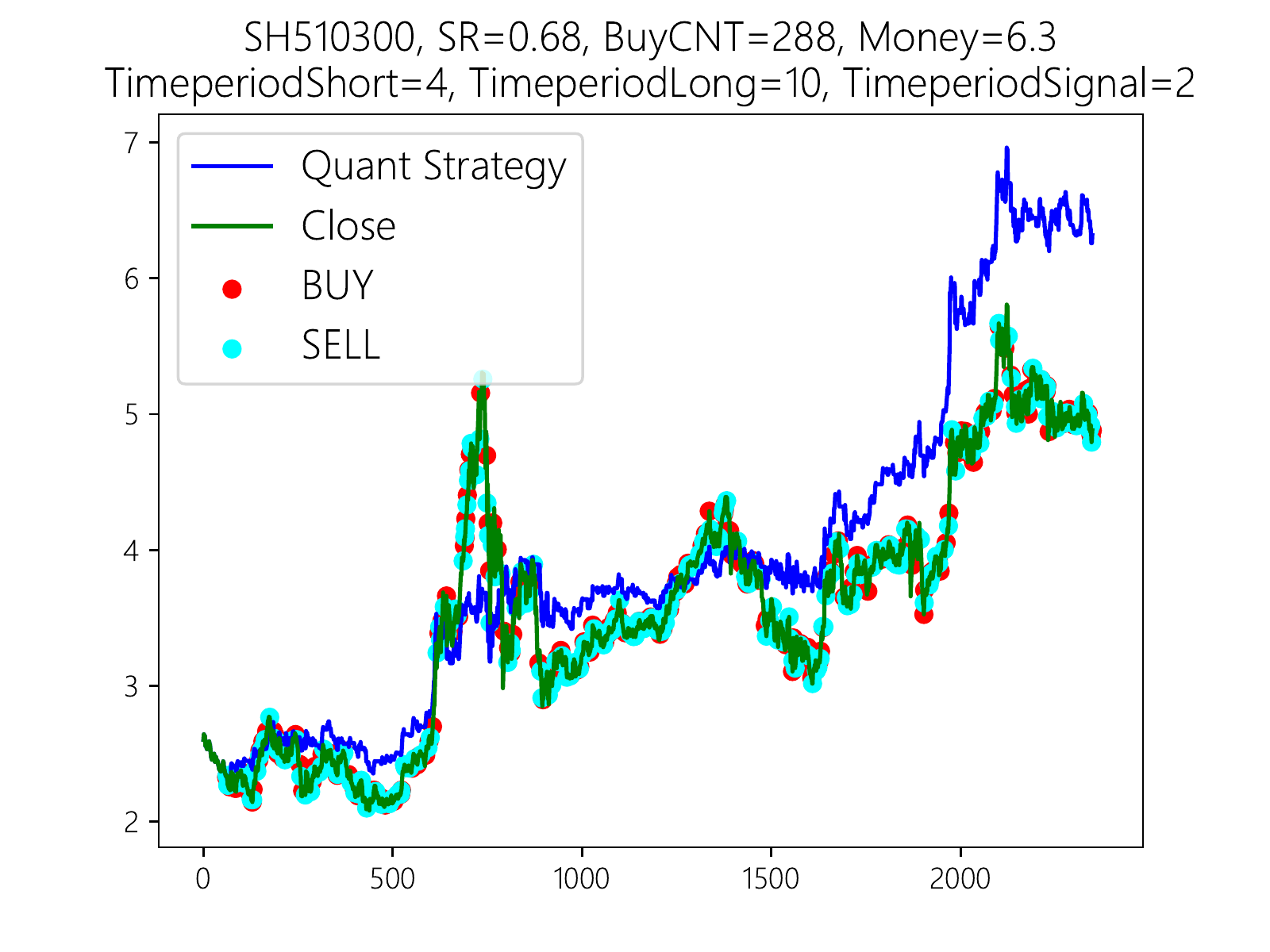}}
	\caption{MACD strategy on SH510300.}
	\label{fig:sh510300-macd-ama-12}
\end{figure}

The result of the MACD strategy on the SH510300 is shown in Figure~\ref{fig:sh510300-macd-ama-12}.
The detailed measures in Figure~\ref{fig:sh510300-macd-ama-1} are given as follows where again the ``IR" is the information ratio whose benchmark is set to be the SH510300 itself. The result shows the MACD work well with a small MDD and an acceptable SR or IR. However, the BuyCNT is large, approximately 4 times per month, which says the turnover is large. It may cause a large fee payment.
\begin{python}
Initial Price:        2.604
Final Price:          12.371646201432752
RR of whole period:   4.751016206387385
RR/year:              1.1804000407529112
RR of year-1: 1.0237116887501392
RR of year-2: 1.0612853834574043
RR of year-3: 1.4779805119148437
RR of year-4: 1.2079752530396133
RR of year-5: 1.1402069617344712
RR of year-6: 1.0224306409466652
RR of year-7: 1.1683210530919812
RR of year-8: 1.4766918267971532
RR of year-9: 1.230244980326647
Total number of buy count:  402
MAX rate:  1.4779805119148437  MIN rate:  1.0224306409466652
MDD:  0.18557388402199687
SR:  1.1494491803824245
IR:  0.46413451717265125
\end{python}

\section{Machine Learning Strategy}
One thing to notice it that, although the Aroon strategy alone is not working well, machine learning techniques such as uniform/linear blending can be employed together with other strategies to develop new algorithms \citep{sill2009feature, lu2017machine}.

Machine learning techniques for time series problems are explored in the field of airplane ticker prediction \citep{lu2017machine}. We here only briefly discuss potential methods, though we do not observe any promising results from this simple idea. However, an extensive test is not applied, and readers are recommended to explore methods in \citep{lu2017machine}, e.g., for both regression or classification ideas, for outlier removal. From the above strategies, several features can be extracted, e.g., the SMA, EMA, AMA, RSI, Aroon of each day, and even the volume or liquidity (total volumes in a specific time frame) on that day.\footnote{More features are discussed in \citep{kakushadze2016performance, kakushadze2016101, tulchinsky2019finding, kakushadze2018151} and references therein.} These features can be treated as inputs of the machine learning ``blackbox". For the output, considering the classification problem, it is reasonable to set the output of that day as 1 (i.e., a buy signal) if the price can achieve a rate of return of $1\%$ (just an example here) in one of the next 5 trading days (again, 5 is just an example); and set the output to be 0 if otherwise (i.e., a sell signal). Then different from the above strategies, the training set can be selected as the first 10 years for the S\&P500 data; and the validation set can be chosen as the last year. In practice, other features or indicators are required to obtain a promising result and we shall not give further details.

\newpage
\vskip 0.2in
\bibliography{bib}

\end{document}